\documentclass[fleqn,usenatbib]{mnras}
\usepackage{newtxtext,newtxmath}
\usepackage[T1]{fontenc}
\usepackage{graphicx}	
\usepackage{amsmath}	
\usepackage{array}

\title{Analysis of physical processes in eruptive YSOs with near infrared spectra and multi-wavelength light curves}

\author[Z. Guo et al.]{Zhen Guo$^{1}$\thanks{E-mail: z.guo4@herts.ac.uk},
P. W. Lucas$^{1}$,
C. Contreras Pe{\~n}a$^{2,1}$,
L. C. Smith$^{3}$,
C. Morris$^{1}$,
\newauthor R. G. Kurtev$^{4,6}$,  J. Borissova$^{4,6}$, J. Alonso-Garc\'{i}a$^{5,6}$, D. Minniti$^{7,8}$, A.-N. Chen{\'e}$^9$
\newauthor M. S. N. Kumar$^{10}$, A. Caratti o Garatti$^{11}$, D. Froebrich$^{12}$, and W. H. Stimson$^{1}$.
\\
$^{1}$Centre for Astrophysics Research, University of Hertfordshire, Hatfield AL10 9AB, UK\\
$^{2}$Department of Physics and Astronomy, University of Exeter, Stocker Road, Exeter, Devon EX4 4SB, UK\\
$^{3}$Institute of Astronomy, University of Cambridge, Madingley Road, Cambridge, CB3 0HA, UK\\
$^{4}$Instituto de F{\'i}sica y Astronom{\'i}a, Universidad de Valpara{\'i}so, ave. Gran Breta{\~n}a, 1111, Casilla 5030, Valpara{\'i}so, Chile\\
$^{5}$Centro de Astronom{\'i}a (CITEVA), Universidad de Antofagasta, Av. Angamos 601, Antofagasta, Chile\\
$^{6}$Millennium Institute of Astrophysics, Santiago, Chile\\
$^{7}$Departamento de Ciencias Fisicas, Universidad Andres Bello, Republica 220, Santiago,
 Chile\\
$^{8}$Vatican Observatory, V00120 Vatican City State, Italy\\
$^{9}$Gemini Observatory/NSF’s NOIRLab, 670 N. A`ohoku Place, Hilo, Hawai'i, 96720, USA\\
$^{10}$ Instituto de Astrof{\'i}sica e Ci{\^e}ncias do Espaco, Universidade do Porto, CAUP, Rua das Estrelas, 4150-762 Porto, Portugal\\
$^{11}$Dublin Institute for Advanced Studies, School of Cosmic Physics, Astronomy and Astrophysics Section, 31 Fitzwilliam Place, Dublin 2, Ireland\\
$^{12}$Centre for Astrophysics and Planetary Science, School of Physical
Sciences, University of Kent, Canterbury CT2 7NH, UK
}

\date{Accepted XXX. Received YYY; in original form ZZZ}
\pubyear{2021}

\begin{document}
\label{firstpage}
\maketitle

\begin{abstract}

The decade-long Vista Variables in the Via Lactea (VVV) survey has detected numerous highly variable young stellar objects (YSOs). We present a study of 61 highly variable VVV YSOs ($\Delta K_s = 1$--5~mag), combining near infrared spectra from Magellan and VLT with VVV and NEOWISE light curves to investigate physical mechanisms behind eruptive events. Most sources are spectroscopically confirmed as eruptive variables (typically Class I YSOs) but variable extinction is also seen. Among them, magnetically controlled accretion, identified by H{\sc i} recombination emission (usually accompanied by CO emission), is observed in 46 YSOs. Boundary layer accretion, associated with FU~Ori-like outbursts identified by CO overtone and H$_2$O absorption, is observed only in longer duration events ($\ge$5~yr total duration). However, even in long duration events, the magnetically controlled accretion mode predominates, with amplitudes similar to the boundary layer mode. Shorter (100--700~days) eruptive events usually have lower amplitudes and these events are generally either periodic accretors or multiple timescale events, wherein large photometric changes occur on timescales of weeks and years. We find that the ratio of amplitudes in $K_s$ and $W2$ can distinguish between variable accretion and variable extinction. 
Several YSOs are periodic or quasi-periodic variables. We identify examples of periodic accretors and extinction-driven periodicity among them (with periods up to 5~yr) though more data are needed to classify some cases.  The data suggest that dynamic interactions with a companion may control the accretion rate in a substantial proportion of eruptive systems, although star-disc interactions should also be considered. 

\end{abstract}

\begin{keywords}
stars: pre-main sequence -- stars: protostar -- stars: variables: T Tauri -- infrared: stars -- stars: Wolf-Rayet -- stars: AGB
\end{keywords}

\section{Introduction}
\label{sec:intro}

Pre-main-sequence (PMS) stellar evolution is shaped by the mass accretion process \citep[see review by][]{Hartmann2016}. As a departure from the steady mass accretion scenario \citep[e.g.][]{Shu1977}, episodic accretion during the PMS evolution was originally proposed to solve the luminosity problem \citep{Kenyon1990, Hartmann1996}, specifically the under-luminosity of low mass Class I YSOs. The mass accretion rate of a YSO would be enhanced by up to 4 orders of magnitudes for relatively short periods \citep{Dunham2012, Audard2014} and the luminosity would be lower at other times. Theoretical models explored several physical mechanisms that might trigger episodic accretion, including outside-in thermal instability introduced by a massive planet \citep{Lodato2004, Clarke2005}, imbalance between accretion controlled by gravitational and magneto-rotational instabilities at different radii within the disc \citep{Zhu2009b}, disc fragmentation \citep{Vorobyov2005, Vorobyov2010} or piling up of disc materials outside the star-disc co-rotation radius \citep{DAngelo2010, DAngelo2012}.  

Photometric variations related to episodic accretion have been observed among Class I and Class II objects with a wide range of amplitudes and timescales \citep[see review by][and references therein]{Audard2014}. In general, most eruptive YSOs were sorted into two groups, EXors and FUors. EXors have amplitudes up to $\Delta V = 5$~mag \citep{Lorenzetti2009} on timescales of a few hundred days, and some of them have repeated outbursts \citep{Herbig1989, Herbig2008, Audard2010}. Spectroscopic studies revealed that EXors are undergoing the magnetospheric accretion process with indicators like broad H{\sc i} recombination emission lines \citep[e.g.][]{Lorenzetti2009}. FUors typically have similarly high amplitudes but much longer-lasting outbursts, with a rapid rise ($\sim$100 days) followed by a century-long decay \citep{Hartmann1996, Kenyon2000}.  Most FUor outbursts have higher luminosity than EXors with high mass accretion rate ($10^{-5}$ -- $10^{-4}$ $M_{\odot}{\rm yr}^{-1}$). The mass accretion process of FUors is not controlled by the stellar magnetic field. As a consequence, their near infrared spectra are dominated by broad H$_2$O and CO absorption bands; Br$\gamma$ emission is not seen and the Paschen lines are often detected in absorption \citep{Connelley2018}. As the prototype of FUors, an extended self-luminous disc is found around FU~Ori, as a consequence of inefficient radiative transfer throughout the geometrically thick disc \citep{Zhu2007, Zhu2009, Hartmann2011}. 

Recently, more eruptive YSOs were discovered not belonging to the EXor and FUor categories. A few YSOs show intermediate amplitude and timescale, while having similar spectral features to FUors, e.g. HBC 722 \citep{Kospal2016} and V371~Ser \citep{Hodapp2012}. On the other hand, some eruptive YSOs with emission-line spectra have variation timescales longer than typical EXors, for example V1647 Ori \citep{Aspin2009} and Gaia19bey \citep{Hodapp2020}. The YSO V346~Nor has unique characteristics, with a FUor-like outburst duration but an outflow-dominated spectrum \citep{Kospal2020}. A more complete sample of eruptive objects is necessary to draw a whole picture of the episodic accretion process throughout the PMS evolution.

The VISTA Variables in the Via Lactea (VVV) survey has imaged the southern Galactic plane and the bulge in the near-infrared at several dozen epochs since 2010 \citep{Minniti2010, Saito2012}. The initial VVV survey was extended as ``{\it VVVX}" in 2016 \citep{Minniti2016}, so $K_s$ (2.15~$\mu$m) light curves of $10^9$ targets have been observed on a timescale of a decade. A series of works have been carried out to study eruptive YSOs in the VVV field\footnote{The VVV field covers the Galactic bulge ($-10^{\circ} < b < 5^{\circ}$) and disc ($|b| < 2^{\circ}$) with longitude $-65^{\circ}< l <10^{\circ}$. {\it VVVX} also covers a large new area but that does not form part of this study}. \citet[][hereafter Paper I]{Contreras2017}, discovered 816 high amplitude variables from the 2010-2014 VVV $K_s$-band light curves ($\Delta K_s > 1$~mag). This work showed that eruptive variability is at least an order of magnitude more common in Class I YSOs than Class II YSOs, see also \citet{Contreras2019}. The spectroscopic follow-up work \citep[][hereafter referred to as Paper II]{Contreras2017b} found that most Class I eruptive variables have outbursts of intermediate duration, typically 1--5~yr, whilst the nature of the spectra (EXor-like emission or FUor-like absorption) in many cases did not fit expectations based on the duration of classical FUor and EXor outbursts in optically bright systems. Optically detected FUors and EXors have Class II or flat-spectrum SEDs suggesting a more evolved nature, though some appear to have massive circumstellar envelopes despite this \citep{Green2013, Feher2017}. 
A new category of eruptive YSOs ``MNors'' was tentatively proposed for these systems having intermediate outburst duration or inconsistent spectroscopic and time domain characteristics according to the classical FUor/EXor scheme. This was a broad classification, including not only sources with ``eruptive'' light curves but also some that showed evidence for long period accretion-driven variability, dubbed ``LPV-YSOs''. \citet[][Paper III]{Guo2020} studied the near-infrared spectroscopic variation of 14 eruptive objects from the Paper II sample, on day-to-day and year-to-year timescales. Both variable mass accretion and variable line-of-sight extinction were found on long-term eruptive Class I objects. In Paper III, we also defined a new type of variable YSO, the ``multiple timescale variable'' (MTV), wherein the amplitudes on short timescales ($t <$~30~d) are comparable to the slower inter-year variations. MTVs have emission line spectra, consistent with the magnetospheric accretion scenario. The physical mechanisms of these month-long variations are likely due to unstable accretion, but this is unclear at present.
 
In this work, we conducted near-infrared spectroscopic follow-up of 38 highly variable YSO candidates from the VVV survey with a variety of light curve types. When added to the previous sample of Paper II, the combined sample size is 61 spectroscopically confirmed YSOs. Here, this large spectroscopic sample, dominated by eruptive YSOs, is used to constrain/infer the accretion mode, measure accretion luminosity and the accretion burst duration, thereby allowing us to study the detailed nature of episodic accretion in embedded protostars. We classify the YSOs in various sub-types via their spectroscopic and multi-wavelength time domain properties. The observational behaviours and underlying physical processes are then further investigated.

This paper is organised as follows: Target selection methods, observational information, and data reduction routines are described in \S\ref{sec:info}. Spectroscopic and photometric results and target classifications are shown in \S\ref{sec:res}. In \S\ref{sec:dis}, we discuss variation mechanisms by studying multi-wavelength light curves, then take a statistical view of eruptive YSOs in the VVV survey and give details of a few individual objects. A conclusion is presented in \S\ref{sec:con}.

\begin{table*} 
\caption{Basic information for all observed targets}
\renewcommand\arraystretch{0.9}
\begin{tabular}{l c c c c r c c c c c c c c c}
\hline
\hline
Name & RA & Dec & $K_{\rm s}$ & $\Delta$$K_{\rm s}$  & $\Delta W2$  & $\Delta K_{\rm s, WISE}$ & $\alpha_{\rm class}$ &  \multicolumn{2}{c}{Variable star} & ${t_{\rm phot}}^c$ & Period & $d_{\rm SFR}$  \\
& J2000 & J2000 & mag & mag & mag & mag & & class$^a$ & class$^b$ & day & Y/ N & kpc\\
\hline
  v14 &  12:12:18.13 &   -62:49:04.5 & 15.56 &  1.64 & 0.44 & 1.16 &  0.61  & LPV-YSO & M & 398 &  Y & 2.5  \\
  v16 &  12:13:29.76 &   -62:41:07.7 & 13.65 &  1.66 & 0.35 & 1.66 & 0.07  & Eruptive  & L &  > 3477 & N & 9.7 \\
  v51 &  13:19:42.87 &   -63:01:01.9 & 14.56 &  2.93 & 0.60 & 2.71  & -  & AGB  & L & 1000 & N & - \\
  v53 &  13:27:02.40 &   -63:06:22.5 & 14.56 &  2.42 & 1.29 & 1.66  &  0.61  & Eruptive & M & 236 & N & - \\
  v84 &  14:09:11.55 &   -61:32:24.3 & 11.99 &  2.76 &  -   &  -    & -   & Nova  & L &  > 500 & N & - \\
 v128 &  14:58:29.67 &   -59:09:40.3 & 14.41 &  2.54 & 0.83 & 1.79 &  0.56  & Eruptive  & L & > 1132 & N & 2.6 \\
 v181 &  15:46:39.17 &   -55:50:28.3 & 12.76 &  3.47 & 1.58 & 1.41  &  0.58 & Eruptive   & M &  183.9 & Y? & - \\
 v190 &  15:47:03.56 &   -54:43:10.2 & 13.53 &  1.44 & 0.67 & 0.80  & -0.43 & LPV-YSO & L & 612 & Y & 5.0 \\
 v237 &  16:10:48.22 &   -51:42:45.0 & 13.95 &  1.92 & 0.16 & 0.91  &  1.02 & Dipper? & L & 1943 & N & 4.2 \\
 v309 &  16:40:58.19 &   -47:06:31.9 & 13.56 &  2.60 & 0.28 & 1.17  &  1.30 & LPV-YSO? & M & 181 & Y? & 12.2 \\
 v319 &  16:45:17.04 &   -46:05:55.4 & 11.66 &  1.56 & 0.75 & 0.79  & -  & Symbiotic & L & 307 & Y & - \\
 v335 &  16:52:09.73 &   -45:52:49.6 & 13.00 &  3.63 & 0.74 & 1.64  &  0.45 & Dipper  & L &   1041 & N & 2.3 \\
 v370 &  16:57:48.65 &   -43:04:42.5 & 11.82 &  0.94 & 0.34 & 0.62  &  - & Symbiotic & L &   419 & Y & - \\
 v371 &  16:59:50.04 &   -43:17:21.6 & 14.56 &  2.09 &  -   &  -    & - & Wolf-Rayet & M &   193 & Y & - \\
 v376 &  16:58:44.44 &   -42:47:36.6 & 12.49 &  1.71 & 0.62 & 0.81  & -0.37 & Fader  & M & >1300 & N & - \\
 v389 &  17:03:17.18 &   -42:25:49.9 & 13.59 &  1.97 & 0.56 & 1.57 &  0.69  & Eruptive? & M & 1036 & N & 2.2 \\
 v467 &  13:01:13.23 &   -62:25:27.9 & 12.79 &  2.64 & 0.85 & 2.37  & -0.24  & LPV-YSO &  L & 1840 & Y & 4.5 \\
 v618 &  15:42:54.67 &   -55:00:52.8 & 12.13 &  1.33 & 0.60 & 0.94  & -0.09 & STV & M & 140 & N & - \\
 v621 &  15:43:12.04 &   -54:23:08.8 & 15.45 &  1.91 & 0.51  & 0.25  & -0.05  & Eruptive & L & > 2898 & N & 6.0 \\
 v636 &  15:51:46.28 &   -53:25:57.3 & 13.60 &  1.43 & 0.62 & 0.96  &  0.43 & LPV-YSO & M &  565 & Y & 2.8 \\
 v713 &  16:33:52.79 &   -46:52:18.5 & 11.75 &  2.03 & 1.60 & 2.03  & -0.04  & Fader & L &  > 1130 & N & 3.2 \\
 Stim1 &  12:57:44.23 &   -62:15:06.4 & 14.20 &  2.45 & 1.70 & 2.45  &  0.63 & Eruptive & L & > 3452 & N & - \\
 Stim5 &  16:12:14.38 &   -51:50:24.6 & 12.78 &  3.81 & 1.71 & 3.81 &  0.9 & Eruptive & L & > 3452 & N & 2.4 \\
 Stim13 &  16:19:10.80 &   -51:03:53.0 & 12.39 &  1.51 & 0.20 & 0.76 & 1.39  & Dipper & L & 1588 & N & 3.1 \\
 DR4\_v5  &  13:29:26.28 & -62:23:26.5 & 14.11 &  3.96 & 0.31 & 0.81 &  0.95  & Dipper & L &  > 1230 & N & - \\
 DR4\_v10  &  14:25:13.98 & -60:20:20.0 & 15.67 &  3.56 & 0.79 & 1.57  &  1.86 & Eruptive & L  &  > 3437 & N & 4.6 \\
 DR4\_v15  &  15:07:11.11 & -58:50:32.9 & 17.41 &  3.60 & 1.41 & 2.15  &  0.56 & Eruptive & L &  > 3456 & N & 4.3 \\
 DR4\_v17  &  15:09:35.66 & -57:35:22.6 & 13.69 &  3.83 & 1.10 & 2.80  &  0.79  & LPV-YSO & L  & 1155 & Y & 4.0 \\
 DR4\_v18  &  15:30:17.93 & -55:34:55.1 & 13.19 &  4.52 & 2.22 & 4.37  & 0.21 & Dipper & L &  > 796 & N & - \\
 DR4\_v20  &  15:44:26.03 & -54:01:38.4 & 17.03 &  3.35 & 0.50 & 0.68  &  0.74  & Eruptive & L & > 2913 & N & 2.5 \\
 DR4\_v30  &  16:22:40.18 & -49:06:26.4 & 13.68 &  3.88 & 1.37 & 2.14  &  0.36  & STV or Eruptive & M & 30 & N & - \\
 DR4\_v34  &  16:29:06.99 & -48:51:16.9 & 14.11 &  3.57 & 0.44 & 1.04  & -0.19  & Eruptive & L & > 2904 & N & 3.2 \\
 DR4\_v39  &  16:46:30.13 & -46:04:39.7 & 13.56 &  4.76 & 1.06 & 2.01  &  0.57 & Dipper & L  & 1445 & N & 3.0  \\
 DR4\_v42  &  16:50:14.77 & -44:03:30.7 & 15.74 &  3.36 & 1.26 & 2.38 &  1.02  & Eruptive & L &  > 900 & N & 5.1 \\
 DR4\_v44  &  16:52:04.42 & -43:33:26.0 & 15.98 &  3.57  & 1.04 & 1.46 &  0.83  & Eruptive & L &  > 2921 & N &  3.5 \\
 DR4\_v55  &  17:29:02.42 & -34:00:36.2 & 13.87 &  3.59 & 2.57 & 2.93  &  1.92  & LPV-YSO & L & 954 & Y &  3.4 \\
 DR4\_v67 &  17:41:31.14 & -31:26:12.1 & 14.03 &  5.30 & 1.45 & 3.37 &  1.05  & Eruptive & L & 2572 & N & 4.5 \\
 DR4\_v89 & 17:50:50.54 & -29:10:48.7 & 14.23 &  3.04 & 0.23 & 0.85  &  0.85  & Eruptive & M & 710 & N & 2.3  \\
 \hline
\hline
\end{tabular}

\label{tab:info}
\flushleft{$a$: Variable classification based on light curves and spectral features. YSOs are classified by categories defined in Paper I . LPV and STV stand for long period variable and short-term variable, respectively. Uncertain classifications are marked with ``?''.\\
$b$: ``M'' as MTV defined in Paper III. MTVs have comparable or larger variation on short timescales (within 1~year) than the inter-year variation.\\ ``L'' represents objects with long term variation.  \\
$c$: $t_{\rm phot}$ represents the typical timescale of $K_s$-band variation. Definition in \S\ref{sec:lc}. \\
 $d_{\rm SFR}$: literature-based distance of star forming regions within 5 arc minutes. \\
The $\alpha_{\rm class}$ of v621 is fit between 2 to 5 $\mu$m.\\}
\end{table*}

\section{Observation and data reduction}
\label{sec:info}
\subsection{Target selection and basic information}
\label{sec:target}
In this paper, 38 targets were selected for follow-up in our 2017 and 2019 spectroscopic campaigns based on their variability in $K_s$ in the VVV survey and evidence for star formation activity in the surrounding field. For the 2017 observations, 24 objects were selected from our previous study of the VVV light curves from 2010 to 2014 (Paper I), which had provided a list of 816 near-infrared variable stars with $\Delta K_s > 1$~mag. These objects were sufficiently bright ($K_s \le 14.5$) in 2017 to allow good quality data to be obtained at intermediate resolution in brief observations. They were chosen to encompass a range of the light curve types defined in Paper I: ``eruptive'', ``dipper'', ``fader'', and ``LPV-YSO'' (long period variable YSO).  This broadens the selection compared to the sample in Paper II, which was heavily weighted towards YSOs with eruptive light curves. Short term variables (``denoted STV'', whether periodic or not) were not selected, these almost always having $\Delta K_s \le 1.5$~mag in Paper I. However, v618, a source previously classified as ``LPV-YSO'' is reclassified as ``STV''. Herein we use a single ``v'' prefix for 21 of these 24 sources, as opposed to ``VVVv'' in Paper I. The remaining three have the prefix ``Stim'', denoting the discoverer W. Stimson, a previous group member in the University of Hertfordshire who identified a total of 14 highly variable stars during completeness testing in Paper I, that were not included in the published list. We give details of these sources in Appendix A. 

For the 2019 campaign, another 15 targets, mostly having eruptive light curves, were selected from a set of  27 YSO candidates found in a search of the VVV 4th Data Release (available at \url{http://vsa.roe.ac.uk}) for high amplitude variable stars of all types with $\Delta K_s > 3$~mag. The 15 targets selected were those for which the available photometry indicated that they would likely be in a bright state during the 2019 spectroscopic campaign, or at least an intermediate state. Fourteen targets have a ``DR4\_v'' prefix, following Lucas et al., (in prep). One object ``Stim5'' also appears in the previous selection and it was observed in both the 2017 and 2019 campaigns, providing a helpful comparison of results from the two different telescopes used.

The light curve classifications of all 38 targets are refined using light curves drawn from the VVV/VIRAC2 catalog, \citep[][Smith et al. in prep]{Smith2018}. For YSOs we follow the method described in Paper I, with the addition of the MTV category (see \S\ref{sec:intro}) More information on light curve analysis is shown in \S\ref{sec:lc}. Information on the targets, including coordinates, mean $K_s$ magnitudes,  variation amplitudes, light curve classifications, and typical duration of variations are listed in Table~\ref{tab:info}.

The evolutionary stages of YSOs are classified by the slope between the near- and mid-infrared spectral energy distributions (SED), as $\alpha_{\rm class}$ defined by $d({\rm log}F_\lambda)/d({\rm log}\lambda)$ \citep{Lada1987}. YSOs at early evolution stages (Classes 0 \& I) have $\alpha_{\rm class} > 0.3$ while Class II objects with a thick disc have $\alpha_{\rm class} < -0.3$. In between, YSOs with  $ -0.3 < \alpha_{\rm class} < 0.3$ are defined as ``flat-spectrum'' objects, as a transitional stage from Class I to Class II \citep{Greene2002, Grossschedl2019}. In this work, the infrared fluxes from 2 to 24 $\mu$m are drawn from public surveys including the $K_s$ bandpass data from the VVV/VIRAC2, IRAC $I1$ to $I4$ data from the {\it Spitzer}/GLIMPSE survey catalog \citep{Benjamin2003}, MIPS24 data from the {\it Spitzer}/MIPSGAL survey \citep{Carey2009, Gutermuth2015} and $W1$ to $W4$ bandpass data from the AllWISE data release of the {\it WISE} survey \citep{Wright2010}. {\it WISE} and {\it Spitzer} data are obtained via online catalogs at the NASA/IPAC Infrared Science Archive ({\it IRSA}\footnote{\url{http://irsa.ipac.caltech.edu}}). In the Galactic plane, WISE photometry in $W3$ (12~$\mu$m) and $W4$ (22~$\mu$m) may be seriously compromised by the bright non-uniform background in star formation regions. Genuine sources detected in $W1$ and $W2$ can have spurious detections in $W3$ and $W4$ due to forced photometry at the same locations. To guard against this, visual inspection was performed on the Spitzer Enhanced Imaging Products (SEIP) images and the {\it unWISE} \citep{Meisner2017} images. In a few cases, where there is no mid-infrared detection in the $W4$ or MIPS~24$\mu$m passbands, $\alpha_{\rm class}$ derived from an SED fit between 2 and 12~$\mu$m.

For measurement of $\alpha_{\rm class}$, contemporaneous $K_s$ and WISE photometry are strongly preferred. To complement the {\it WISE} photometry obtained in 2010, we only use VVV $K_s$ data from 2010 to fit the SED slope. A comparison between $K_s$ and $W2$ amplitudes will be discussed in \S\ref{sec:neowise}. The measured $\alpha_{\rm class}$ values for YSOs are listed in Table~\ref{tab:info}. Two objects, v84 and v621, lack archival $W3$ and $W4$ photometric detections. We visually inspected images from SEIP and ALLWISE and no source is seen at the given coordinates. From $K_s$ to $W2$ and/or from {\it Spitzer} $I1$ to $I4$ bands, both objects have a steeply declining SED toward longer wavelengths, in line with Class III objects or  main-sequence stars. Based on the $\alpha_{\rm class}$ measurement, 73\% (24/33) of our targets are Class I objects, 21\%  (7/33) are flat-spectrum objects and 6\% (2/33) are disk-bearing Class II YSOs. Another 5 sources were classified as post-main-sequence objects based on their spectral features (see~\S\ref{sec:specclass}).

\begin{figure} 
\includegraphics[width=3.2in,angle=0]{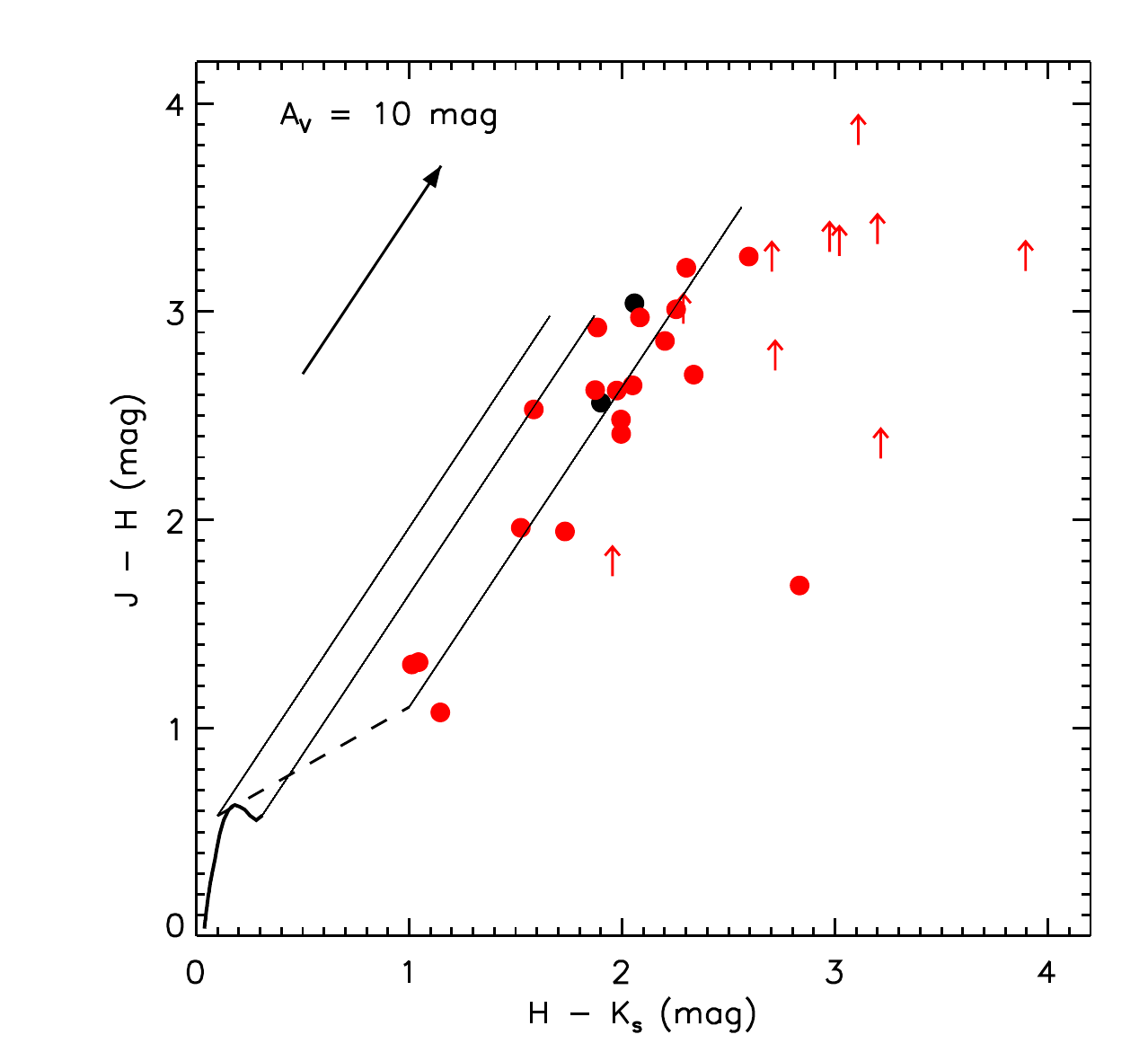}
\caption{Colour-colour diagram for sources with $H$ and $K$ magnitudes in multi-band epochs of the VVV/VVVX survey. Objects lack of $J$ detections, marked by upward arrows, are artificially set $J$ as 20 mag. Class I and flat-spectrum objects identified by $\alpha_{\rm class}$ are shown by the red dots. Dashed line is the intrinsic locus of Classical T Tauri Stars (CTTS, or Class II objects) from \citet{Meyer1997}. The solid curve represents the empirical colour of main sequence stars \citep{Pecaut2013}. The solid straight lines are reddening vectors assuming $R_V = 3.1$ \citep{WangS2019}.}
\label{fig:ccd}
\end{figure}

There are two facts for readers to note here. First, targets in our work lie in the Galactic mid-plane and so their $\alpha_{\rm class}$ are slightly biased towards high values due to foreground extinction. Second, the variable nature of our targets may also bias the measurement of the $\alpha_{\rm class}$ value. For example, an ongoing near-infrared outburst would reduce $\alpha_{\rm class}$ if the energy is not efficiently reprocessed at far infrared wavelengths. Conversely, a dimming due to extinction would cause a higher $\alpha_{\rm class}$ value. Since most of the eruptive systems were in a quiescent state in 2010, the measurement of $\alpha_{\rm class}$ for those objects is still a useful indicator of evolutionary stage. Similarly, the dippers (whose classification suggests an extinction event) underwent their dips after 2010. However, as we will see, the spectra are typically disc-dominated and show H$_2$ emission lines (an indicator of Class I status, \citealt{Greene1996}). This implies an earlier stage of evolution than classical EXors for example. A near-infrared colour-colour diagram is shown in Figure \ref{fig:ccd}.  The $J$-, $H$- and $K_s$ magnitudes are obtained from several multi-band photometric epochs from VVV/VVVX. For some eruptive systems, the $J$ magnitude is only detected from the bright epoch in 2015 that close to the spectroscopic epochs. To keep consistency, we hereby favoured the bright epochs for sources with more than one available epochs. For the sources lacking a $J$-bandpass detection, the $J$ magnitude is artificially set to 20, the typical detection limit of VVV. These sources are marked by upward-pointing arrows in Figure~\ref{fig:ccd}. Through comparison with the location of reddened classical T Tauri stars \citep{Meyer1997}, we see that most of Class I and flat-spectrum YSOs ($\alpha_{\rm class} > -0.3$) in this work have a $K_s$-bandpass excess.

\subsection{Spectroscopic observations}

In the 2017 campaign, spectra of 24 targets were obtained using the Echelle mode of Folded-port Infra-Red Echellette (FIRE) spectrograph \citep{Simcoe2013} on the Magellan Telescope over two consecutive nights (May 29th to 30th) at Las Campanas Observatory, Chile. The FIRE spectrograph covers the near-infrared waveband from 0.8 $\mu$m to 2.45 $\mu$m. A slit width of either 0.6\arcsec~ or 0.75\arcsec~ was used, depending on the seeing conditions, giving resolution $R \sim 6000$ or $R \sim 4500$, respectively. The data reduction procedures are similar to those described in Paper III using the FIREHOSE V2.0 pipeline  \citep{Gagne2015} plus customised programs based on the IDL platform.

The 2019 spectroscopic campaign was conducted using the slit mode of the X-shooter spectrograph  \citep{Vernet2011} on the ESO Very Large Telescope (VLT), at Paranal Observatory, Chile. Data were taken in Service Mode (program ID 0103.C-0622(A)). Both the visible and near-infrared arms were used simultaneously during the observations and the exposure time is adjusted for individual target based on the expected brightness. In this work, we only present the near-infrared spectra since a visible spectrum was detected in only one source (DR4\_v30). A slit width of 0.6\arcsec~ was used for all targets, giving resolution $R \approx 8000$ in the $K_s$ bandpass. Blind offsetting was applied for both spectrographs for most targets since they were typically fainter than 20th magnitude in the acquisition images, taken in the $J$ bandpass for FIRE and the $z$ bandpass for X-shooter. Observation date, exposure time, and spectrograph information for individual targets are listed in Table~\ref{tab:obs}. 

\subsection{Reduction of X-shooter spectra}

Spectra were extracted by the X-shooter pipeline as constructed in the ESO Recipe Flexible Execution Workbench software \citep[{\it Reflex};][]{Freudling2013}. The data reduction workflow first creates master calibration frames, including bias, flat and dark fields. Then the location of the spectra (taken at two separate nod positions) and the wavelength solution are generated by a 2D mapping algorithm using arc lamp frames. Sky backgrounds are extracted under the STARE reduction mode of the pipeline, which provides an additional step of wavelength calibration for individual targets. We noticed a constant 83 km s$^{-1}$ shift between the wavelength solution provided by the pipeline and the wavelengths of the OH sky lines between 2.1 and 2.5~$\mu$m. This shift was manually corrected during the data reduction. The flux calibration was derived from telluric standard stars observed on the same night as the scientific targets. Telluric absorption lines are further corrected by the {\it molecfit} software \citep{Smette2015, Kausch2015}.

An accurate absolute flux calibration could not be performed because most targets were acquired by blind offsetting and because observation of these relatively bright infrared stars was queued for relatively poor seeing conditions by Paranal standards (typically 1.0--1.5\arcsec~ seeing at the zenith). We present the pipeline produced fluxes for the X-shooter spectra nonetheless. The synthesised flux of Xshooter targets through the $K_s$ filter is shown by a blue dot in the relevant panels in Figure \ref{fig:lc_sum_2}. The reader should be aware that the integrated flux may be less than the real flux due to imperfect target acquisition on the slit and changeable observing conditions. Examples of reduced spectra are presented in Figure~\ref{fig:spec_example}.

\subsection{Spectroscopic measurements and CO model}

Three principal spectroscopic features are measured in this work: the CO overtone bands beyond 2.29~$\mu$m, molecular hydrogen (H$_2$) emission lines, and the Br$\gamma$ line at 2.166~$\mu$m. Heliocentric velocities and the sun's motion relative to the Local Standard of Rest is calculated by the data reduction pipelines and are corrected before measuring spectral features. All measurements are conducted by custom-written programs in the IDL platform. The measurements of line features followed the Gaussian fitting method developed in Paper III, and the results are shown in Table \ref{tab:lines}. The typical uncertainty of the radial velocity measurement is about 1 pixel on the array detector, which corresponds to $\sim 12$ km s$^{-1}$ for both detectors. In individual sources,  H$_2$ emission, originating from the stellar wind or outflow, often has a blue-shifted radial velocity relative to the Br$\gamma$ and CO emission that trace the mass accretion process and the rotating disc. The radial velocity difference between H$_2$ and Br$\gamma$ lines reaches 100 km s$^{-1}$ in a few sources.

The $^{12}$CO overtone features are measured by fitting synthetic models to the continuum-subtracted spectra. The CO model used in this work is based on previous works (\citet{Kraus2000}, and Appendix E in Paper II). By considering the ro-vibrational energy states of the CO molecule \citep{Dunham1932, Farrenq1991}, this model calculates the spectral intensity of CO band heads for a range of temperatures and column densities. In addition, the line-of-sight Keplerian rotation velocity are now added to the previous model, for a range of disc inclination angles. The CO models are then applied to fit the first two bands  ($v = 2 - 0$ and $v = 3 - 1$). As mentioned in Paper II, observed CO profiles usually have broader band heads than the models, and high Keplerian velocity from an accretion disc was proposed as an explanation of the broadened emission profile \citep{Najita1996, Davies2010}. In this work, we noticed that the unexpected broadening of the CO emission is only seen on the first ($ v = 2 - 0 $) band for most YSOs, while the second band ($ v = 3 - 1 $) is usually well fit by the model. This broadening issue could not be fully solved by adding a more realistic disc model including the temperature gradient in the inner disc \citep[e.g.][]{Wheelwright2010, Ilee2013}. We believe that there are some missing spectral features in the first band in our model. Hence, in our analysis and discussion we use the CO model fitting only for radial velocity measurements (which are reliable because there are many well-fitted lines in the CO bands) and we do not quote the fitted temperatures and column densities.

\begin{table*}
\renewcommand\arraystretch{1.4}
\centering
\caption{Spectral Characteristics of Br$\gamma$ line, H$_2$ (2.12 $\mu$m) line and CO band heads feature.}
\begin{tabular}{l l c c  c c  c c c c}
\hline
\hline
Object & Instrument & \multicolumn{2}{c}{Br$\gamma$} & \multicolumn{2}{c}{ H$_2$ (2.12 $\mu$m) }  & \multicolumn{2}{c}{CO} & \multicolumn{2}{c}{$d_{k}$}\\
\hline
& &RV (km s$^{-1}$) & EW (\AA) & RV (km s$^{-1}$) & EW (\AA)  & RV (km s$^{-1}$) & EW (\AA) & $d_{\rm near}$ (kpc) & $d_{\rm far}$ (kpc) \\
\hline
\multicolumn{2}{l}{Young stellar objects:} \\
 v14 & FIRE &-34.1 & -5.8 $\pm$ 1.1 &    -   &      -       &    -     &        -     &         $\star$$1.88^{+1.08 }_{ -0.87}$    &  $6.09^{+0.93 }_{ -1.13}$    \\
 v16 & FIRE &  6.3 &  0.4 $\pm$ 0.2 &    -   &      -       &    2.2 &   12.5 $ \pm$   7.0  & - & $\star$$8.20^{+1.20}_{-1.11}$\\
v53 & FIRE & 93.9 & -3.6 $\pm$ 0.5 &    -   &      -       &   51.5 &  -42.6 $ \pm$   5.6 & - & $\star$$13.90^{+1.74 }_{-1.24}$ \\
 v128 & FIRE &  61.3 &  -1.2 $\pm$ 1.0 &  -45.2 & -0.9 $\pm$  0.2 &    -     &        -     &        -        &  -   \\
 v181 & FIRE &  5.0 & -4.4 $\pm$ 0.8 &  -48.4 & -2.0 $\pm$  0.8 &   18.1 &  -32.1 $ \pm$   8.7 & - & $15.03^{+1.44 }_{ -1.08}$  \\
 v190 & FIRE &-46.0 & -2.1 $\pm$ 0.9 & -135.2 & -2.8 $\pm$  0.4 &  -66.8 &  -10.3 $ \pm$   6.1 & $\star$$3.95  ^{+0.42 }_{ -0.38}$ & $10.01  ^{+0.35 }_{ -0.57} $ \\
 v237 & FIRE & -1.3 & -0.5 $\pm$ 0.3 &    -   &      -       &  -85.1 &    7.8 $ \pm$   7.4 & $4.96  ^{+0.74 }_{ -0.83}$ & $\star$$9.69  ^{+0.82 }_{ -0.82}$  \\
 v309 & FIRE &-47.2 & -1.7 $\pm$ 0.8 & -135.3 & -8.5 $\pm$  0.4 &  -35.3 &  -31.7 $ \pm$   5.1 & $\star$$ 2.83  ^{+0.67 }_{ -0.92}  $ & $ 12.66  ^{+0.94 }_{ -0.78}$ \\
 v335 & FIRE &-54.8 & -4.7 $\pm$ 0.3 &  -41.9 & -0.4 $\pm$  0.2 &  -60.0 &  -13.0 $ \pm$   5.2 & $\star$$4.23  ^{+0.73 }_{ -0.79} $ & $11.32  ^{+0.90 }_{ -0.60} $ \\
 v376 & FIRE &-11.3 & -0.9 $\pm$ 0.3 &    -   &      -       &    -     &        -         &   $\star$$1.24^{+0.99 }_{ -0.85} $   &  $14.67^{+1.26 }_{ -1.26}$ \\
 v389 & FIRE &-52.4 & -3.4 $\pm$ 0.9 &  -51.2 & -3.9 $\pm$  0.3 &    -     &        -         &   $4.28^{+0.73 }_{ -0.80}$   & $11.83^{+0.78 }_{ -0.78}$    \\
 v467 & FIRE &-12.7 & -0.5 $\pm$ 0.4 &  -66.5 & -1.7 $\pm$  0.2 &   -3.6 &   -9.9 $ \pm$   5.7 & $0.20  ^{+0.94 }_{ -0.19}$ & $\star$$8.98  ^{+1.28 }_{ -0.85}$ \\
  v618 & FIRE &  -   &      -       &  -38.3 & -6.0 $\pm$ 1.5 &    -     &        -      &        -       &  -   \\
 v621 & FIRE & 13.4 & -7.1 $\pm$ 0.9 &    -   &      -       &    -     &        -       &        -     &  $14.74^{+1.44 }_{ -1.00}$ \\
 v636 & FIRE &  -   &      -       &  -43.6 & -3.3 $\pm$  0.3 &    -     &        -      &        -       &  -   \\
 v713 & FIRE &-76.4 & -3.4 $\pm$ 0.7 &  -47.3 & -2.8 $\pm$  0.2 &  -49.5 &  -15.3 $ \pm$  13.6 & $\star$$3.51  ^{+0.68 }_{ -0.81}$ & $11.87  ^{+0.89 }_{ -0.67}$ \\
Stim1 & FIRE & -2.0 & -3.3 $\pm$ 0.3 &    -   &      -       &   17.8 &  -18.7 $ \pm$   3.1 & - & $10.50  ^{+1.14 }_{ -1.06}$ \\
 Stim5 & FIRE &-97.1 & -2.9 $\pm$ 0.7 & -127.5 & -2.2 $\pm$  0.4 &  -91.2 &  -34.0 $ \pm$   5.2 & $\star$$5.14  ^{+0.83 }_{ -0.73} $ & $9.52 ^{+0.72 }_{ -0.92} $\\
 Stim5 & XSHOOTER &  -91.5 &  -2.6 $\pm$  2.0 & -159.2 & -14.2 $\pm$   0.1 &  -78.7 &  -22.1 $ \pm$  14.2 & $\star$$4.49^{+0.88}_{-0.66}$ & $10.15^{+0.71}_{-0.93}$ \\
 Stim13 & FIRE &-40.1 & -1.3 $\pm$ 0.9 &  -51.0 & -2.9 $\pm$  0.5 &    -     &        -       &   $\star$$2.68^{+0.75 }_{ -0.75}$ &  $12.04^{+0.86 }_{ -0.73} $   \\
 DR4\_v5  & XSHOOTER &    -    &        -         &  -79.0 &  -9.1 $\pm$   0.5 &  -27.9 &  -25.8 $ \pm$  12.2 & $\star$$1.82^{+1.12}_{-0.96}$ & $8.26^{+1.03}_{-1.17}$\\
 DR4\_v10  & XSHOOTER & -105.3 &  -4.0 $\pm$  0.8 &  -84.4 &  -1.7 $\pm$   0.3 &  -78.2 &   -8.9 $ \pm$   6.1 & $\star$$4.25^{+1.01}_{-0.54}$ & $\star$$7.30^{+0.62}_{-0.91}$\\
 DR4\_v15  & XSHOOTER &    -    &        -         &  -90.5 &  -3.1 $\pm$   0.1 &    -     &        -      & - & -   \\
 DR4\_v17  & XSHOOTER &  -98.1 &  -1.2 $\pm$  0.4 & -126.3 &  -2.1 $\pm$   0.4 &  -73.1 &  -12.2 $ \pm$   6.5 & $\star$$4.34^{+0.89}_{-0.84}$ & $8.62^{+0.78}_{-0.95}$\\
 DR4\_v18  & XSHOOTER & -100.0 &  -3.0 $\pm$  0.4 &  -83.8 &  -1.0 $\pm$   0.2 &  -90.2 &   -6.9 $ \pm$   5.7 & $\star$$5.18^{+0.74}_{-0.82}$ & $8.25^{+0.89}_{-0.67}$\\
 DR4\_v20  & XSHOOTER &    -    &        -         &  -48.4 &  -2.5 $\pm$   0.5 &  -83.7 &   40.6 $ \pm$   9.2 & $\star$$4.67^{+0.97}_{-0.67}$ & $9.26^{+0.69}_{-0.98}$\\
 DR4\_v30  & XSHOOTER & -143.3 &  -1.1 $\pm$  0.1 &    -   &        -         & -126.8 &   -1.9 $ \pm$   0.6 & $\star$$6.65^{+0.39}_{-0.78}$ & $8.35^{+0.86}_{-0.38}$\\
 DR4\_v34  & XSHOOTER &  -58.4 &  -3.8 $\pm$  0.2 & -121.0 &  -3.8 $\pm$   0.3 &  -59.4 &  -28.4 $ \pm$   3.5 & $\star$$3.95^{+0.54}_{-0.90}$ & $11.21^{+0.90}_{-0.60}$\\
 DR4\_v39  & XSHOOTER &    -    &        -         &  -77.3 & -35.3 $\pm$   0.9 &  -72.5 &   -7.5 $ \pm$  16.0 & $\star$$4.63^{+0.81}_{-0.64}$ & $10.98^{+0.64}_{-0.87}$\\
 DR4\_v42  & XSHOOTER & -111.0 &  -4.8 $\pm$  0.9 & -117.9 & -23.3 $\pm$   0.2 & -114.7 &  -17.4 $ \pm$  17.6 & $6.56^{+0.66}_{-0.62}$ &  $9.21^{+0.65}_{-0.65}$ \\
 DR4\_v44  & XSHOOTER &  -85.0 &  -2.1 $\pm$  0.3 &    -   &        -         &  -48.9 &  -25.4 $ \pm$   6.3 & $\star$$3.87^{+0.73}_{-0.81}$  & $11.98^{+0.93}_{-0.77}$\\
 DR4\_v55  & XSHOOTER &    -    &        -         &  -34.7 & -30.5 $\pm$   1.3 &    -     &        -      & - & -   \\
 DR4\_v67 & XSHOOTER &    -    &        -         &  -43.4 & -12.1 $\pm$   0.1 &    -     &        -      & - & -   \\
 DR4\_v89 & XSHOOTER &    -    &        -         &  -51.4 &  -5.2 $\pm$   0.3 &    -     &        -       & - & -  \\
\hline
\multicolumn{2}{l}{Post main sequence objects:} \\
 v51 & FIRE &-34.3 &  0.9 $\pm$ 0.5 &    -   &      -       &   37.3 &    7.8 $ \pm$   5.6 & - & $\star$$12.46^{+1.48 }_{-1.06}$ \\
  v84 & FIRE & -218.9; 208.0 & -5.8 $\pm$ 0.7 &    -   &      -       &    -     &        -     &        -      &  -     \\
  v319 & FIRE &  128.6   &  1.1 $\pm$ 0.4   &    -   &      -       &  119.8 &   56.9 $ \pm$   6.1 & - & - \\
 v370 & FIRE & 26.1 & -0.9 $\pm$ 0.4 &    -   &      -       &  -10.2 &   49.9 $ \pm$  10.8 & $\star$$1.45  ^{+0.74 }_{ -1.07} $ & $\star$$14.61  ^{+1.41 }_{ -1.16} $\\
 v371 & FIRE & -88.2 & -30.7 $\pm$ 3.7 &   -   &      -       &  - &   -  &    $5.79^{+0.72 }_{ -0.72}$  &  $10.21^{+0.66 }_{ -0.76}$    \\

\hline
\hline
\end{tabular}
\flushleft{The spectral resolutions of the FIRE and XSHOOTER spectra are about 50 km s$^{-1}$ and 37 km s$^{-1}$, respectively.  The uncertainty of the radial velocity measurement is equivalent to the pixel size of the array detector, which is about 12 km s$^{-1}$ in both spectrographs. \\
$\star$: Kinematic distances agree with PM distance solutions}
\label{tab:lines}
\end{table*}

\subsection{Distance measurements}
\label{sec:dists}

Wherever possible, kinematic distances, $d_k$, were derived from the radial velocities measured in the CO overtone band fits described above. For objects without CO features, we used the radial velocity of the Br$\gamma$ line to derive the kinematic distance because H$_2$ lines more often arise in a blue-shifted wind or outflow. The uncertainty of the radial velocity measurement is approximately 1 array pixel, about 12 km s$^{-1}$ in both spectrographs. The kinematic distances are measured via the ``Monte Carlo simulation mode'' of the online tool provided by \citet{Wenger2018}, which includes the uncertainties in the Galactic rotation model. The constants in the calculations are adopted from \citet{Reid2014}, including the circular rotation speed of the solar system ($\Theta_0 = 240 \pm 8\, {\rm km\, s}^{-1}$), the distance to the Galactic centre ($R_0 = 8.34 \pm 0.16 $ kpc), and the Galactic rotation curve.  The kinematic distances and uncertainties are listed in Table \ref{tab:lines}. For targets lacking Br$\gamma$ and CO features and targets for which the measured radial velocity is inconsistent with the Galactic rotation model (e.g. halo stars) distances not given in Table \ref{tab:lines}. In addition, we noticed that the difference between the radial velocities of Br$\gamma$ line and CO bandhead emission sometimes significantly exceeds the uncertainty (12 km s$^{-1}$). Therefore, the kinematic distance derived by the radial velocity of Br$\gamma$ line alone is highly uncertain and needs to be confirmed by other methods, such as a literature-based distance to the surrounding star formation region, $d_{\rm SFR}$. 

Most of our targets are heavily embedded distant Class I YSOs, very faint in the optical waveband, so Gaia measurements are not available. Here, the measured proper motions (PMs) of individual objects are provided in the VIRAC2\footnote{The VIRAC2 PMs, though calibrated using the {\it Gaia} second data release, are preliminary. The final absolute calibration, based on the {\it Gaia} ``eDR3'' reference frame, is still in progress but any changes are expected to be very small.} catalogue, which were used to break the ambiguity between the near and far kinematic distance solutions. The PM-based probability density function for distance is calculated by a Monte-Carlo method built on the Python platform, using one of two models of the Milky Way disc rotation and the peculiar velocity dispersion, adopted for YSOs and non-YSOs treated as Galactic ``thin disc'' stars respectively \citep[the latter drawing on][]{Bensby2013}. For a given PM measurement, a distance solution is computed by randomly sampling from the peculiar velocity distribution and the PM error distribution using Gaussian errors.\footnote{The approach is an evolution of that of \citet{Smith2018}. Details of the adopted rotation curves and peculiar velocity distributions used for YSOs and thin disc stars in the sample will be given in a separate paper (C. Morris et al., in prep). A relatively broad peculiar velocity spread (20 km s$^{-1}$ in the U and V components) is used for YSOs in order to allow for uncertainties in the rotation curve, i.e. folding the two effects into a single parameter.}. The probability histogram is then generated by binning 1 million samples for each star. Kinematic distances with consistent with the sometimes rather wide range of distances allowed by the PM-based distance solutions are marked by the $\star$ symbol in Table \ref{tab:lines}. Several targets lack consistent $d_{k}$ and PM distance solutions, which could be due to underestimated errors in the PM measurements for faint objects, large orbital motions in binary systems, intrinsic blue-shifts of expanding CO shells around Asymptotic Giant Branch (AGB) stars, or a Galactic halo location.

\subsection{Light curve Analysis}
\label{sec:lc}
The 2010--2015 $K_s$ light curves of 21 targets with the ``v'' prefix were previously published in Paper I, based on aperture photometry. In this work, we present a new set of VVV/VIRAC2 $K_s$ light curves for all 61 targets in the combined sample (see online supplementary information, examples are shown in Table~\ref{tab:lc_example}). The light curves, contain $K_s$ profile fitting photometry from 2010 to 2019. The profile fitting photometry was carried out with DoPHOT \citep{Schechter1993, Garcia2012}. The absolute photometric calibration of the VIRAC2 catalogue is based on a new procedure designed to mitigate issues pointed out by \citep{Hajdu2020} that arise from blending of multiple VVV sources in 2MASS data for very crowded star fields in the inner Galactic bulge. 

We applied the $\chi-$value, an output of DoPHOT, from the VIRAC2 catalogue as a selection criteria in order to remove less reliable data points. Detections with $\chi-$value~$> 3$ are excluded from the final light curves. As a second step, the relation between $K_s$ magnitudes and photometric errors ($\delta K_s$) is fitted by a 3rd order polynomial function derived from all sources in this work. All sources share the same ${K_s}$ vs. $\delta K_s$ relationship. Detections with $\delta K_s$ deviating from this relationship by more than 3$\sigma$ are rejected from the final light curves. The typical photometric error is 0.020~mag at $K_s = 11$~mag and 0.069~mag at $K_s = 16$~mag. In the final light curves (shown in Figure~\ref{fig:lc_sum}~and~\ref{fig:lc_sum_2}) the data are averaged in 1 day bins because this work focuses on longer-term variability rather than intra-night changes. Even though the error reported by DoPHOT is small, the magnitudes above $K_s =$ 11.25~mag, could suffer from a not completely linear response from the detector.

The photometric period of individual object is measured using the generalised Lomb-Scargle periodogram in IDL \citep{Zechmeister2009}. More details are shown in \S\ref{sec:period}. The variation timescale ($t_{\rm phot}$) is defined as follows. For periodic variables,  $t_{\rm phot}$ is the photometric period (see Table \ref{tab:info}). For other YSOs, $t_{\rm phot}$ is the time difference between two photometric minima for sources with eruptive light curves and the time between maxima for ``dippers''. For objects with ongoing outbursts and decays, we indicate in Table \ref{tab:info} that $t_{\rm phot}$ is a lower limit. The nature of photometric variations are summarised in Table \ref{tab:info}.

\section{Results}
\label{sec:res}
In this section, we will present the spectral features of our targets, period analysis, and classifications of variables based on both spectroscopic and photometric information.

\subsection{Spectroscopic features of accretion modes of YSOs}

The spectroscopic features of YSOs are shaped by mass accretion process \citep{Hartmann1996, Hartmann2016}.  In general, there are two types of accretion natures in YSOs which are distinguishable by certain spectral features. Magnetospheric accretion and boundary layer accretion were introduced in \S3.1 of Paper III and references therein. Here we briefly summarise the spectral features of these two accretion scenarios.

Magnetospheric accretion is widely observed on disc-bearing YSOs. Emission features are generated around the accretion shock on the stellar surface, such as H{\sc i} recombination lines (e.g. Pa$\beta$ and Br$\gamma$) and Ca {\sc ii}  triplet (8498, 8542, and, 8662 \AA). Emission fluxes of Ca {\sc ii} and H{\sc i} recombination lines have tight empirical correlations with the mass accretion rate \citep{Muzerolle1998b, Alcala2014}. Originating from the chromosphere and the disc, the Na{\sc i} emission doublet at 2.206 $\mu$m is closely correlated with Br$\gamma$ emission (Paper III). Recent near-infrared surveys found CO overtone emission bands, although excited in the inner gaseous disc, have a positive correlation with Br$\gamma$ emission on both non-eruptive \citep[][]{Connelley2010} and eruptive YSOs \citep[][and Paper III]{Kospal2011EX}, indicating that CO emission is associated with a high rate of magnetospheric accretion, perhaps due to irradiation by hot spots on the star arising from the accretion columns \citep{Hodapp2020}.

Boundary layer accretion, mostly seen in FUors, shows strong H$_2$O and CO absorption bands arising in cool upper layers of its self-luminous accretion disc \citep[see review by][]{Audard2014}. In Paper III, we found stronger CO absorption is associated with a brighter near-infrared continuum which is consistent with the physical expectation that the disc's vertical temperature gradient is steeper when the accretion rate is higher and more energy is being released in the disc's mid-plane. 
Weak photospheric CO absorption features are seen in YSOs and they are associated with other absorption lines (e.g. Na{\sc i} and Ca{\sc i}) \citep[][and Paper II]{Johnskrull2001}. In addition, strong CO absorption is also observed in the photosphere of very low mass YSOs \citep[e.g.][]{Lucas2001, Allers2007, Jose2020}, which are much less luminous than FUor outbursts and hence should be much less distant. Leaving aside these cases of photospheric absorption, the strong CO overtone absorption in a distant eruptive YSO is an efficient indicator of boundary layer accretion process. For the following results and discussions, YSOs are termed ``FUor-like'' when CO absorption features are detected.

The near-infrared $H_2$ emission lines are indicators of stellar winds and outflows \citep{Chrysostomou2008, Greene2010}.  Although seen in non-variable sources like protoplanetary nebulae, H$_2$ lines are recognised as signatures of YSOs. After inspecting samples of eruptive YSO spectra \citep[][and Paper III]{Connelley2018}, we found the strength of H$_2$ emission lines is independent of variation in other spectral features. The absence of H$_2$ emission in most FUors might be a veiling effect of the hot self-luminous disc (see discussions of v332 in \S4.2.3 of Paper III) or rather the absence of collimated winds/jets during the FUor burst.

\begin{table}
\renewcommand\arraystretch{0.9}
\centering
\caption{Spectral Characteristics and classifications}
\begin{tabular}{l p{0.35cm}<{\centering} p{0.35cm}<{\centering} c p{0.35cm}<{\centering} c p{0.35cm}<{\centering} c}
\hline
\hline
Name & Br$\gamma$ & Pa$\beta$ & Na{\sc i} & H$_2$$^*$ &  CO & H$_2$O & Spec Class \\
\hline
\multicolumn{4}{l}{ H$_2$ emission only}\\
\multicolumn{4}{l}{ Outflow dominated}\\
v618 & - & - & - & E & -  & - & Emission line \\
 DR4\_v15  & - & - & - & E & - & - & Emission line \\
 DR4\_v55  & - & - & - & E  & - & - & Emission line\\
 DR4\_v67  & - & - & - & E & - & - & Emission line\\
 DR4\_v89  & - & - & - & E & - & -& Emission line \\
\hline
\multicolumn{7}{l}{ H$_2$, Br$\gamma$, and/or CO bandhead emission}\\
v14 & E & - & - & - & - & - & Emission line \\
v53 & E & - & E  & -  & E & & Emission line \\
v128 & E & -  & - & E & - & A & Emission line \\
v181 & E & E & - & E & E & - & Emission line \\
v190 & E & - & E & E & E & - & Emission line\\
v309 & E & - & E & E & E & - & Emission line \\
v335 & E & E & E & E  & E & -& Emission line \\
v376 & E & - & - & -  & - & - & Emission line  \\
v389 & E & - & - & E & - & - & Emission line \\
v467 & E & - & - & E  & E & - & Emission line \\
v621 & E & - & - & - & ? & - & Emission line  \\
v636 & E? & - & - & E & -  & - & Emission line \\
v713 & E & E & - & E  & - & - & Emission line  \\
Stim1 & E & E & E & - & E & - & Emission line  \\
Stim13 & E & - & - & E  & - & - & Emission line  \\
Stim5 & E& - & E & E & E & - & Emission line \\
DR4\_v5  & E & - & - & E & E & - & Emission line\\
DR4\_v10  & E & - & - & E & E & - & Emission line\\
DR4\_v17  & E & - & - & E & E & -& Emission line \\
DR4\_v18  & E & E & E & E & E & -& Emission line \\
DR4\_v30  & E & E & - & - & E & -& Emission line \\
DR4\_v34  & E & E & E & E & E & -& Emission line \\
DR4\_v39  & - & - & - & E & E & -& Emission line \\
DR4\_v42  & E & - & - & E & E & -& Emission line \\
DR4\_v44  & E & E & E & - & E & -& Emission line \\
\hline
\multicolumn{4}{l}{CO absorption} \\
v16 & A & - & - & - & A & A & FUor-like \\
v237 & E & - & - & - & A & - & FUor-like \\ 
DR4\_v20  & - & - & - & E & A & A  & FUor-like\\
\hline
\multicolumn{4}{l}{Post main sequence objects} \\
v51 & - & - & - & -  & A & A & AGB\\
v319 & E & - & - & -  & A & - & Symbiotic \\
v370 & E & - & - & -  & A & A & Symbiotic\\
 v84 & E & -  & E & E  & - & - & Nova\\
 v371 & E & - & - & -  & - & - & Wolf-Rayet \\
 \hline
\hline
\end{tabular}
\flushleft{Emission lines or band features are marked by `E' while absorption features are marked by `A'. Non detections are shown by `-'. \\
$'*'$ denoted H$_2$ 1 - 0 S(1) line at 2.12 $\mu$m.\\
 In \S~\ref{sec:var_t}, YSOs only have H$_2$ emission in their spectra are classified as ``outflow dominated'' sources.}
\label{tab:sumlines}
\end{table}

\subsection{Spectroscopic classification}
\label{sec:specclass}
Here we identify non-YSOs in the sample and then sort the variable YSOs into two categories based their spectral features. YSOs with $^{12}$CO $(\Delta\nu = 2)$ overtone absorption beyond 2.29 $\rm\mu$m are classified as ``FUor-like'' objects, and YSOs with emission features (e.g. H$_2$, Br$\gamma$, Na{\sc i}, or CO emission) are classified as ``emission line spectra'' YSOs. The spectral characteristics of our targets are summarised in Table \ref{tab:sumlines}. 

\subsubsection{Emission line YSOs}

33 of the 38 new spectroscopic targets observed in this work are identified as YSOs.
Br$\gamma$ emission is clearly detected in 23 of them and there is a marginal detection ($\sim 1.5 \sigma$) in one additional source, v237. Other indicators of magnetospheric accretion, including Pa$\beta$ emission, Na{\sc i} emission and CO overtone emission, are only observed among objects with Br$\gamma$ emission, with the exception of DR4\_v39. H$_2$ emission lines are seen in 24 YSOs. Following the classification methods introduced above, there are 30 YSOs classified as ``emission line objects''. This category includes 24 YSOs with indicators of magnetospheric accretion and five YSOs with ``outflow-dominated'' spectra, these displaying only H$_2$ emission, sometimes accompanied by [Fe{\sc ii}] emission. One object (v621) has debatable spectral features as broad Br$\gamma$ and He~I emission lines, which are often observed on nova. However, the light curve of v621 had a quick rise and then stayed on a brightness plateau for at least 6.5 years, which has not seen among novae. Hence we still classify v621 as an eruptive YSO. 

\subsubsection{FUor-like YSOs}

CO overtone absorption is observed in six of the new spectroscopic targets, three of which are identified as FUor-like YSOs. Two of the three YSOs also display water vapour absorption bands. When we include FUor-like VVV sources from Paper II, there are six YSOs with FUor-like spectra in the combined sample. These are v16, v237, DR4\_v20, v322, v721 and v717. Of the three new FUor-like YSOs listed in Table \ref{tab:sumlines}, one (v237) has marginally detected Br$\gamma$ emission, as noted above, one (v16) has marginally detected Br$\gamma$ absorption and one (DR4\_v20) has no Br$\gamma$ detection but displays clear H$_2$ emission lines. Due to the large photometric variability of the three and their large distances (see Table \ref{tab:info}), there is no possibility of confusion between FUor-like YSOs and very young brown dwarfs having CO and water vapour absorption in the photosphere.

\subsubsection{non-YSOs}

Among the six new spectroscopic targets displaying $^{12}$CO overtone absorption, three also show the $^{12}$CO second overtone ($\Delta \nu = 3$) absorption bands in the $H$ bandpass and/or strong $^{13}$CO ($\nu=2-0$) first overtone absorption. These three are then identified as AGB stars (v51, v319 and v370, \S\ref{sec:AGB}). In the case of v51, since there were only two dips in the light curve, it may have a 2500~day period. However, the brightness is very different before and after the first dip so it may be aperiodic. The other two AGB stars have periods of a few hundred days, which are typical for Mira variables.

In addition, double-peaked and broad Br$\gamma$ emission associated with He{\sc i} or He{\sc ii}  emission lines are detected in two sources. The spectrum of v371 is very similar to that of a Wolf-Rayet type star, with strong He{\sc ii} emission lines blended with Br$\gamma$ emission and broad N{\sc iii} emission at 2.11~$\mu$m \citep{Mauerhan2009}. We name v371 as WR80-1 following the up-to-date naming convention of massive stars \citep{Rosslowe2015}\footnote{\url{http://pacrowther.staff.shef.ac.uk/WRcat}}. A 193~d quasi-period is found in the VVV light curve of this star. Although periodic variation is widely detected in Wolf-Rayet type stars, the 193~d period is longer than the timescale of wind-dominated variation \citep[WR 1, 16.9~d;][]{Chene2010} and is much shorter than the multi-year-long dust making process \citep{Williams2019}. We therefore suggest that v371 is perhaps in a binary system. 

V84 has a blue SED ($\alpha = - 2.3$) and the light curve shows a fading trend with $\Delta K_s = 2.76$~mag in the first 600~d of the VVV survey. It remained at a low brightness level thereafter. Strong double-peaked He{\sc i} and Br$\gamma$ emission lines are detected, suggesting a fast rotating envelope ($v\sim 200$~km s$^{-1}$). Both photometric and spectroscopic features of v84 are consistent with previously detected novae, such as v514 in Paper II, V1493 Aql \citep{Venturini2004}, and V2491 Cyg \citep{Naik2009}. Towards the end of the VVV VIRAC2 light curve, the brightness of v84 shows signs of rising again slightly. It might therefore be a recurrent nova, though significant photometric variations are also seen in classical novae long after the outburst.

\subsubsection{The combined sample}

Including sources from Paper II, 75 high-amplitude VVV variable stars have near-infrared spectra. Following the spectroscopic classification above, 61 sources are confirmed as YSOs: 6 FUor-like objects and 55 objects emission line objects. Of the 55 emission line objects, 46 YSOs have signatures of magnetospheric accretion, including 45 having HI recombination line emission and one source (DR4\_v39) lacking this feature but having CO overtone emission. The remaining eight emission line YSOs have outflow-dominated spectra so their accretion modes are not yet determined. According to the spectral indices calculated in \S\ref{sec:target} and Paper II, about 2/3 of these spectroscopically confirmed YSOs are Class I objects (40/61), some of them are flat-spectrum objects (17/61), and 4 of them are Class II objects.
 
\begin{table}
\renewcommand\arraystretch{0.9}
\centering
\caption{Periodic variables from VVV light curves}
\begin{tabular}{l c c c c c}
\hline
\hline
Name & Period (d) & Period (d) & $\Delta K_s$& $\delta_{K_s}$  & Class \\
\hline
 & this work & Paper I &  (mag)  & (mag)  \\
 \hline
 \multicolumn{2}{l}{YSOs from this work:} \\
v14 &   397 &  124 &  1.64& 0.21 &  Q \\
v190 &   611 &  125 &  1.44& 0.22 &  Q \\
v467$^{a}$ &  1840 &  - &  2.64& 0.24 & P \\
v636 &   565 &  609 &  1.43& 0.21 &  Q \\
DR4\_v17  &  1154 & - & 3.03& 0.12 & P \\
DR4\_v55  &   953 &  - & 3.59& 0.34 & P \\
v181$^{b}$ &   184 &  - &  3.49& 0.55 & Q \\
v309$^{b}$ &   181 &  191 &  2.60& 0.43 &  Q \\
\hline
\multicolumn{2}{l}{YSOs from Paper II:} \\
v32 & 1439 & - & 1.80 & 0.18 & P \\
v65 & 30.2 & 30 & 0.98 & 0.21 & Q \\
v94 & 436 & - & 1.27 & 0.24 & Q \\
v480 & 306 & 303 & 1.51 & 0.24 & Q \\
v630 & 409 & - & 1.57 & 0.31 & Q \\
\hline
\multicolumn{2}{l}{Post main sequence objects:} \\
v319 &   307 &  - &  0.86& 0.07 & P \\
v370 &   419 &  445 &  0.94& 0.11 & Q \\
v371 &   193 &  186 &  2.09& 0.28 &  Q \\
\hline
\end{tabular}
\flushleft{ $\delta_{K_s}$: Fitting residual on the phase-folded light curve. \\
Class: Periodic (P) or Quasi-periodic (Q) variable. \\
a: Only two full periods were detected in the VVV light curve. \\ 
b: Sources with suspicious half-year periods and doubtful periodicity.}
\label{tab:periods}
\end{table}

\begin{figure} 
\includegraphics[width=3.in,angle=0]{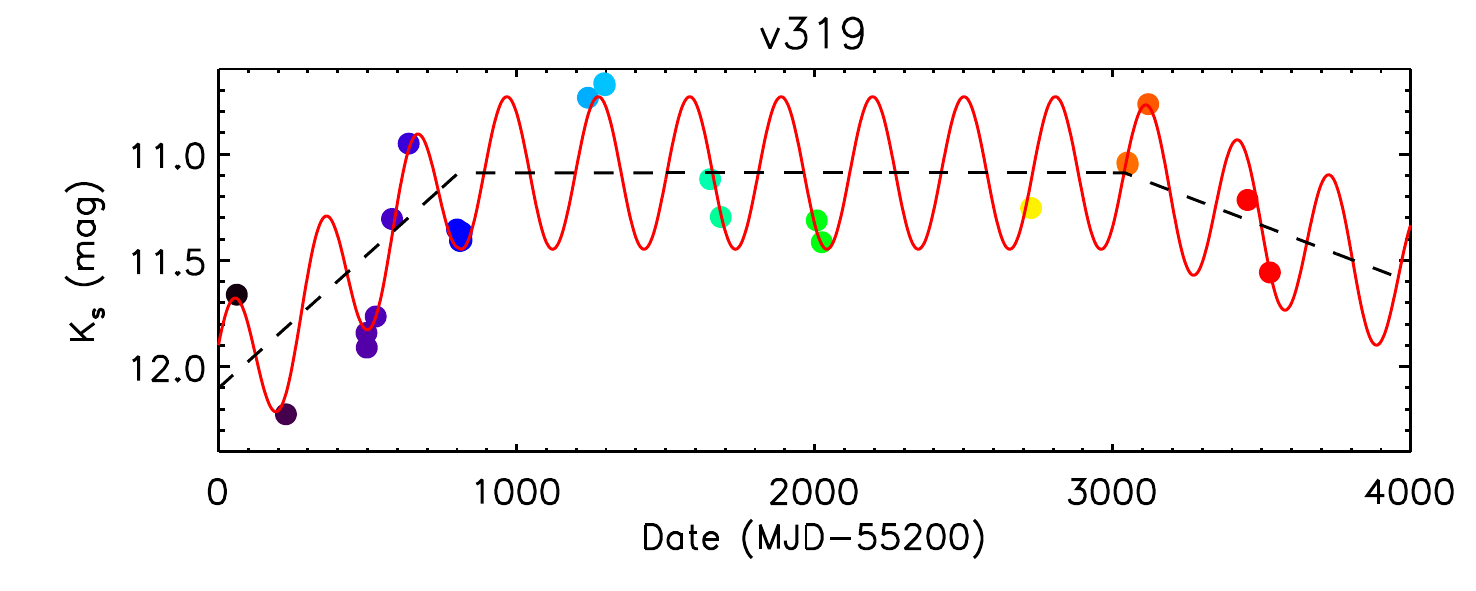}
\includegraphics[width=3.in,angle=0]{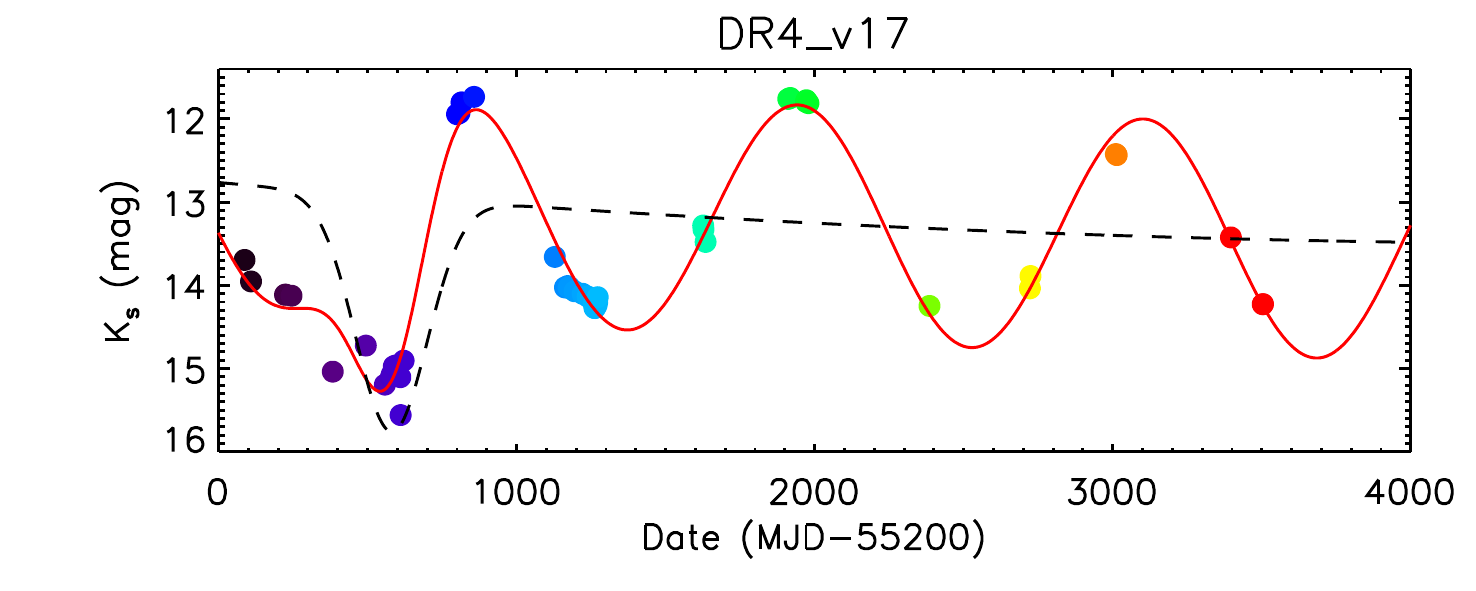}
\caption{The $K_s$-band VVV light curve of v319 and DR4\_v17. Data points are colour coded by the observation time. The long term trends are fitted by the dashed lines, and the overall variations are fitted red solid lines.}
\label{fig:lc319}
\end{figure}

\begin{figure*} 
\includegraphics[height=2.7in,angle=0]{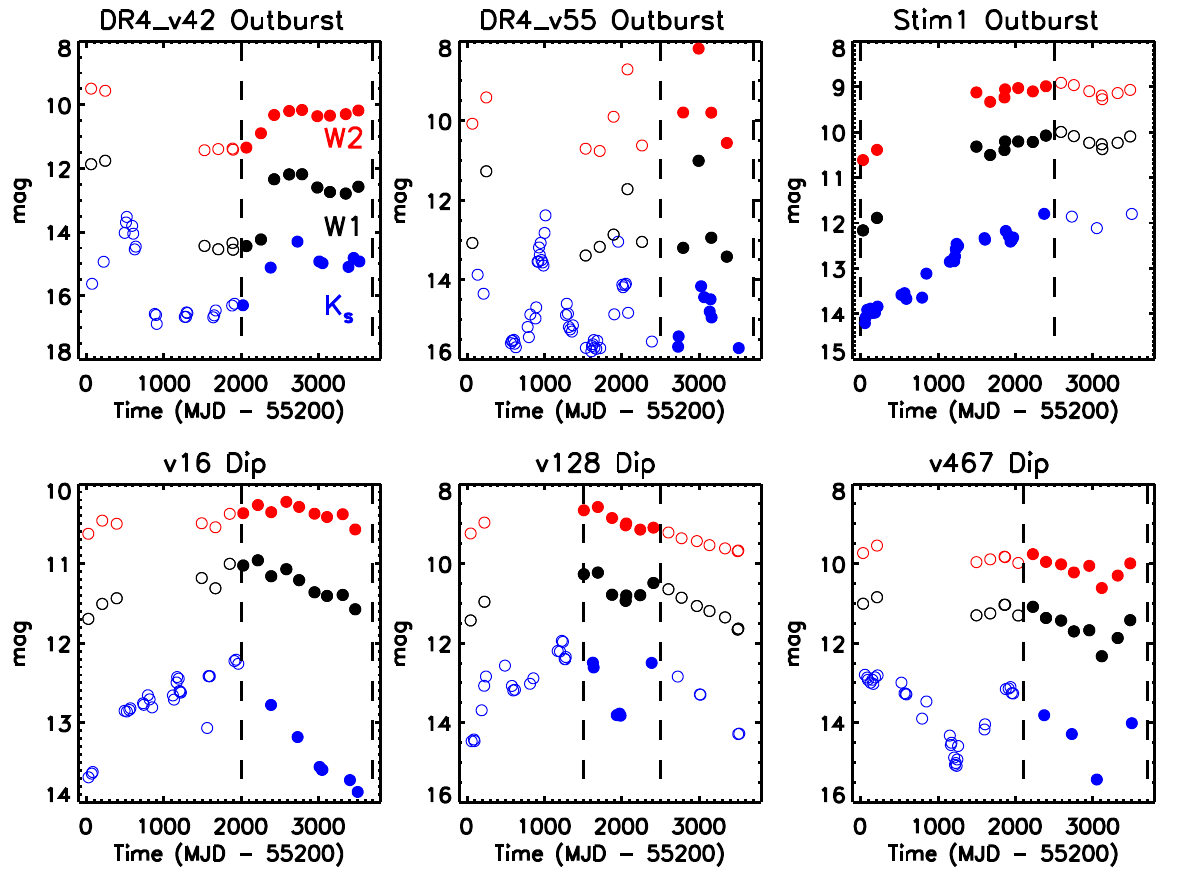}
\includegraphics[height=2.8in,angle=0]{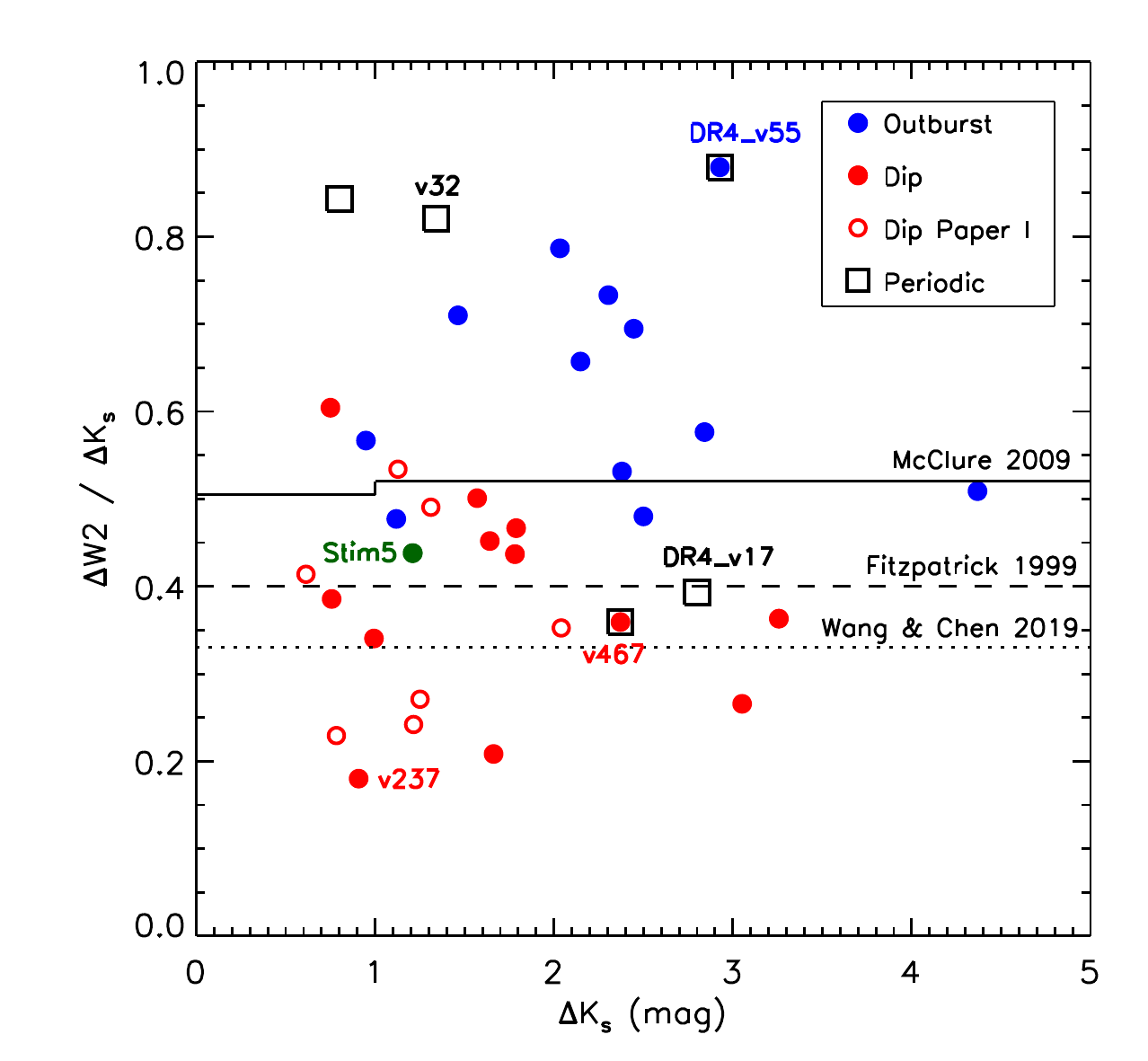}

\caption{{\it Left:} Examples of multi-band light curves and visually inspected morphologies. Light curves within the time range of certain dip or outburst are shown by filled circles. The bandpass of the light curves are labelled in the {\it upper left} panel. {\it Right:} The ratio between $W2$- and $K_s$-band amplitude of individual events that are classified as dip (red) and outburst (blue). 
Red open circles represent dips obtained from variable YSO candidates in Paper I. Long-term periodic YSOs are shown by squares. Sinusoidal variables (v32 and DR4\_v17) are labelled separately. In addition, Stim5 is marked by the green dot and the amplitudes are measured between two spectroscopic epochs. The $\Delta W2$/$\Delta K_s$ ratios predicted from analytical \citep[$R_V = 3.1$,][]{Fitzpatrick1999} and empirical \citep{McClure2009, WangS2019} extinction curves are shown by horizontal lines.}
\label{fig:amp_neo_vvv}
\end{figure*}

\subsection{Periodic or Quasi-periodic variables}
\label{sec:period}
Most periodic variations on YSOs are associated with stellar or disc rotation. For instance, star spots on Class II/III objects have sinusoidal variation on timescales from days to weeks \citep[e.g.][]{Rebull2018, Guo2018b}. Periodic extinction by asymmetric inner disc structures are mostly seen on Class II objects on timescales from one week to months \citep[e.g.][]{Bouvier2007}. Although most episodic accretion bursts (FUor- and EXor-type) are unpredictable in Class 0/I objects, {three YSOs in the literature display periodic variability attributed to a large and periodic change in the accretion rate. Periodic accretion events seen in the Class I YSOs} LRLL~54361 \citep[25~d,][]{Muzerolle2013} and V347~Aur \citep[160~d,][]{Dahm2020} are presumably triggered by undetected stellar companions. A $\sim$530~d periodic accretion flare is observed on the Class 0 object EC53 (V371~Ser) by both near-infrared and millimetre surveys \citep{Hodapp2012, YHLee2020}, which is interpreted as a consequence of piled-up material in the inner disc.


Periodic VVV variables were identified in Paper I from 2010 -- 2015 light curves by phase dispersion minimisation (PDM) method. In this work, we applied generalised Lomb-Scargle periodogram \citep{Zechmeister2009} to VIRAC2 version light curves of spectroscopic followed up variables in this work and spectroscopic confirmed YSOs from Paper II. Light curve data are averaged in 1~day bins to avoid the over-sampling issue. The fitting period is constrained to be in the range of 10 to 2000~days, based on the cadence of the light curves and the duration of the VVV survey. The best-fit period is identified by the highest peak in the periodogram. Two particular targets, v319 and DR4\_v17 have light curves that are combinations of long term ($t > 500$~days) trends and periodic variations. Hence, their power spectra were generated after fitting and removing the long term trends (see Figure~\ref{fig:lc319}). Finally, we visually inspected the phase-folded light curves to confirm the periods.

In total, 16 objects have periods in their light curves ranging from 30 to 1840 days. All measured periods passed the 1\% false alarm possibility test \citep[method from][]{Guo2018a}. The residual to the fit in the phase-folded light curves is used to categorise these objects as periodic or quasi-periodic. The phase-folded light curves are fitted with the following analytic functions  (for the purpose of this categorisation only): Gaussian, polynomial burst, and sinusoidal curves, examples of which are given in Figure~\ref{fig:phase_fold}. Objects are defined as periodic if the standard deviation of the fitting residual is smaller than 10\% of the amplitude in $K_s$. Under this definition, five objects are identified as periodic variables and 11 objects as quasi-periodic variables (listed in Table \ref{tab:periods}). The longest period is seen on v467 ($P = 1840 $~d), which shows two dips in the unfolded light curve. The periodicity of this source is yet to be confirmed as only two full periods were observed within the VVV survey.

Two quasi-periodic variables have periods around 180~days, which are suspicious as they might be false periods introduced by the observational cadence (i.e. one to three epochs close together in each year from 2016--2019). Both of them have large fitting residuals ($\delta K_s > 0.4$~mag). In particular, the unfolded light curve of v181 is a combination of of a sharp dip ($t \sim 100$~d) and a gradual long-term variation. It is possible that the observed light curve features arise from a poorly sampled periodic variation, but the evidence is fairly weak. In this case, the 184~day period on v181 is highly doubtful so list it as ``eruptive'' rather than ``LPV-YSO'' in Table~\ref{tab:info}.

Eight YSOs have periods detected by the PDM method in Paper I, including six with similar periods in the two works.  Two objects have longer periods measured in this work than the results from Paper I. The period of v14 measured in Paper I corresponds to the second highest peak in the periodogram generated in this work. Among the other eight objects without periods in Paper I, DR4\_v17 and DR4\_v55 are new sources first studied in this work, v32 and v467 have periods longer than 1000~days, and the rest were classified as aperiodic in Paper I. 

Based on the spectroscopic features, 13 periodic or quasi-periodic variables are identified as YSOs, and all of them are emission line YSOs. In the following section, we will discuss the near-to-mid infrared amplitude of long-term periodic YSOs (v32, DR4\_v17, and DR4\_v55), in order to identify periodic accretion bursts. Separately, one periodic and two quasi-periodic objects are identified as post-main-sequence stars: two AGB stars (see \S\ref{sec:AGB}) and a Wolf-Rayet star. 

\section{Discussion}
\label{sec:dis}

\subsection{Near-infrared colour variation}
\label{sec:neowise}

The variation mechanisms of eruptive objects can be revealed by infrared colour changes. For example, the optical to near-infrared SEDs of most optically bright EXors during the outbursts are well fitted by additional thermal emission from the accretion disc \citep[e.g.][]{Lorenzetti2012}. In other cases, wavelength-dependent flux changes can be used to infer the radial progression of an outburst through the disc, e.g. inward in the case of the recent FUor Gaia~17bpi \citep{Hillenbrand2018} or outward in the case of WISE~1422-6115 \citep{Lucas2020}. In this section, we use {\it WISE} $W1$ (3.4~$\mu$m) and $W2$ (4.6~$\mu$m) time series photometry from the IRSA archive\footnote{A few targets lack {\it WISE} data in the AllWISE and NEOWISE catalogs due to blending. The $W1$ and $W2$ light curves of v14, v621, DR4\_v15, DR4\_v20 and DR4\_v89 were extracted from the time series of unWISE image stacks \citep[unWISE,][]{Lang2014a, Meisner2018} by running the {\sc Crowdsource} crowded field photometry pipeline in an efficient forced photometry mode \citep{Schlafly2019}, using the coordinates given in the deep unWISE catalogues available at \url{https://catalog.unwise.me/} which were derived from running {\sc Crowdsource} on the five-year unWISE image stacks \citep{Meisner2019}.}, taken from the AllWISE MultiEpoch Photometry Table for 2010 \citep{Wright2010} and the ongoing NEOWISE Reactivation database from late 2013 onward \citep{Mainzer2014}. There was a 3.5 year gap between the AllWISE and NEOWISE data. The cadence of the WISE light curves is about 180 days (after averaging of the several scans taken over a period of a few days at each epoch). A few examples of the multi-band light curves are shown in Figure~\ref{fig:amp_neo_vvv}. 

\begin{figure*}
\includegraphics[width=2.3 in,angle=0]{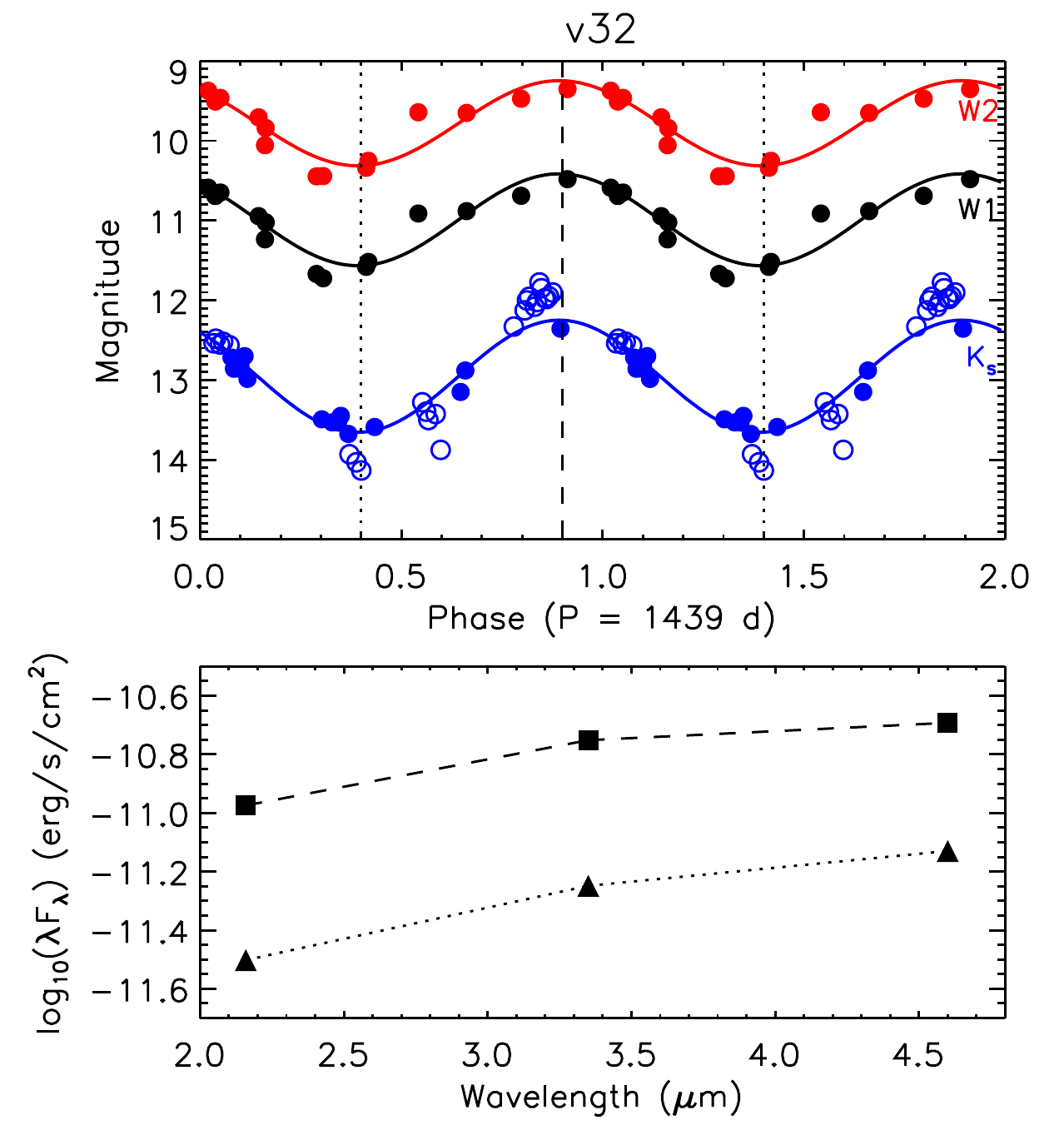}
\includegraphics[width=2.3 in,angle=0]{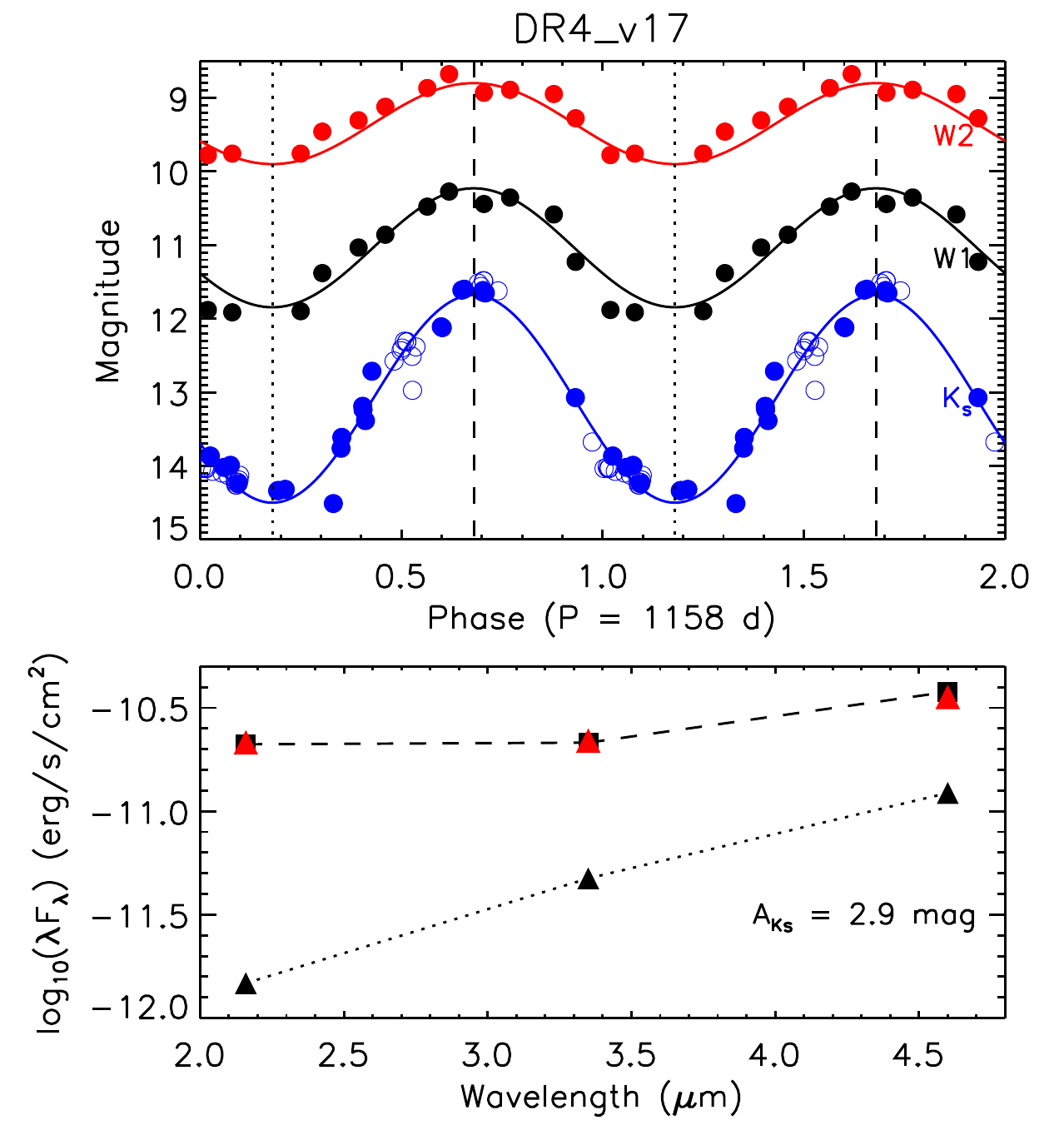}
\includegraphics[width=2.3 in,angle=0]{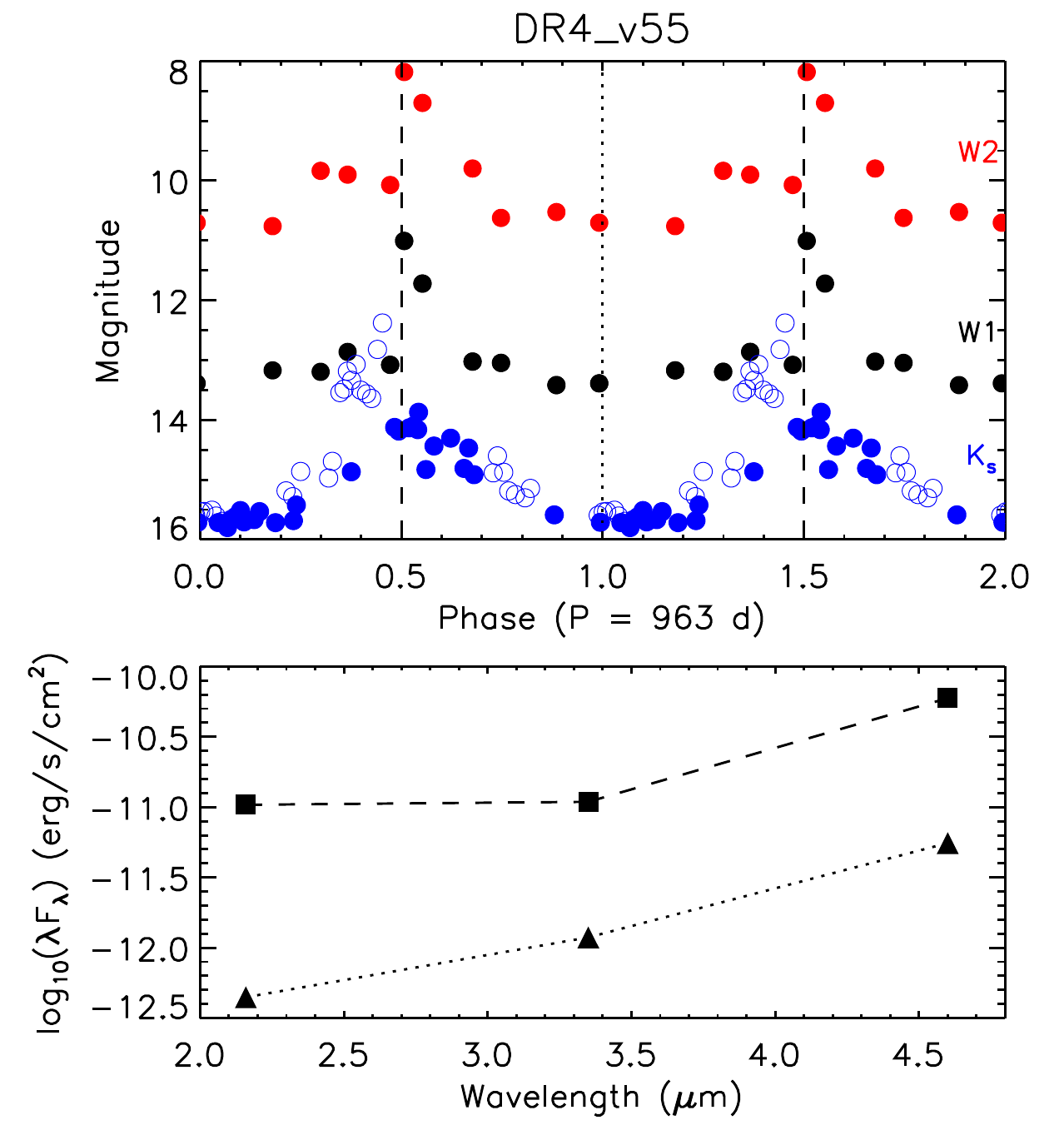}

\caption{ {\it Upper:} Phase-folded multi-band light curves of 3 emission line YSOs: v32, DR4\_v17 and DR4\_v55. $K_s$ data lying outside the time coverage of {\it WISE} photometry are shown by open circles. Dotted and dashed lines mark out the photometric minimum and maximum. Sinusoidal curves were fit to objects v32 and DR4\_v17. {\it Bottom:} The three point ($K_s$, $W1$, $W2$) SEDs during photometric maximums ($\blacksquare$) and minimums ($\blacktriangle$). In the case of DR4\_v17, red triangles are overplotted after de-reddening the SED at photometric minimum by $A_{K_s} = 2.9$~mag. We see that the dereddened SED matches the SED measured at photometric maximum. Error bars have similar sizes to the symbols and are therefore not plotted.}
\label{fig:irsed}
\end{figure*}

Here we measure the VVV $K_s$ ($\Delta K_s$) and {\it WISE} $W2$ ($\Delta W2$) variations in portions of the light curves of the variable YSOs in our sample, to attempt to distinguish between accretion-driven and extinction-driven changes on timescales from one to nine years. From the sample of 61 spectroscopically confirmed YSOs from Paper II and this work, 34 have $W2$-band light curves with long-term photometric variations, including 5 FUor-like objects. This sample excludes multiple time scale variables and periodic variables with $P < 700$ days, to avoid sampling issues of low cadence VVV and {\it WISE} light curves. Almost all of these 34 YSOs are thought to be eruptive variables undergoing episodic accretion. In a few cases this is not proven, e.g. four (v406, DR4\_v15, DR4\_v55 and DR4\_v67) were classified as having outflow-dominated spectra, though the absence of accretion signatures can be attributed to observation during quiescence in some cases. One additional source DR4\_v5 is classified as a Dipper in Table 1 but the spectrum shows clear signatures of a high accretion rate. The $K_s$, $W1$, and $W2$ light curves of the 34 YSOs are visually inspected and sorted into subgroups where an outburst or a dip can be identified in a certain time window. Examples are shown in Figure~\ref{fig:amp_neo_vvv}. These outbursts or dips are sometimes relatively brief or low in amplitude in comparison to the total variation and the nine year light curve duration. Eruptive YSOs are known to have photometric variations due to extinction as well as episodic accretion, e.g. V1515~Cyg and VSX~ J205126.1+440523 \citep{Clarke2005, Kospal2011}, so it is not surprising to observe dips within the light curves of eruptive YSOs that could be due to changing extinction.

YSOs without suitable dips or outbursts sampled by the WISE and VVV data are excluded from Figure~\ref{fig:amp_neo_vvv}. Periodic variables with sinusoidal light curves are labelled as such since the variation could be equally well regarded as a dip or a burst in many cases. We chose the ratio between $K_s$- and $W2$ amplitudes, instead of between $W1$ and $W2$, to distinguish accretion and extinction driven variations. The wide wavelength range between the $K_s$ and $W2$ bandpasses provides sufficient amplitude differences to measure changing line-of-sight extinction, according to the near- to mid-infrared extinction laws \citep[e.g.][]{Fitzpatrick1999, McClure2009}. In Figure~\ref{fig:amp_neo_vvv}, we show the measured $\Delta K_s$ and $\Delta W2$ for individual dips and outbursts within the time range of the {\it WISE} data. $\Delta K_s$ and $\Delta W2$ are defined within the window of each dip or burst as appropriate. While the VVV and WISE measurements are not contemporaneous, the amplitude of these multi-year dips and outbursts is captured fairly well by the data. We therefore have confidence in the overall trends revealed by these metrics even though individual stars are uncertain. In addition, the $\Delta W2$/$\Delta K_s$ ratio of seven dippers from Paper I are also shown in Figure~\ref{fig:amp_neo_vvv}. These objects are yet to be confirmed as YSOs by spectroscopic follow-up.

All of the 35 sources plotted in Figure~\ref{fig:amp_neo_vvv} have a bluer-when brighter, redder-when-fainter behaviour in the $K_s-W2$ colour. This behaviour of the infrared colours is typical of the of optically bright EXors where an outburst takes place in the inner disc, causing it to become hotter and bluer \citep{Lorenzetti2012}. However, Figure~\ref{fig:amp_neo_vvv} shows that dips and outbursts typically have different $\Delta W2 / \Delta K_s$ ratios. Dippers have smaller $\Delta W2 / \Delta K_s$ ratios, ranging between 0.18 to 0.50. Several dips with very low $\Delta W2 / \Delta K_s$ ratios can be explained by a source of obscuration that hides the near-infrared emitting-region of the disc more than the mid-infrared emitting region. Similar obscuration behaviour were recently observed on a Class II object AA Tau, in which the mid-infrared variation is anti correlated with optical variation \citep{Covey2021}. By contrast, the $\Delta W2 / \Delta K_s$ of most outbursts are larger than 0.4. The boundary between two morphologies,  $0.4 - 0.5$, is consistent with the infrared extinction laws \citep[e.g.][]{Fitzpatrick1999, McClure2009, WangS2019}. Although some overlap is seen around the boundary, distinct distributions of the $\Delta W2 / \Delta K_s$ ratio are seen for dips and outbursts, which can be attributed to changes in extinction and accretion rate, respectively.

Since the VVV and {\it WISE} light curves were not obtained simultaneously, the measurement of $\Delta W2 / \Delta K_s$ for individual YSOs is imprecise, especially for variables with low amplitude or strong variation on a short timescale. However, colour variations of periodic variables having long periods are more precisely measured in the phase-folded light curves.  Figure~\ref{fig:irsed} presents phase-folded multi-band light curves of v32, DR4\_v17, and DR4\_v55 with their SEDs at photometric minima and maxima measured within the time coverage of the {\it WISE} light curves. All three objects have emission line spectra, though the CO overtone emission and Br$\gamma$ emission are relatively weak (or absent in the case of DR4\_v55, which shows H$_2$ emission only). In each case the spectra were taken in the faint part of the cycle. For DR4\_v17, the SED measured at photometric maximum is well reconstructed by applying a de-reddening function to the data taken at the minimum ($A_{K_s} = 2.9$~mag, $R_v = 3.1$, \citealt{Fitzpatrick1999}). By contrast, DR4\_v55 and v32 have similar amplitudes in  $K_s$ and in $W2$, indicating their variations are likely linked to accretion bursts modulated by periodic perturbations. For these variable YSOs with periods around 1000 days, the corresponding Keplerian orbital radius is around 2~au for a solar mass star\footnote{$r \sim 2 (P / 1050\,{\rm d})^{2/3}(M_*/\rm M_{\odot})^{1/3}$ au}. 

The study in this section provides a statistical view of the infrared colour variation of YSOs with variability on long timescales. Most dippers have greater infrared colour variation, $\Delta(K_s-W2)$, than accretion dominated eruptive YSOs, which is in line with the infrared dust extinction law ($A_{W2}/A_{K_s} = 0.4 - 0.5$). A few YSOs in this work have both variable line-of-sight extinction and variable mass accretion process, so that $\Delta W2 / \Delta K_s$ ratios is time dependent. For example, Stim5 (green dot in Figure~\ref{fig:amp_neo_vvv}) began an outburst in 2010 that was ongoing throughout the nine year light curve, with $\Delta W2 / \Delta K_s = 0.55$. It is therefore classified as an eruptive YSO. However, between 2017 and 2019 there was a dip with $\Delta W2 / \Delta K_s = 0.43$. Multi-epoch spectroscopy provides a more reliable interpretation of the variable mechanism: we discuss the variability of Stim5 in \S\ref{sec:Stim5}.

\begin{figure*}
\centering
\includegraphics[width=4.2in,angle=0]{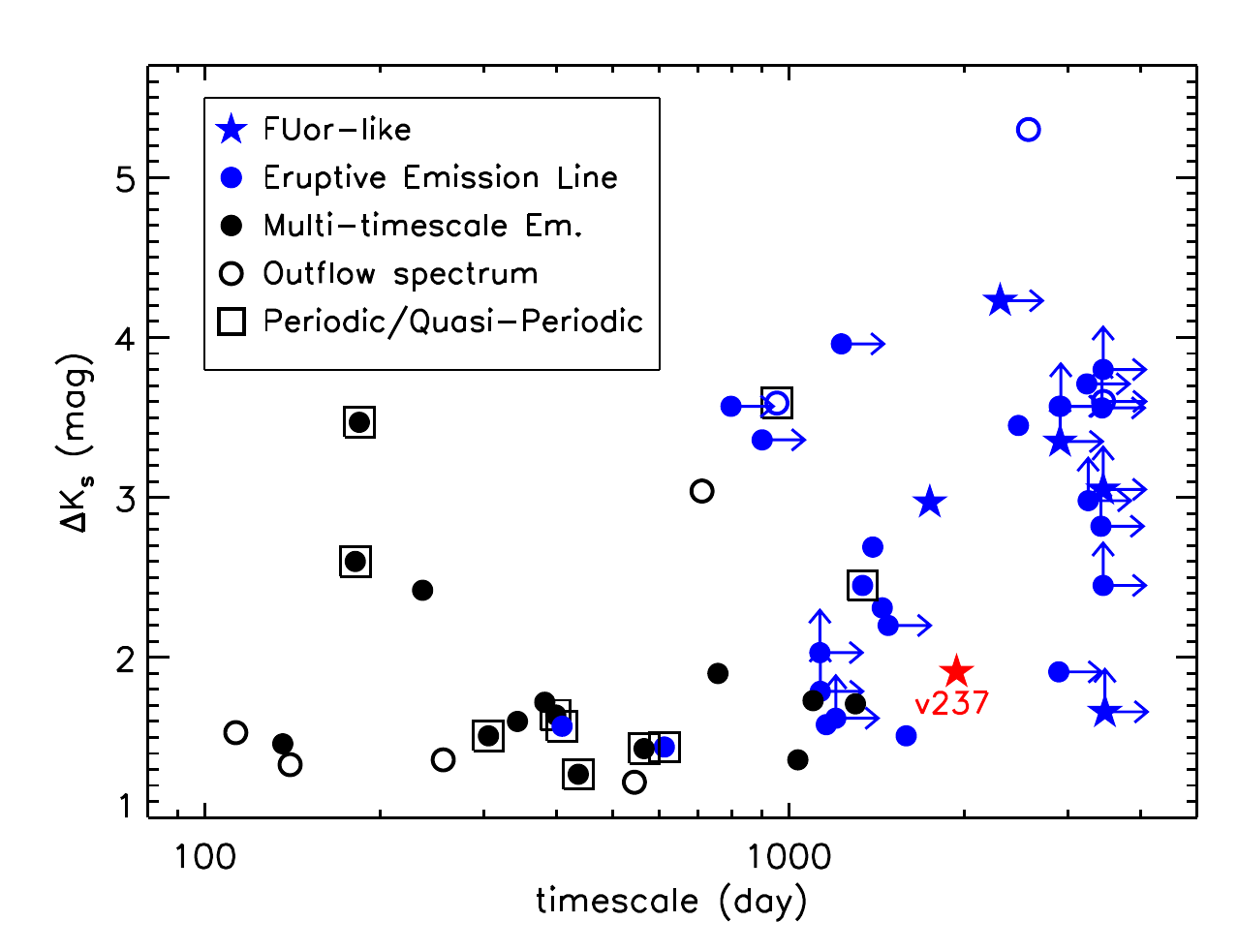}
\caption{$K_s$-band amplitude ($\Delta K_s$) and duration of spectroscopic confirmed eruptive YSOs in the VVV survey. FUor-like YSOs are shown by stars. Emission line YSOs are shown by circles, including YSOs with eruptive light curve (blue colour), MTV (black colour), YSO with magnetospheric accretion scenario (filled circle) and YSOs with outflow dominated spectra (open circle). Periodic and quasi-periodic variables are marked by open squares. FUor-like YSO v237 is shown by a red star as its light curve is dominated by extinction dips.}
\label{fig:timescale}
\end{figure*}

\subsection{Variation timescales of eruptive YSOs}
\label{sec:var_t}
We now consider the timescale of accretion-driven outbursts broken down into the different spectroscopic and light curve categories, in order to gain insight into the physical mechanisms causing these events. As mentioned earlier, the timescale refers to the total duration of an outburst from the start of the event until the return to quiescence, or to the period in the case of the periodic systems. Later in this section we also discuss the rise-times of the longer duration events. Classical FUor and EXor outbursts have quite distinct duration. Most optical EXor outbursts in the literature lasted less than 1000 days \citep{Herbig1989}, while the three classical FUors have remained in outburst for several decades \citep{Hartmann1996}. In Paper II, the ``MNor'' category was tentatively proposed for YSO outbursts of intermediate duration.

Individual objects are placed in the following categories based on their spectra and their light curves.
\begin{itemize}
    \item YSOs with FUor-like spectra
    \item Eruptive YSOs with emission line spectra
    \item Multiple timescale YSOs with emission line spectra
    \item YSOs with outflow-dominated emission line spectra
    \item Periodic or quasi-periodic YSOs with emission line spectra
\end{itemize}

For the first three of categories in the list, there is good spectroscopic evidence for accretion-driven outbursts, either from a FUor-like spectrum or from signatures of magnetospheric accretion. The outflow-dominated systems are less clear-cut: stars in this category show strong H$_2$ emission and sometimes [Fe{\sc ii}] emission but lack spectral signatures of magnetospheric accretion or boundary layer accretion. The effect of geometry might play a role on these ``bona-fide outbursts''. In a nearly edge-on system, the emission from the inner disc is heavily obscured while the $H_2$ emission lines from a jet can dominate the infrared spectrum. The measured line-of-sight extinction of these objects is discussed in \S\ref{sec:extinction} and Appendix~\ref{sec:extinction_appendix}. The timing of the spectroscopy also matters. The three YSOs (DR4\_v15, DR4\_v55, and DR4\_v67) show a long term outburst in their light curves but have outflow-dominated spectra that were taken during the quiescent state. In Figure \ref{fig:timescale} we exclude YSOs that only have dips in light curves attributed to changing extinction, including the periodic variables v467 and DR4\_17, see \S\ref{sec:neowise}. Other periodic variables are retained, some of which have clear evidence for accretion-driven variability. The categories of individual YSOs are shown in Table~\ref{tab:lum}.

The duration and $\Delta K_s$ values of 52 variable YSOs are presented in Figure \ref{fig:timescale}. Two variable stars with timescales around 30 days are not presented (v65 and DR4\_v30). The key results are: (i) all six of the FUor-like YSOs have timescales $\ge$1900~days but (ii) there is no clear separation between them and the emission line objects in duration or amplitude. In fact, the majority of long duration outbursts show emission line spectra.

The multi-year-long photometric variations in the FUor-like YSO v237 (marked by red star in Figure~\ref{fig:timescale}) are due to variable line-of-sight extinction rather than changes in accretion rate. Similar variations have been observed in well-studied FUors such as V1515~Cyg \citep{Clarke2005}. Ongoing accretion bursts are seen in two FUor-like YSOs (v721 and DR4\_v20) with $\Delta K_s > 3$~mag and $t > 3400$~d. Both the $K_s$ amplitude and duration of these two FUor-like YSOs are lower limits since their light curves began to rise before the VVV survey started in 2010. 

As the only FUor-like YSO where the entire outburst was captured in this work, v322 (see \S4.2.2 in Paper III) exhibited a fast outburst (2.4~mag in 200~d) followed by a rapid initial decay at a rate of 1.56~mag~yr$^{-1}$ that then slowed down. The rapid initial decay of v322 is rarely seen among FUor-like YSOs and is comparable to some EXors. One previously discovered ``FUor-like'' Class 0 object V371~Ser (also known as EC~53) has similar low amplitude and multi-year photometric behavior to v322, and it also has $H_2$ emission lines like v322 \citep{Hodapp2012, Connelley2018, YHLee2020}. 

To investigate the pre-outburst brightness of eruptive objects, we gathered the 2MASS photometry \citep{Skrutskie2006} from the IRSA online server. The results of individual sources are presented in Appendix~\ref{sec:2mass}.

\subsubsection{MTVs}

MTVs were defined in Paper III to have significant variability on short timescales ($t \sim 50$~d) as well as large inter-year variability. In this work, the MTVs are found to be YSOs with similar variation on inter-year and intra-year timescales, and lack of a strong longer term decline or outburst in their VVV light curves. In some cases (e.g. DR4\_v30, v181), there are long term trends in the light curves but their amplitude is smaller than the inter-year variation. Many photometric variations of MTVs are only partially captured by the VVV/VVVX light curves due to their sparse sampling. Hence, the variation amplitudes and timescales of MTVs are highly uncertain. In Figure~\ref{fig:timescale} we see that most MTVs (except v181) have relatively short timescales for their inter-year variations, consistent with the above definition, and the amplitudes are lower than is seen in most of the longer-timescale variables. 

Among 20 MTVs, five have outflow-dominated spectra, a higher proportion than that found among longer timescale emission line YSOs (2/30). It is not a very surprising result because MTVs lack of major outbursts which lead to emission features like Br$\gamma$ and CO band heads. The variable nature of outflow-dominated sources are likely linked to inner disc structures, either resulting from unstable mass accretion or variable line-of-sight extinction. Six MTVs have quasi-periodic variations. The spectral indices of MTVs are similar to longer timescale variables, as -0.37 $<$ $\alpha_{\rm class}$ $<$ 1.30. Most of these MTVs are listed as ``eruptive candidates'' in Table~\ref{tab:lum} as no major (long term) eruptive event were observed. 

\subsection{Extinction and luminosity measurements}
\label{sec:extinction}

In this work, the line-of-sight extinctions ($A_{K_s}$) of YSOs are measured by two methods, either by their locations on the near-infrared colour-colour diagram or by the flux ratio between $H_2$ 1-0 Q(3) (2.42 $\mu$m) and 1-0 S(1) (2.12 $\mu$m) lines, where possible. The details of two measurement methods and further discussions are presented in Appendix~\ref{sec:extinction_appendix}. The measured $A_{K_s}$ from both methods are listed in Table~\ref{tab:lum}. One should notice that the measured $A_{K_s}$ is the integrated extinction along the line-of-sight to the star, including interstellar extinction between the sun to the star forming region, the extinction within the molecular cloud and the circumstellar extinction. 

Nine YSOs in the sample have outflow-dominated spectra. The $A_{K_s}$ value is measured for all nine YSOs, mostly by the $H_2$ flux ratio, five of the nine have intermediate extinction, and four have $A_{K_s}$ ranging from 2.20 to 3.79 mag, significantly higher than the extinction found in other emission line objects measured by both methods. We infer that in erupting systems where Br$\gamma$ and CO overtone emission are not seen, this is sometimes due to heavy obscuration of the inner regions of the system. This suggests that high disc inclination is playing a role (i.e. close to edge-on orientation). Evolutionary stage may also be a factor but the spectral indices of the four highly extinguished outflow-dominated sources span a wide range ($\alpha_{SED}$ = 0.11, 0.33, 1.56 for DR4\_v89, DR4\_v15 and DR4\_v55, respectively, whilst v473 lacks a $W4$ detection).

The in-outburst infrared luminosity ($L_{\rm IR}$) and the rise in luminosity from the quiescent state ($\Delta L_{\rm IR}$) of eruptive YSOs are measured using the extinction-corrected $2-22 \mu$m SED. 
The extinction correction is conducted based on the $A_{K_s}$ value measured above, in which the $H_2$ line ratio method is preferred, and the empirical near-to-mid infrared extinction law is from \citet{McClure2009}. For preference, the distances used in the calculation are radial-velocity-based kinematic distances ($d_k$, listed in Table~\ref{tab:lines}) and the distance ambiguity is broken using the VIRAC2 proper motions, see $\S$\ref{sec:dists}. Where $d_k$ is unavailable, we use the distance to the star forming region ($d_{\rm SFR}$) given in Table~\ref{tab:info} or the distance determined in Paper II \footnote{We exclude v621 from this measurement due to its unknown distance. The $d_k$ value, based on the Br$\gamma$ line, disagrees with the PM solution. No distance has been published for the star forming region.}.
To measure the quiescent luminosity we use the $K_s$ and $W1$ to $W4$ magnitudes from VVV and ALLWISE, these data being taken in 2010 when the sources were in a faint state. To calculate the in-outburst luminosity,
$L_{\rm IR}$, we use the $K_s$, $W1$, and $W2$ magnitudes measured in the bright state. The mid-infrared luminosity increased during outburst can be affected by many mechanisms \citep{MacFarlane2019} so we estimate the in-outburst $W3$ and $W4$ magnitudes by assuming that the $W2$-$W3$ and $W3$-$W4$ colours measured in the ALLWISE survey remain constant whilst the $W2$ flux increases. Luminosity is not calculated for YSOs lacking $W3$ and $W4$ detections in ALLWISE, such as the FUor-like source v322. This leaves us with $L_{\rm IR}$ values for five FUor-like YSOs. In a few systems the luminosity is a lower limit because the extinction value, $A_{K_s}$, is a lower limit. These luminosity calculations are obviously uncertain but they allow us to make comparisons between FUor-like YSOs and emission line YSOs in the sample and eruptive YSOs in the literature.

\begin{figure}
    \centering
    \includegraphics[width=3.3in,angle=0]{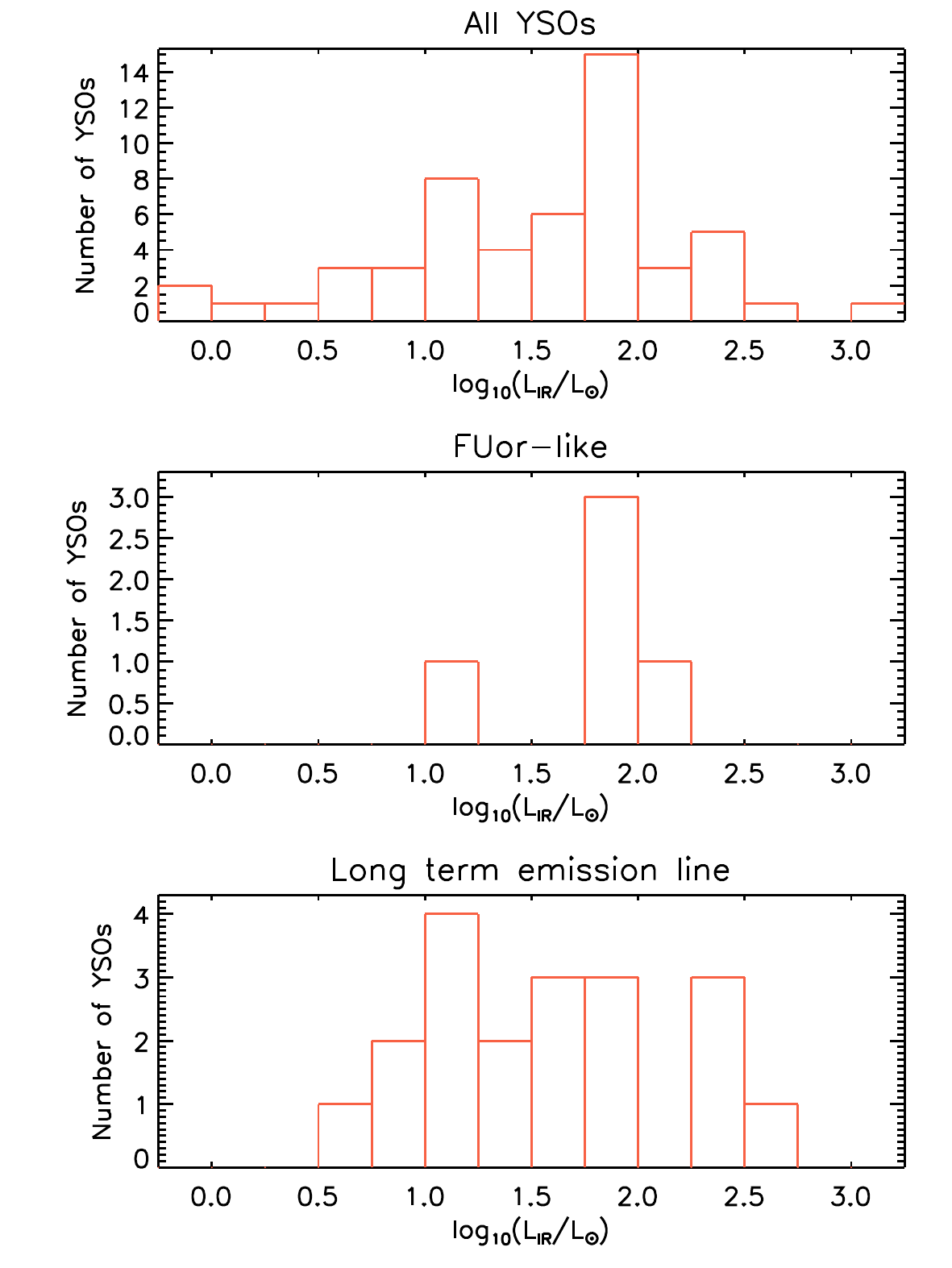}
    \caption{Histograms of the infrared luminosity ($L_{\rm IR}$, between $2 - 22 \mu$m ) of YSOs in this work. All YSOs, FUor-like objects and eruptive YSOs with emission line spectra are presented separately. YSOs lack of {\it WISE} detections are not shown in this figure.}
    \label{fig:bol_lum}
\end{figure}

The distribution of $L_{\rm IR}$ of FUor-like objects and eruptive YSOs with emission line spectra are presented in Figure~\ref{fig:bol_lum}. 
In our sample, FUor-like objects have $L_{\rm IR}$ ranging from 10 to 143 L$_{\odot}$, which is similar to previously identified FUors \citep[e.g.][]{Connelley2018}. Long-term eruptive emission line YSOs have a similar range of luminosities to the FUor-like YSOs.
In many sources the quiescent luminosity is not measured, e.g. if the source was in outburst in 2010. However, the rise in luminosity, $\Delta L_{\rm IR}$, is approximately equal to $L_{\rm IR}$, the in-outburst luminosity, where this is measured. $L_{\rm IR}$ is therefore a rough estimate of the accretion luminosity during the bright state, in the measured wavelength range
\footnote{In Paper II, the accretion luminosity of emission line YSOs was estimated using $L_{\rm Br\gamma}$, which was strongly model-dependent and could be heavily biased by the veiling effect from the continuum.}.

In long term magnetospheric accretion controlled outbursts, $L_{\rm IR}$ is typically $10^1$ to $10^2$ L$_{\odot}$. This is a little higher than many previously discovered EXors in nearby star forming regions, such as EX Lupi \citep{Aspin2010} and others \citep{Lorenzetti2012}. This is likely to be a selection effect, owing to the larger distances to the YSOs presented here. Our results indicate that accretion rates are typically of the same order for the boundary layer mode and the magnetospheric mode in embedded Class I YSOs. This contrasts with nearby optically visible systems where EXors typically have lower luminosities and accretion rates than FUors. The measured extinctions, luminosities and classifications of the 61 spectroscopically confirmed YSOs are listed in Table~\ref{tab:lum}.

\subsection{Classification}

In this section, we refine the classification of eruptive YSOs based on their photometric and spectroscopic features. In Paper II, ``MNor'' was defined as a tentative subgroup of eruptive YSOs. ``MNors'' have mixed characteristics of classical EXors and classical FUors, and with Class I or flat-spectrum SEDs. In terms of outburst duration (often intermediate, i.e. 1--5~yr) or spectral features (emission or absorption), ``MNors'' are opposite to those expected from the outburst timescale. Under this definition, ``MNor'' contains both emission line YSOs having outbursts longer than typical ``EXors'' and FUor-like objects without decades-long photometric eruptions. Among eruptive YSOs in the VVV survey, we found that ``MNors'' are more common than either ``FUors'' or ``EXors'' among eruptive YSOs with a large variety of duration and amplitude. The term is therefore not especially useful, except to highlight that the FUor/EXor categories defined for optically bright systems cannot encompass what is seen in embedded systems. Hence, in this work, eruptive YSOs are sorted by accretion mode and light curve properties instead of traditional categories. The category of FUors is fairly well defined by their unique spectral features, long duration and the outburst luminosity, whilst there are some differences in the rise time of individual objects. On the other hand, we favour ``emission line objects'' to ``EXors'' owing to the wide range in duration of eruptive events controlled by the magnetospheric accretion mode. The final classification is presented in Table~\ref{tab:lum}.

Among the 16 variables with the longest duration ($t > 2000$~d) in our sample, nine have indicators of the magnetospheric accretion process and two are dominated by outflows. We considered whether this result might be affected by disc orientation. Discs viewed close to face-on (i.e. at small inclinations) are more likely to display emission line features arising in the innermost disc and in magnetospheric accretion columns, whereas these can be hidden from view at high inclinations. We do not see significantly lower extinction in emission line objects than FUor-like sources, which indicates that our result is not strongly biased by the disc orientation (see \S\ref{sec:extinction_appendix}). We conclude that magnetospheric accretion can be maintained during long-duration outbursts and that this is more common than boundary layer accretion in Class I YSOs. 

Both slow-rising and fast-rising outbursts were observed among emission line objects. Slow-rising eruptive YSOs (e.g. Stim1 and DR4\_v15) are possibly triggered by thermal-instability where the instability is formed at the inner disc without a dynamical perturbation and then propagates inside-out \citep{Lin1985, Bell1994}. On the other hand, outbursts with fast eruptive nature (e.g. v621 and DR4\_v34), could be consequences of disc fragments being accreted onto the protostar \citep[e.g.][]{Vorobyov2006}.

\subsection{AGB stars and symbiotic systems: v51, v319 and v370}

In this work, we identified three AGB stars via the $^{12}$CO and $^{13}$CO absorption bands. Notably, two AGB stars, v319 and v370, exhibit both Br$\gamma$ emission and CO absorption. In addition, the WISE $W1$-$W2$ colours of these two stars are bluer than solitary AGB stars. We hereby identify v319 and v370 as D-type symbiotic stars \citep[e.g.][]{Kenyon1984, Corradi2008}. More detailed description of these three AGB stars are provided in Appendix~\ref{sec:AGB}.

\section{Summary and Conclusions}
\label{sec:con}

In this work, we report results of spectroscopic follow-up for 38 high amplitude variable stars ($\Delta K_s > 1$~mag) discovered in the VVV survey, including 33 that are spectroscopically confirmed as YSOs. We add them to the sample of 28 confirmed YSOs from Paper II for a combined sample of 61 highly variable YSOs. Based on spectral index, 56 of them are Class I and flat-spectrum objects. We publish the VIRAC2 version of the VVV $K_s$ light curves from 2010 to 2019 for all 61 VVV YSOs (available in the online supplemental material, examples in Table~\ref{tab:lc_example}). The sample includes some YSOs with long duration outbursts  (typically 2--4 mag amplitude), some with periodic variability, some MTVs and some with extinction-dominated variability. There are also a few post-main-sequence variable stars. 

Our main findings are as follows.
\begin{itemize}
\item Spectroscopically, the 61 variable YSOs are sorted as 55 emission line YSOs and 6 FUor-like objects. Among emission line YSOs, 46 sources display emission features linked to the magnetospheric accretion process (Pa$\beta$ or Br$\gamma$) usually accompanied by CO overtone emission and sometimes Na~{\sc i} emission. The remaining nine emission line sources show only lines associated with a molecular outflow or disc wind ($H_2$ or [Fe {\sc II}]). The FUor-like objects have $^{12}$CO overtone absorption bands and water absorption in the $H$ and $K_s$ bandpasses, associated with the boundary layer accretion process. 

\item The dominant timescales of variability seen in the YSO sample range from months to a decade for different sources. FUor-like spectra are only seen during outburst events with duration $\ge$5~yr. 

\item Signatures of magnetospheric accretion were observed among most eruptive YSOs with long duration outbursts. Therefore, the stellar magnetic field is still in control during these outbursts, which have similar infrared amplitudes and luminosities to the FUor-like sources. This contrasts with past results for long duration optically selected events, where FUor-like spectra are almost always seen \citep[e.g.][]{Audard2014}. There is no clear difference in the measured extinction for the two spectral categories, indicating that the spectral classification is not driven primarily by the disc inclination angle and veiling. 

\item {Similar in-outburst luminosities were seen among emission line and FUor-like eruptive YSOs which suggests that they may have similar accretion rates.}

\item Periodic accretors comprise a significant minority amongst eruptive protostars. Using the Lomb-Scargle method, photometric periods are detected in several emission line YSOs, though three or four of these 
require more data to confirm or disprove the periodic behaviour.
 For the four longest period systems ($P > $~950~d), two systems are periodic accretors and two systems as cases of periodic extinction identified by the $\Delta{W2}/\Delta{K_s}$ colour ratio. While a determination could not be made for the other nine quasi-periodic YSOs. The finding of periodic accretors suggests that an orbiting companion or star-disc interactions are be able to modify the mass accretion rate.  At present, the two cyclically accreting YSOs, v32 and DR4\_v55, only have spectra taken in the quiescent state, where relatively weak CO overtone and Br$\gamma$ emission features were observed comparing to other eruptive spectra taken in outburst. Multiple epochs of spectroscopy and multi-wavelength photometry will be needed to study the cyclical accretion and to distinguish between accretion and extinction in the other periodic systems. 

\item MTVs have relatively short duration outbursts ($<1$--3~yr). First noted in Paper III as YSOs that vary strongly on timescales of weeks and years, these systems show repeated bursts or dips and they have typical amplitudes below 2~mag in $K_s$. The rapid component of the variability is likely linked to an unstable accretion process, star-disc interactions or inner disc structures causing variable extinction. By contrast, frequent intra-year high amplitude variations are not often seen among long-term eruptive YSOs.

\item Variable line-of-sight extinction contributes to the photometric variation in a few sources. Such changes are sometimes seen on the ``bright plateau'' in the light curve of eruptive YSOs (e.g. Stim5). The extinction dips are recognised by light curve morphologies and are further identified by a smaller $\Delta{W2}/\Delta{K_s}$ ratio than is seen in outbursts. As noted above, the long period variability in two of the emission line YSOs is extinction driven, presumably by asymmetric disc structures. The light curve of one FUor-like source, v237, is also dominated by extinction dips in the 2010--2019 time interval. The combined 2MASS and VVV light curve suggests that v237 has been in a high state of accretion for at least two decades.  

\item In this work, five sources are identified as post-main-sequence stars. The quasi-periodic variable v371 is a newly discovered Wolf-Rayet star in a binary system, which is named as WR80-1 in accordance with the naming convention. Two AGB stars with periodic/quasi-periodic light curves are classified as D-type symbiotic stars with Br$\gamma$ emission features arising from the accretion process of their companion stars. 

\end{itemize}

\begin{table*}
\renewcommand\arraystretch{1.0}
\centering
\caption{}
\begin{tabular}{l r c c c c c c c c c}
\hline
\hline
Name & Spectral Class & Final Classification & Distance & $L_{\rm IR}$ & $\Delta L_{\rm IR}$ & $A_{K_s}^{1}$ & $A_{K_s}^{2}$ & Timescale & $\Delta K_s$ & $\alpha_{\rm class}$\\
 & & & kpc & $\log$(L$_\odot$) & $\log$(L$_\odot$) & mag  & mag & day & mag & \\
\hline
     v14 &  Emission Line  &  Eruptive Candidate &  1.9$^{a}$ &  0.16 & -0.11 &  0.00 &   -   &  398 &  1.6 &  0.61 \\
     v16 &   FUor-like     &  FUor-like  &  8.2$^{a}$ &  1.84 &  1.23 &  0.18 &   -   & 3476 &  1.7 &  0.07 \\
     v20 &  Emission Line  &  Eruptive with Magnetosphere &  2.5$^{c}$ &  1.97 &  1.65 &  1.40 & 1.71 & 1100 &  1.7 &  0.47 \\
     v32 &  Emission Line  &  Eruptive with Magnetosphere &  1.9$^{a}$ &  0.75 &  0.50 &  0.49 &   -   & 1337 &  2.5 &  0.08 \\
     v53 &  Emission Line  &  Eruptive Candidate & 13.9$^{a}$ &  2.19 &  1.99 &  0.00 &   -   &  236 &  2.4 &  0.61 \\
     v63 &  Emission Line  &  Eruptive Candidate &  3.5$^{b}$ &  0.74 &  0.36 &  0.89 &   -   &  136 &  1.5 & -0.24 \\
     v65 &  Emission Line  &  Eruptive Candidate &  3.5$^{b}$ &  1.08 &  0.80 &  1.00 &   -   &   30 &  1.4 & -0.27 \\
     v94 &  Emission Line  &  Eruptive with Magnetosphere &  3.5$^{b}$ &  1.56 &  1.29 &  0.00 &   -   &  436 &  1.3 &  0.17 \\
    v118 &  Emission Line  & Eruptive Candidate &  2.2$^{c}$ &  0.91 &  0.88 &  0.19 &   -   &   50 &  3.5 &  0.12 \\
    v128 &  Emission Line  &  Eruptive with Magnetosphere &  2.6$^{b}$ &  1.30 &  1.06 & >1.00 &   -   & 1131 &  1.8 &  0.56 \\
    v181 &  Emission Line  &  Eruptive Candidate &   -  &  -   &   -   &  0.79 &   -   &  183 &  3.5 &  0.58 \\
    v190 &  Emission Line  &  Eruptive Candidate &  4.0$^{a}$ &  1.25 &  0.57 &  1.28 & 2.50 &  612 &  1.4 & -0.43 \\
    v193 &  Emission Line  &  Non-eruptive &  4.5$^{a}$ &  1.92 &  1.79 & >2.01 &   -   &  763 &  1.4 &  0.35 \\
    v237 &   FUor-like     &  Pre-VVV-eruptive? &  5.0$^{a}$ &  1.91 &  1.34 & >0.96 &   -   & 1943 &  1.9 &  0.88 \\
    v270 &  Emission Line  &  Eruptive with Magnetosphere &  4.6$^{a}$ &  1.79 &  1.61 &  1.55 &   -   & 3237 &  3.7 &  0.87 \\
    v309 &  Emission Line  &  Eruptive with Magnetosphere &  2.8$^{a}$ &  2.35 &  1.99 & >1.42 & 1.45 &  181 &  2.6 &  1.30 \\
    v322 &   FUor-like     &  FUor-like  &  3.8$^{a}$ &   -   &   -   &  0.45 &   -   & 1743 &  3.0 &  0.82 \\
    v335 &  Emission Line  &  Non-eruptive &  4.2$^{a}$ &  1.86 &  1.51 &  0.71 &   -   & 1041 &  3.6 &  0.45 \\
    v374 &  Emission Line  &  Eruptive with Magnetosphere &  2.9$^{b}$ &  2.47 &  2.02 &  1.04 &   -   & 3254 &  3.0 &  0.56 \\
    v376 &  Emission Line  &  Eruptive with Magnetosphere &  1.2$^{a}$ & -0.04 & -0.37 &  0.90 &   -   & 1300 &  1.7 & -0.37 \\
    v389 &  Emission Line  &  Eruptive with Magnetosphere &  4.3$^{a}$ &  1.80 &  1.46 & >0.91 &   -   & 1036 &  1.4 &  0.68 \\
    v405 &  Emission Line  &  Non-eruptive &  2.3$^{b}$ &  1.85 &  1.64 & >0.15 &   -   &  733 &  2.2 &  1.10 \\
    v406 &     Outflow     &  Non-eruptive &  2.3$^{b}$ &  1.57 &  1.25 &  0.92 &   -   &  796 &  1.7 &  0.80 \\
    v452 &  Emission Line  &  Eruptive with Magnetosphere &  2.8$^{c}$ &  1.12 &  1.05 &  0.82 &   -   & 2472 &  3.5 &  0.21 \\
    v467 &  Emission Line  &  Non-eruptive (Periodic) &  9.0$^{a}$ &  1.86 &  1.58 &  0.98 &   -   & 1840 &  2.6 & -0.27 \\
    v473 &     Outflow     &  Outflow Dominated &  3.7$^{b}$ &  1.69 &  1.21 &  -  & 2.21 &  113 &  1.5 &  0.89 \\
    v480 &  Emission Line  &  Eruptive Candidate &  3.7$^{b}$ &  0.30 & -0.07 &  0.00 &   -   &  306 &  1.5 & -0.81 \\
    v562 &  Emission Line  &  Eruptive with Magnetosphere &  2.9$^{b}$ &  0.97 &  0.82 &  0.96 &   -   & 1392 &  2.7 & -0.02 \\
    v618 &     Outflow     &  Outflow Dominated &   -  &  -   &   -   &  0.77 & 1.29 &  140 &  1.3 & -0.09 \\
    v621 &  Emission Line  &  Eruptive with Magnetosphere &   -  &  -   &   -   &  1.26 &   -   & 2898 &  1.9 & -0.05 \\
    v625 &  Emission Line  &  Eruptive Candidate &  2.3$^{b}$ &  0.64 &  0.41 & >0.00 &   -   &  343 &  1.6 &  0.17 \\
    v628 &     Outflow     &  Outflow Dominated &  2.3$^{a}$ &  -   &   -   &  0.45 &   -   &  256 &  1.4 & -0.47 \\
    v630 &  Emission Line  &  Eruptive with Magnetosphere &  1.9$^{a}$ &   -   &   -   &  0.95 &   -   &  409 &  1.6 &  2.70 \\
    v631 &  Emission Line  &  Eruptive with Magnetosphere &  2.3$^{b}$ &  1.61 &  1.52 &  0.73 &   -   & 3422 &  2.8 &  0.62 \\
    v632 &  Emission Line  &  Eruptive Candidate &  2.5$^{b}$ &  1.32 &  0.84 &  0.14 &   -   &  382 &  1.7 & -0.05 \\
    v636 &  Emission Line  &  Eruptive Candidate &  2.8$^{b}$ &  1.07 &  0.77 &  1.19 &   -   &  565 &  1.4 &  0.43 \\
    v662 &  Emission Line  &  Eruptive with Magnetosphere &  3.1$^{b}$ &  1.11 &  0.82 & >0.00 &   -   & 1159 &  1.6 &  0.53 \\
    v665 &  Emission Line  &  Eruptive with Magnetosphere &  4.3$^{a}$ &  1.93 &  1.44 &  1.41 &   -   & 1203 &  1.6 &  0.58 \\
    v699 &  Emission Line  &  Eruptive with Magnetosphere &  4.9$^{a}$ &  1.87 &  1.26 & >1.05 & 1.94 &  756 &  1.9 &  0.91 \\
    v713 &  Emission Line  &  Eruptive with Magnetosphere &  3.5$^{a}$ &  1.58 &  1.40 &  0.19 &   -   & 1130 &  2.0 & -0.04 \\
    v717 &   FUor-like     &  FUor-like  &  6.1$^{a}$ &  2.16 &  1.77 &  1.63 &   -   & 2300 &  4.2 &  0.23 \\
    v721 &   FUor-like     &  FUor-like  &  4.9$^{a}$ &  1.98 &  1.80 &  0.85 &   -   & 3452 &  3.0 &  0.55 \\
    v800 &  Emission Line  &  Eruptive with Magnetosphere &  1.4$^{b}$ &  2.45 &  2.36 &  1.18 &   -   & 1479 &  2.2 &  0.94 \\
    v815 &     Outflow     &  Outflow Dominated &  3.1$^{b}$ &  1.28 &  0.87 &  -  & 0.48 &  544 &  1.2 &  1.08 \\
   Stim1 &  Emission Line  &  Eruptive with Magnetosphere & 10.5$^{a}$ &  2.47 &  2.35 &  0.40 &   -   & 3452 &  2.5 &  0.63 \\
   Stim5 &  Emission Line  &  Eruptive with Magnetosphere &  5.1$^{a}$ &  1.41 &  1.33 &  0.92 & 0.44 & 3452 &  3.8 &  0.90 \\
    Stim13 &  Emission Line  &  Eruptive with Magnetosphere &  2.7$^{a}$ &  2.73 &  1.61 & >0.40 & 1.22 & 1588 &  1.5 &  1.33 \\
  DR4\_v5 &  Emission Line  &  Eruptive with Magnetosphere &  1.8$^{a}$ &  1.23 &  0.69 &  1.14 &   -   & 1230 &  4.0 &  0.96 \\
 DR4\_v10 &  Emission Line  &  Eruptive with Magnetosphere &  4.2$^{a}$ &  0.87 &  0.57 & >1.06 &   -   & 3437 &  3.6 &  1.86 \\
 DR4\_v15 &     Outflow     &  Outflow Dominated &  4.3$^{b}$ &  2.21 &  2.04 &  -  & 3.29 & 3456 &  3.6 &  0.56 \\
 DR4\_v17 &  Emission Line  &  Non-eruptive (Periodic) &  4.3$^{a}$ &  1.92 &  1.79 &  1.05 &   -   & 1155 &  3.8 &  0.78 \\
 DR4\_v18 &  Emission Line  &  Eruptive with Magnetosphere &  5.2$^{a}$ &  1.68 &  1.60 &  0.54 &   -   &  796 &  3.6 &  0.21 \\
 DR4\_v20 &   FUor-like     &  FUor-like  &  4.7$^{a}$ &  1.02 &  0.78 & >0.60 &   -   & 2913 &  3.3 &  0.74 \\
 DR4\_v30 &  Emission Line  &  Eruptive Candidate &  6.7$^{a}$ &  2.31 &  2.21 &  1.02 &   -   &   30 &  3.9 &  0.36 \\
 DR4\_v34 &  Emission Line  &  Eruptive with Magnetosphere &  4.0$^{a}$ &  1.94 &  1.17 &  1.30 &   -   & 2904 &  3.6 & -0.19 \\
 DR4\_v39 &  Emission Line  &  Eruptive with Magnetosphere &  4.6$^{a}$ &  1.55 &  1.30 &  0.90 & 1.16 & 1445 &  2.3 &  0.51 \\
 DR4\_v42 &  Emission Line  &  Eruptive with Magnetosphere &  6.6$^{a}$ &  1.84 &  1.74 & >0.07 & 1.11 &  900 &  3.4 &  1.26 \\
 DR4\_v44 &  Emission Line  &  Eruptive with Magnetosphere &  3.9$^{a}$ &  1.17 &  0.89 & >1.05 &   -   & 2921 &  3.6 &  0.83 \\
 DR4\_v55 &     Outflow     &  Outflow Dominated &  3.4$^{b}$ &  3.04 &  2.99 &  -  & 3.79 &  954 &  3.6 &  1.97 \\
 DR4\_v67 &     Outflow     &  Outflow Dominated &  4.5$^{b}$ &  2.01 &  1.94 &  1.20 & 1.19 & 2572 &  5.3 &  1.05 \\
 DR4\_v89 &     Outflow     &  Outflow Dominated &  2.3$^{b}$ &  1.49 &  0.93 &  -  & 2.39 &  710 &  3.0 &  0.85 \\
\hline
\end{tabular}
\flushleft{``Eruptive candidate'' represents a YSO with eruptive spectrum but also lack of a significant long term variation.\\
1: $A_{K_s}$ measured by the location on near-infrared colour colour diagram. \\
2: $A_{K_s}$ measured by the flux ration between the $H_2$ 1-0 Q(3) and 1-0 S(1) emission \\
{a: Kinematic distance measured by the radial velocity of emission features.} \\
{b: Distance estimated by the nearby star-forming regions. c: Distance estimated from the SED fitting in Paper II.}
}
\label{tab:lum}
\end{table*}

\section*{Acknowledgements}

ZG, PWL, and CJM acknowledge support by STFC Consolidated Grants ST/R00905/1, ST/M001008/1 and ST/J001333/1 and the STFC PATT-linked grant ST/L001403/1. We thank the staff of the Magellan Telescopes and the European Southern Observatory for their work in operating the facilities used in this project. This work has made use of the University of Hertfordshire's high-performance computing facility (\url{http://uhhpc.herts.ac.uk}).

The contribution of CCP was funded by a Leverhulme Trust Research Project Grant

We gratefully acknowledge data from the ESO Public Survey program ID 179.B-2002 taken with the VISTA telescope, and products from the Cambridge Astronomical Survey Unit (CASU). This paper includes data gathered with the 6.5 meter Magellan Telescopes located at Las Campanas Observatory, Chile during Chilean programes: CN2014A-16 and CN2015A-78. D.M. and J.A.-G. thank support from the BASAL Center for Astrophysics and Associated Technologies (CATA) through grant AFB170002 and FONDECYT Regular grant No. 1170121. 

J.B., R.K., and J.A.-G. are supported by ANID, Millennium Science Initiative ICN12\_009, awarded to the Millennium Institute of Astrophysics (MAS). J.A.-G. acknowledges support from Fondecyt Regular 1201490.

A.C.G. acknowledges funding from the European Research Council (ERC) under the European Union's Horizon 2020 research and innovation programme (grant agreement No.\, 743029).

Supported by the international Gemini Observatory, a program of NSF's NOIRLab, which is managed by the Association of Universities for Research in Astronomy (AURA) under a cooperative agreement with the National Science Foundation, on behalf of the Gemini partnership of Argentina, Brazil, Canada, Chile, the Republic of Korea, and the United States of America.

MSNK  acknowledges  the  support from  FCT  -  Funda\c{c}\~{a}o para a Ci\^{e}ncia  e  a  Tecnologia  through  Investigador contracts  and exploratory project   (IF/00956/2015/CP1273/CT0002).

This research has made use of the NASA/IPAC Infrared Science Archive, which is funded by the National Aeronautics and Space Administration and operated by the California Institute of Technology.

\section{Data Availability}
The WISE, Spitzer and 2MASS data underlying this article are publicly available at \url{https://irsa.ipac.caltech.edu/Missions/wise.html}, \url{https://irsa.ipac.caltech.edu/Missions/spitzer.html}, 
and \url{https://irsa.ipac.caltech.edu/Missions/2mass.html} respectively.
The VVV and VVVX data are publicly available at the ESO archive \url{http://archive.eso.org/cms.html}. The relevant reduced products for the most recent VVVX epochs have not yet been publicly released but are available on request to the first author. The lasted VVV/VVVX $K_s$-band light curves of the spectroscopic follow-up samples are provided in Table~\ref{tab:lc_example}. The reduced spectra are provided at \url{http://star.herts.ac.uk/~pwl/Lucas/GuoZ/VVVspec/}.

\bibliographystyle{mnras}
\bibliography{reference}

\appendix

\section{Variable YSOs discovered by W. Stimson}

In Paper I, the co-author W. Stimson carefully searched two VVV disc tiles (d064 and d083) for highly variable stars ($\Delta K_s \ge $1~mag) in order to estimate the completeness of the larger search in that work, where strict data quality criteria were used. The search by Stimson found 14 variable stars that were not in the list of 816 variable sources published in Paper I. Spectra for three of these variable stars, Stim1, Stim5 and Stim13, are presented in the present work. No details of these 14 stars were given in Paper I so we list here their coordinates and classifications. We use the abbreviation “Stim” to distinguish these variable stars from sources found in our other searches. The classifications draw on the 2010—2019 VIRAC2 light curves, the infrared colours available from WISE, {\it Spitzer}-GLIMPSE and VVV/VIRAC2 and a search for evidence for star formation activity within a 5 arcminute radius of each source using the SIMBAD database and the WISE colour images \citep[see][for details of the approach with these ancillary data]{Lucas2017}.

\begin{table*}
\centering
\caption{Details of the 14 variable stars found by Stimson}
\begin{tabular}{l c c c c} 
\hline
Name & R.A. & Declination & Class & Light Curve Properties\\ 
\hline
Stim1 & 12:57:44.23 & -62:15:06.4 & YSO &  Eruptive \\
Stim2 & 12:58:29.27 & -61:47:25.0 & YSO &  Eruptive \\
Stim3 & 13:05:30.03 & -62:29:18.7 & YSO &  STV \\
Stim4 & 13:06:37.34 & -62:35:04.3 & YSO &  Dipper \\
Stim5 & 16:12:14.41 & -51:50:24.4 & YSO &  Eruptive \\
Stim6 & 16:13:12.28 & -51:39:03.7 & YSO &  Dipper, MTV \\
Stim7 & 16:14:45.48 & -50:57:49.0 & YSO &  Fader, MTV \\
Stim8 & 16:14:51.45 & -51:45:41.9 & AGB &  P = 860 d \\
Stim9 & 16:16:08.12 & -51:41:31.5 & AGB &  P = 591 d \\
Stim10 & 16:16:13.16 & -51:03:56.8 & EB &  P = 1.22058 d \\
Stim11 & 16:17:32.79 & -51:16:55.9 & YSO &  Eruptive, MTV \\
Stim12 & 16:17:59.50 & -51:57:05.7 & YSO &  Dipper, MTV \\
Stim13 & 16:19:10.80 & -51:03:53.0 & YSO &  Dipper \\
Stim14 & 16:21:42.16 & -51:23:29.6 & Microlensing & - \\
\hline
\end{tabular}
\label{tab:stimson}
\end{table*}

The periods of the two AGB stars and the eclipsing binary (EB) were measured with the PDM2 code of Stellingwerf \citep{Stellingwerf1978}, giving a very precise result in the case of the EB.  Stim14 is listed as a microlensing event, based on visual inspection of the light curve for this isolated blue star. However, the event is not well sampled so this interpretation is not certain. The classification of the YSOs is secure in 8/10 cases. Only Stim2 and Stim12 lack firm evidence for location in a star formation region but their light curve properties and the red $I1-I2$ colours measured by {\it Spitzer}-GLIMPSE, support a YSO interpretation.

\section{VVV light curves}
This section presents the VVV/VVVX $K_s$ light curves of the 38 sources observed in this work. The phase-folded light curves of periodic and quasi-periodic objects in our series are presented in Figure~\ref{fig:phase_fold}. The VIRAC2 $K_s$ light curves of the 61 spectroscopic confirmed YSOs, and the 5 post-main-sequence objects observed in this work are listed in Table~\ref{tab:lc_example}. The infrared magnitudes, including $J$, $H$, $K_s$ from VVV/VVVX, $W1$--$W4$ from {\it ALLWISE}, and $I1$ to {\it MIPS24} from {\it Spitzer} are presented in Table~\ref{tab:info_sed}. Plots of the VIRAC2 light curves of the YSOs from Paper~II are presented in Appendix C and are included in the online Supplementary info.

\begin{table*} 
\caption{VVV/VVVX VIRAC2 light curves for the full spectroscopic sample}
\renewcommand\arraystretch{1.}
\begin{tabular}{l c c c c c c}
\hline
\hline
Name & GLON & GLAT & Observation time & $K_s$ & $\delta K_s$ & $\chi$-value \\
& deg & deg & MJD & mag & mag & \\
\hline
DR4\_v5 &  307.318755 &    0.156974 &  55271.2096 & 14.16 &  0.01 &   1.0 \\
DR4\_v5 &  307.318755 &    0.156974 &  55271.2105 & 14.17 &  0.01 &   0.8 \\
DR4\_v5 &  307.318755 &    0.156974 &  55293.0966 & 13.51 &  0.25 &   1.5 \\
DR4\_v5 &  307.318755 &    0.156974 &  55293.0987 & 13.13 &  0.01 &   0.9 \\
DR4\_v5 &  307.318755 &    0.156974 &  55293.0990 & 13.15 &  0.02 &   1.0 \\
\hline
\hline
\end{tabular}
\flushleft{The full table is provided in the online supplementary files.}
\label{tab:lc_example}
\end{table*}

\begin{table*} 
\caption{Infrared magnitude of the 38 observed sources in this work}
\begin{tabular}{l | c c c | c c c c | c c c c c | r}
\hline
\hline
Name & $J$ & $H$ & $K_s$ & W1 & W2 & W3 & W4 & I1 & I2 & I3 & I4 & 24 $\mu$m & $\alpha_{\rm class}$ \\
\hline
v14 & 17.68 & 16.72 & 15.57 & 13.27 & 12.04 &  -   &  -   & 12.76 & 11.91 & 11.31 & 10.32 &  5.99 &  0.61 \\
v16 & 16.14 & 14.70 & 13.65 & 11.54 & 10.56 &  6.74 &  -   & 11.76 & 10.84 & 10.14 &  9.10 &  6.21 &  0.07 \\
v51 & 19.20 & 16.86 & 14.42 & 12.34 & 10.68 &  9.08 &  6.25 & 11.58 & 10.55 &  9.58 &  8.91 &  5.99 &  0.08 \\
v53 & 17.78 &  -   & 14.54 & 10.98 &  9.65 &  7.72 &  -   & 11.47 & 10.05 &  9.27 &  8.39 &  4.96 &  0.61 \\
v84 & 17.36 & 14.06 & 12.29 & 10.63 &  9.88 &  -   &  -   & 11.55 & 11.04 & 10.66 & 10.36 &  -   & -1.45 \\
\hline
\hline
\end{tabular}
\label{tab:info_sed}
\flushleft{$J, H$ magnitude are from the 2010 multi-band epoch of the VVV survey. The $J$ and $H$ magnitude of following sources are adopted from the 2015 epoch: v53, v319, v335, v389, v621, Stim5, DR4\_v5, DR4\_v10, DR4\_v20, DR4\_v34, DR4\_v44, DR4\_v67. The full table is provided in the online supplementary files.}
\end{table*}

\begin{figure*} 
\centering
\includegraphics[width=7in,angle=0]{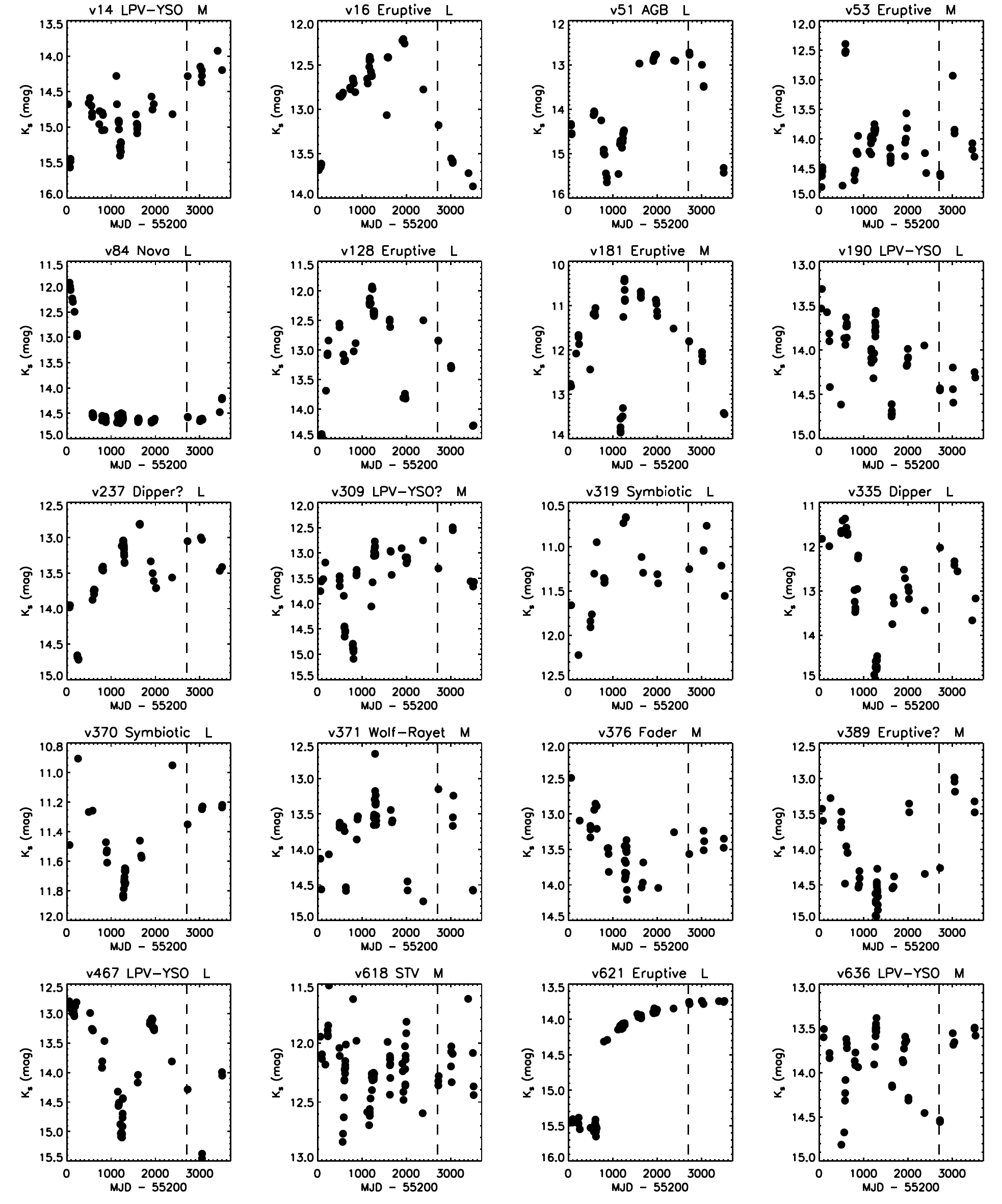}
\caption{$K_s$-band light curves from the VVV survey. The variable classes listed in Table~\ref{tab:info} are shown on the title of each plot. Vertical dashed lines mark out the observation time of spectroscopic epochs. The blue dots represent integrated $K_s$-band magnitudes from spectra taken in the 2019 epoch.}
\label{fig:lc_sum}
\end{figure*}

\begin{figure*} 
\centering
\includegraphics[width=7in,angle=0]{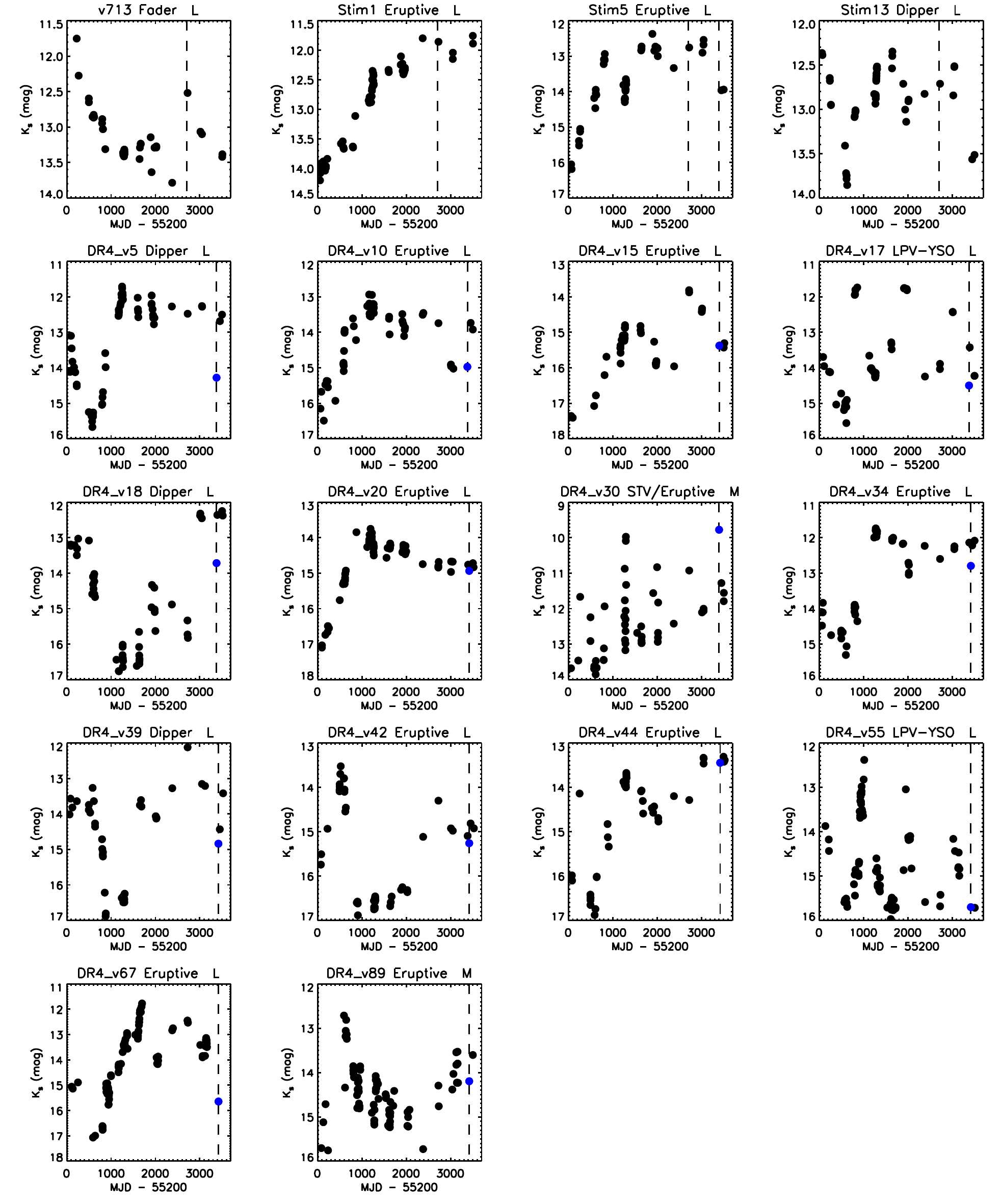}
\caption{Figure~\ref{fig:lc_sum} continued}
\label{fig:lc_sum_2}
\end{figure*}

\begin{figure*} 
\centering
\includegraphics[width=6.7in,angle=0]{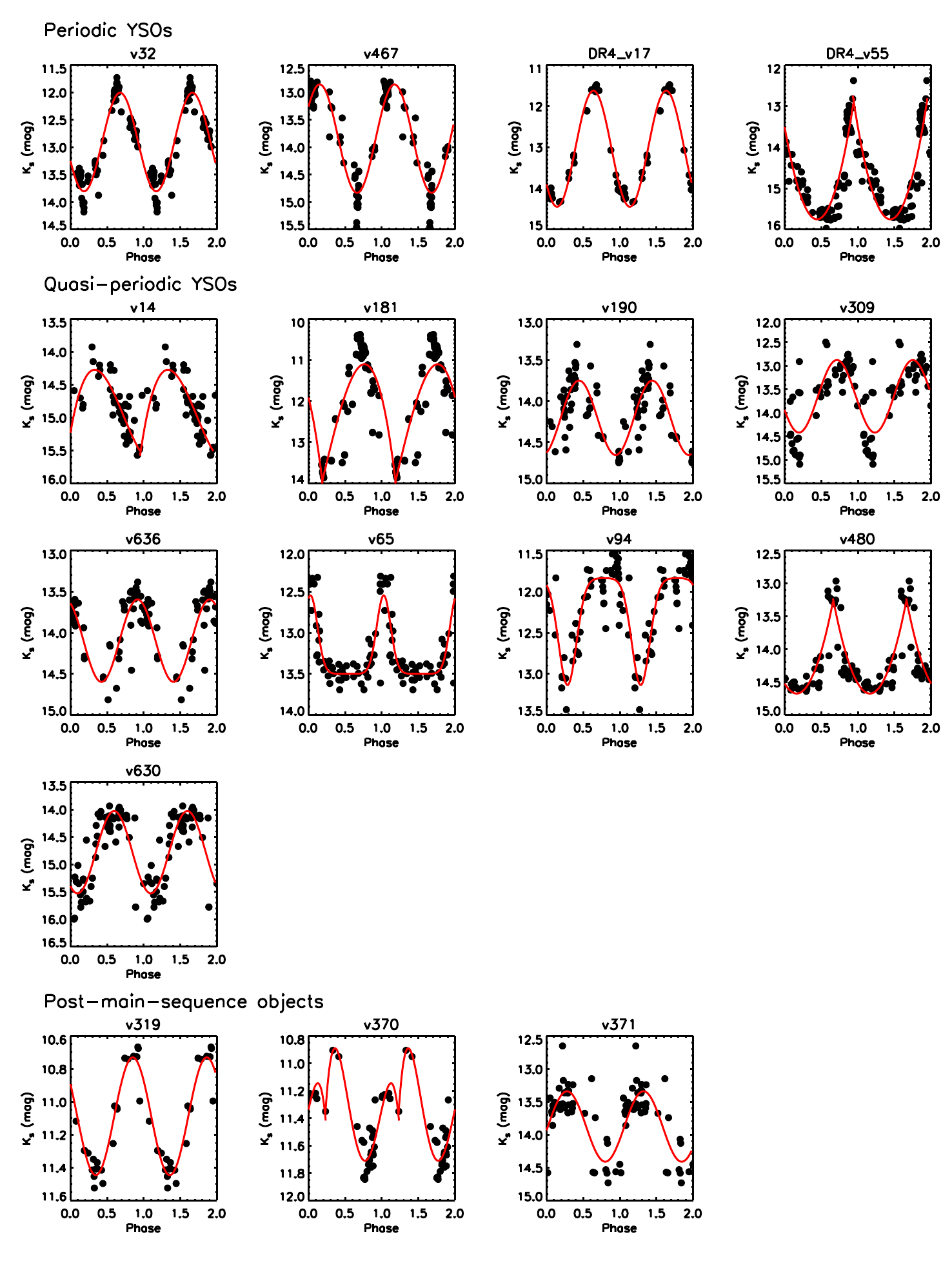}
\caption{Phase-folded $K_s$-band VVV light curves of periodic and quasi-periodic objects found in this work. Analytical fitting curves are shown in red, including Gaussian (v65 and v94), Polynomial (DR4\_v55, v14, v480 and v370), and Sinusoidal functions.}
\label{fig:phase_fold}
\end{figure*}

\begin{table} 
\caption{Information of Spectroscopic Observation}
\renewcommand\arraystretch{0.9}
\begin{tabular}{l c c c c c}
\hline
\hline
Name & Observation & Instrument & Slit Width & N$_{\rm exp}$ & $t_{\rm exp}$ \\
& Date & & ($''$) & & (s) \\
\hline
v14 & 2017-05-31 &  FIRE  &  0.60  & 4 & 253.6  \\
v16 & 2017-05-31 &  FIRE  &  0.60  & 4 & 158.5 \\
v51 & 2017-06-01 &  FIRE  &  0.60  & 4 & 158.5  \\
v53 & 2017-06-01 &  FIRE  &  0.60  & 4 & 253.6  \\
v84 & 2017-06-01 &  FIRE  &  0.60  & 4 & 253.6  \\
\hline
\hline
\end{tabular}
\flushleft{Details of spectroscopic observations. \\
N$_{\rm exp}$: number of exposure. $t_{\rm exp}$: exposure time.\\
The full table is provided in the online supplementary files.}
\label{tab:obs}
\end{table}

\section{VVV light curves of sources observed in Paper II}
Here, we present the VIRAC2b version of light curves of the variable young stellar objects observed in Paper II. 

\begin{figure*} 
\centering
\includegraphics[width=7in,angle=0]{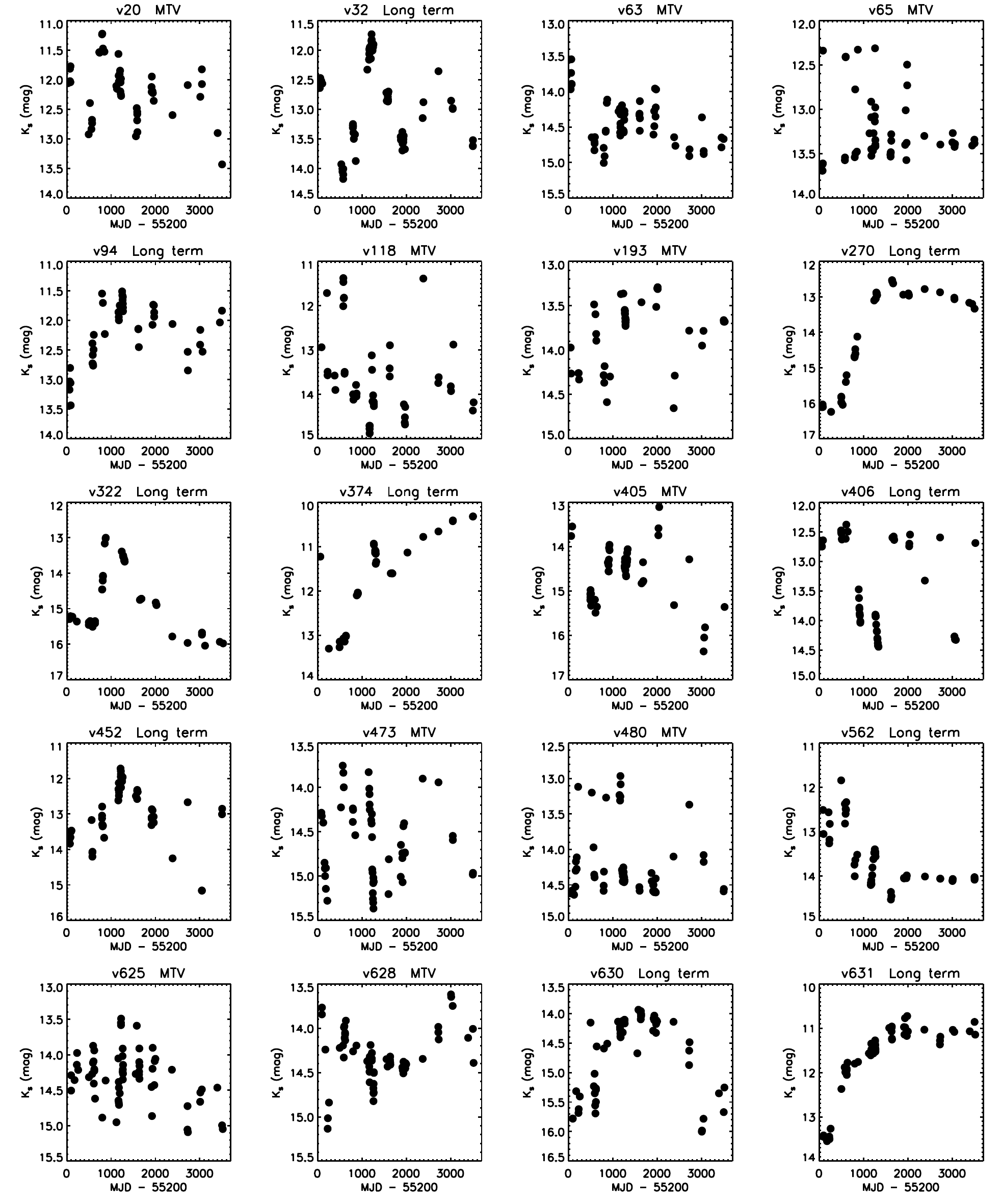}
\caption{The VIRAC2b version of the $K_s$-band light curves from the VVV survey. Sources shown in this figure are spectroscopically confirmed YSOs in Paper II.}
\label{fig:lc_sum_paper2}
\end{figure*}

\begin{figure*} 
\centering
\includegraphics[width=7in,angle=0]{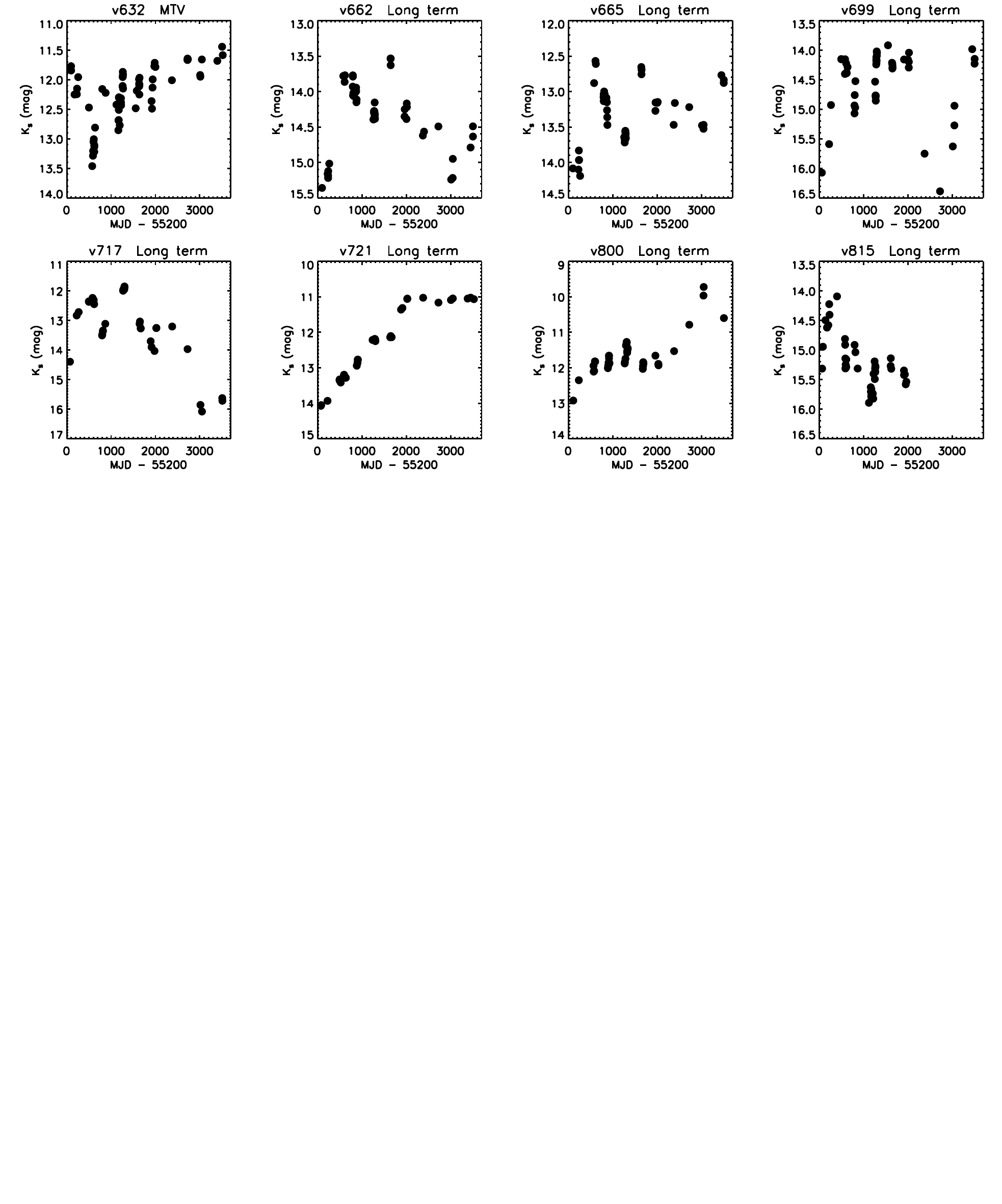}
\caption{Figure~\ref{fig:lc_sum_paper2} continued}
\end{figure*}

\section{2MASS data for individual sources}
\label{sec:2mass}
\begin{figure*}
\centering
\includegraphics[width=7in,angle=0]{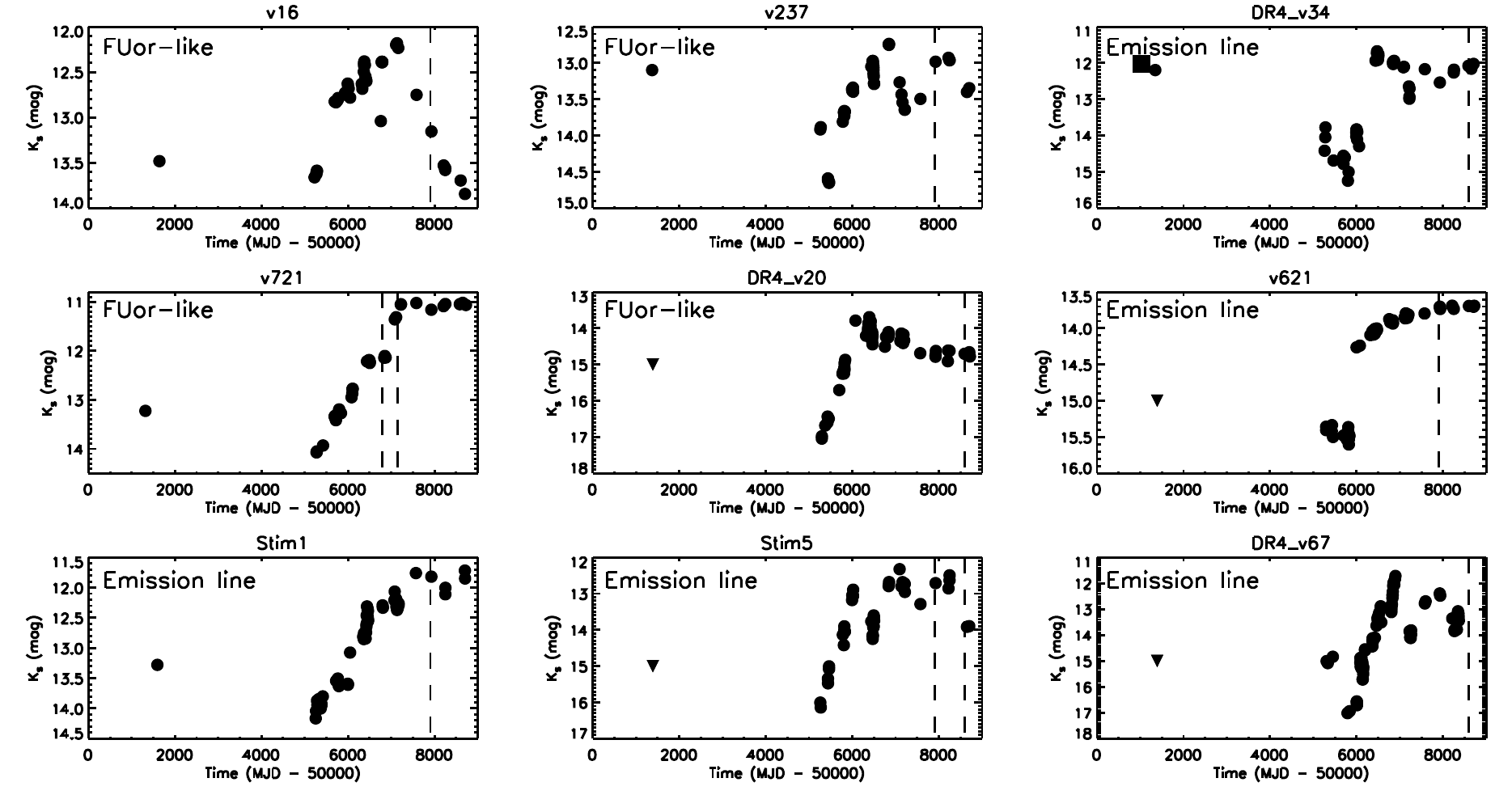}
\caption{2MASS and VVV $K_s$-band light curves of long-term variable YSOs. The spectroscopic category of individual object is labeled on the plots. YSOs without 2MASS detection are shown by lower limits ($K_s > 15$~mag). Spectroscopic epochs are marked by dash lines. This figure is available in the online material.}
\label{fig:2mass}
\end{figure*}

A few 2MASS (MJD~$<$~52000) and VVV joint light curves are shown in Figure~\ref{fig:2mass}, including 4 FUor-like objects and 5 emission line YSOs. According to the transformation equation provided in \citet{GonzalezFernandez2018}, the difference between the 2MASS and VISTA photometric system is negligible ($\sim$0.05~mag) comparing to the photometric amplitude. The extinction-dominated FUor-like YSO v237 had a similar $K_s$ flux in 2MASS and in the out-of-dip portion of the VVV light curve so it has likely remained in outburst for about two decades.

The light curve of v16 appears to show an outburst with total duration $>$3800~days. However, the infrared colour variation of v16 during the fading potion is consistent with changing extinction.

The 2MASS photometry of v721, a slowly rising FUor-like YSO, was $K_s = 13.2$~mag, about 1~mag brighter than faintest magnitude taken by VVV. This suggests that the amplitude of the v721 outburst is not much larger than 3~mag. 
The quiescent state of DR4\_v20 was much fainter than the 2MASS detection limit. The light curve morphology and variation amplitude of three eruptive emission line YSOs (Stim1, Stim5, and DR4\_v67) are comparable to the FUor-like YSO v721. In particular, both v721 and Stim1 experienced a slow rise, at a rate of 0.4 -- 0.6 mag~yr$^{-1}$, while the intra-year variability of Stim1 is slightly larger than v721. However, Stim1 has spectral features of the magnetospheric accretion process. 

The emission line object DR4\_v34 had a photometric minima in 2010. Since 2013, it rapidly raised to a brightness plateau that has similar $K_s$ flux as in the 2MASS. Although the morphology of the light curve looks like an extinction dip around 2010, the VVV colour variation between 2010 and 2015 epoch ($\Delta J = 1.45$ mag and $\Delta K_s = 1.12$ mag) does not agree with the extinction curve ($A_J \sim 3.1 A_{K_s}$). Hence we still classify it as an eruptive object. 

\section{The measurement of extinction}
\label{sec:extinction_appendix}
\begin{figure}
    \centering
    \includegraphics[width=3.2in,angle=0]{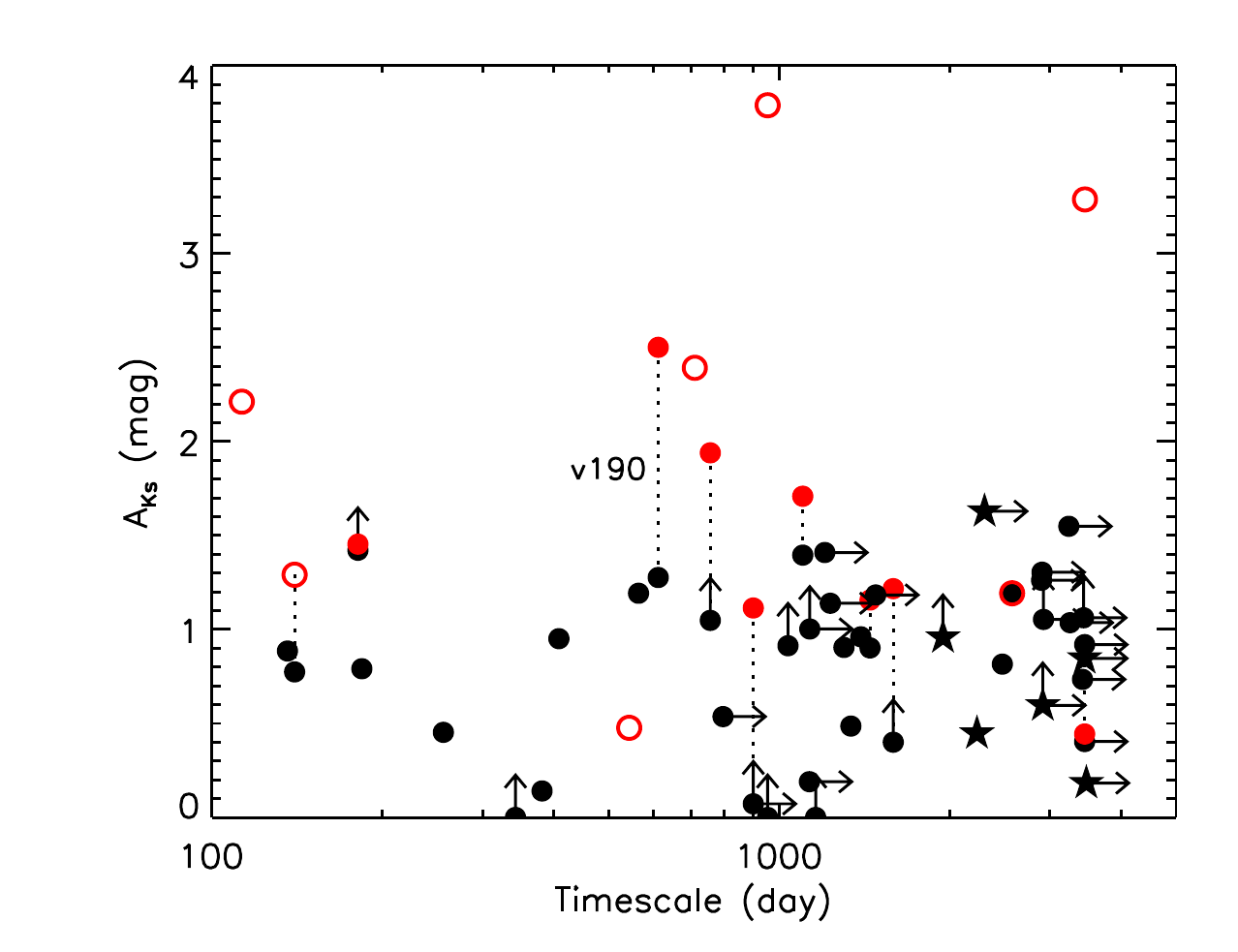}
    \caption{Extinction vs. outburst duration. $A_{K_s}$ measurements using the colour-colour diagram are shown in black, while measurements via the $H_2$ line flux ratio are shown by red. Measurements of the same source are linked by vertical dotted lines. FUor-like YSOs are shown by stars, emission line YSOs are shown by circles, and YSOs with outflow dominated spectra are shown by open circles. Ongoing eruptive events are shown with lower limits on timescale. YSOs undetected in $J$ have lower limits on extinction. We see that FUor-like YSOs and emission line YSOs have similar extinctions. This figure is available in the online material.}
    \label{fig:reddening}
\end{figure}

In this work, the line-of-sight extinction ($A_{K_s}$) is measured by two methods, as their location on the colour-colour diagram and the flux ratio of $H_2$ emission lines where possible. The near-infrared colours during the outburst stage are preferred. The $A_{K_s}$ is measured via de-reddening the location of individual targets on the colour-colour diagram to the Classical T Tauri star locus \citep{Meyer1997}. The extinction estimated by this method is not accurate for targets with outstanding infrared excess. Another method applied the flux ratio between $H_2$ 1-0 Q(3) (2.42~$\mu$m) and 1-0 S(1) (2.12~$\mu$m) lines, as the theoretical value is independent to the excitation conditions \citep[Q(3)/S(1) = 0.7;][]{Turner1977}. We measured the Q(3)/S(1) ratio on 15 YSOs with sufficient Q(3) flux, then compared it against the theoretical value to obtain the reddening factor between two wavelengths. Finally, the $A_{K_s}$ is derived by the power-law extinction function with $\alpha = -2.07$ \citep{WangS2019}.  The extinction and duration of eruptive YSOs are shown in  Figure~\ref{fig:reddening}. Sources lack of measurements are not shown. We see that there is no clear difference between the line-of-sight extinction of FUor-like objects and emission line YSOs. 

Both methods have their own limitations and uncertainties. The accuracy of $H_2$ flux ratio method is strongly rely on the accuracy of the line flux measurement. Large uncertainties are seen on the flux measurement of the Q(3) line, as it is heavily veiled by the continuum and the CO overtone emission. Hence, this method was only carried out on objects with enough signal to noise ratio on their Q(3) emission. The typical error of the flux measurement is 8\% on Q(3) lines, which leads to a 0.4~mag error on $A_{K_s}$. In addition, extinction from the inner disc, between the stellar surface and the emission region of $H_2$, will not be measured by this method. 

The colour-colour-diagram method is not rely on spectral features, hence it can be applied on many sources. However, this method uses the empirical locus of CTTS, while a large fraction of our objects are Class I sources with infrared excess. The location of YSOs on the colour-colour diagram is also affected by the variable mass accretion process (Figure~21 in Paper I). Although one expected the colour variation introduced by episodic accretion should have a different vector than the reddening, a large scattering of colour variation is seen among eruptive YSOs, which further contributes to the uncertainty of the extinction measurement. This method also requests $J$ and $H$-band photometry which are unavailable for most heavily extincted objects. 

The measured $A_{K_s}$ is time dependent especially on objects with variable line-of-sight extinction. To measure the $A_{K_s}$, the $J$, $H$, $K_s$ photometry in the outburst stage, normally closer to the spectroscopic epochs, were used. However, several sources, either lack of $J$-band detection in later epochs or having rapid photometric variation, have notable variation between the multi-band photometric and spectroscopic epochs. For example, a quasi-periodic object v190 had the 3-band VVV photometric epoch at MJD 55244 and a spectroscopic epoch at MJD 57905, with a separation of 0.35 in phase (P = 611 d). The $A_{K_s}$ measured by the 3-band photometry is 1.28 mag and by $H_2$ line flux ratio in the fainter epoch is 2.50 mag, with a difference of 1.22 mag, while the $\Delta K_s$ between two epoch is 0.91 mag. We compared the measured $A_{K_s}$ on targets with more than one VVV multi-band epoch. Large epoch-to-epoch difference ($\Delta A_{K_s} > 0.2$ mag) is only seen among a small number of targets, and most of them are STVs and MTVs.

To conclude the $A_{K_s}$ measurement, both methods have their uncertainties and absolute accurate measurement is not achievable. However, the general picture shows two important results: FUor-type objects have similar line-of-sight extinction as most emission line objects, and some outflow-dominated objects have very large $A_{K_s}$ suggesting geometric effect is involved.

\section{Year-to-year variability of Stim5}
\label{sec:Stim5}
\begin{figure*}
\centering
\includegraphics[width=3.2in,angle=0]{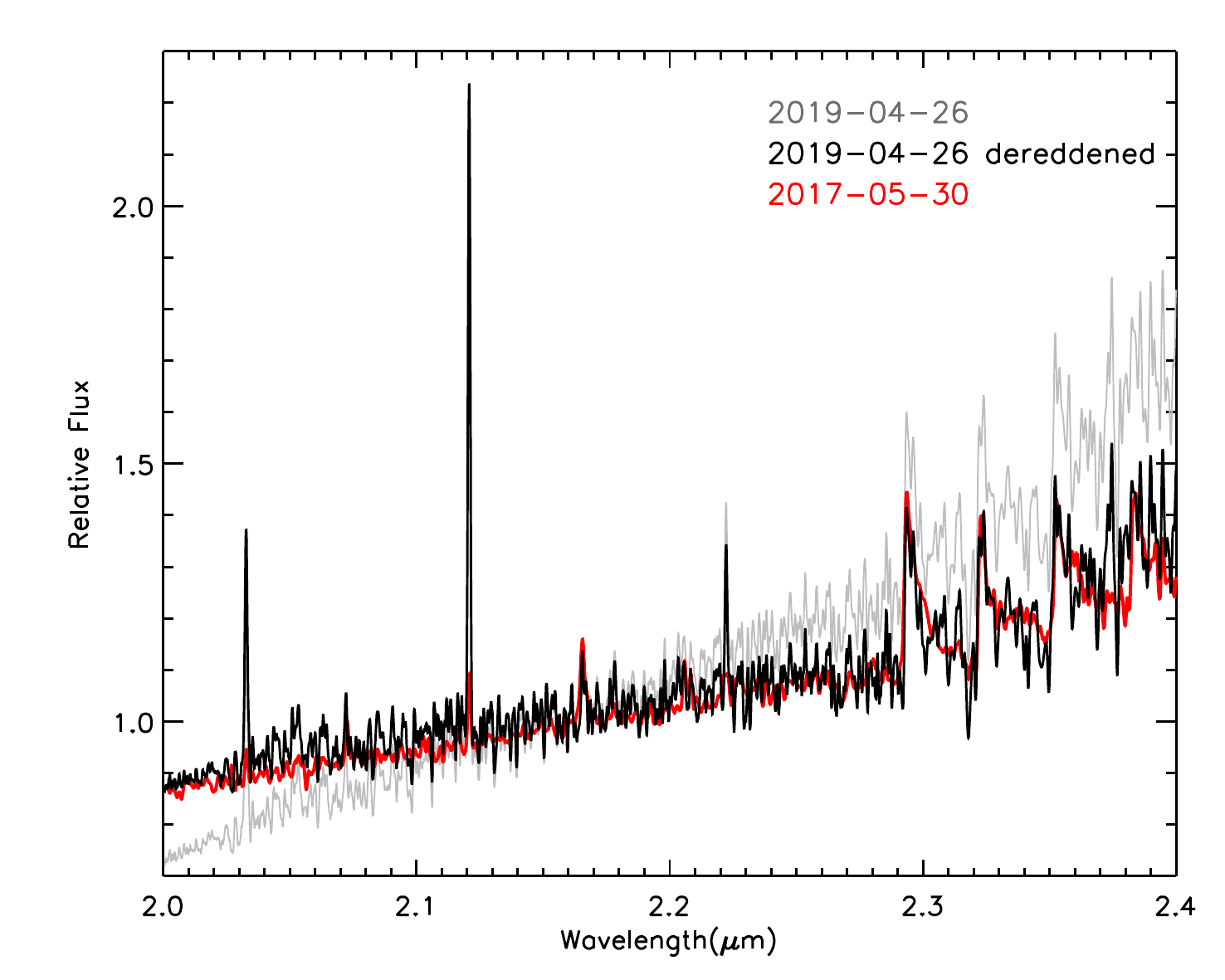}
\includegraphics[width=3in,angle=0]{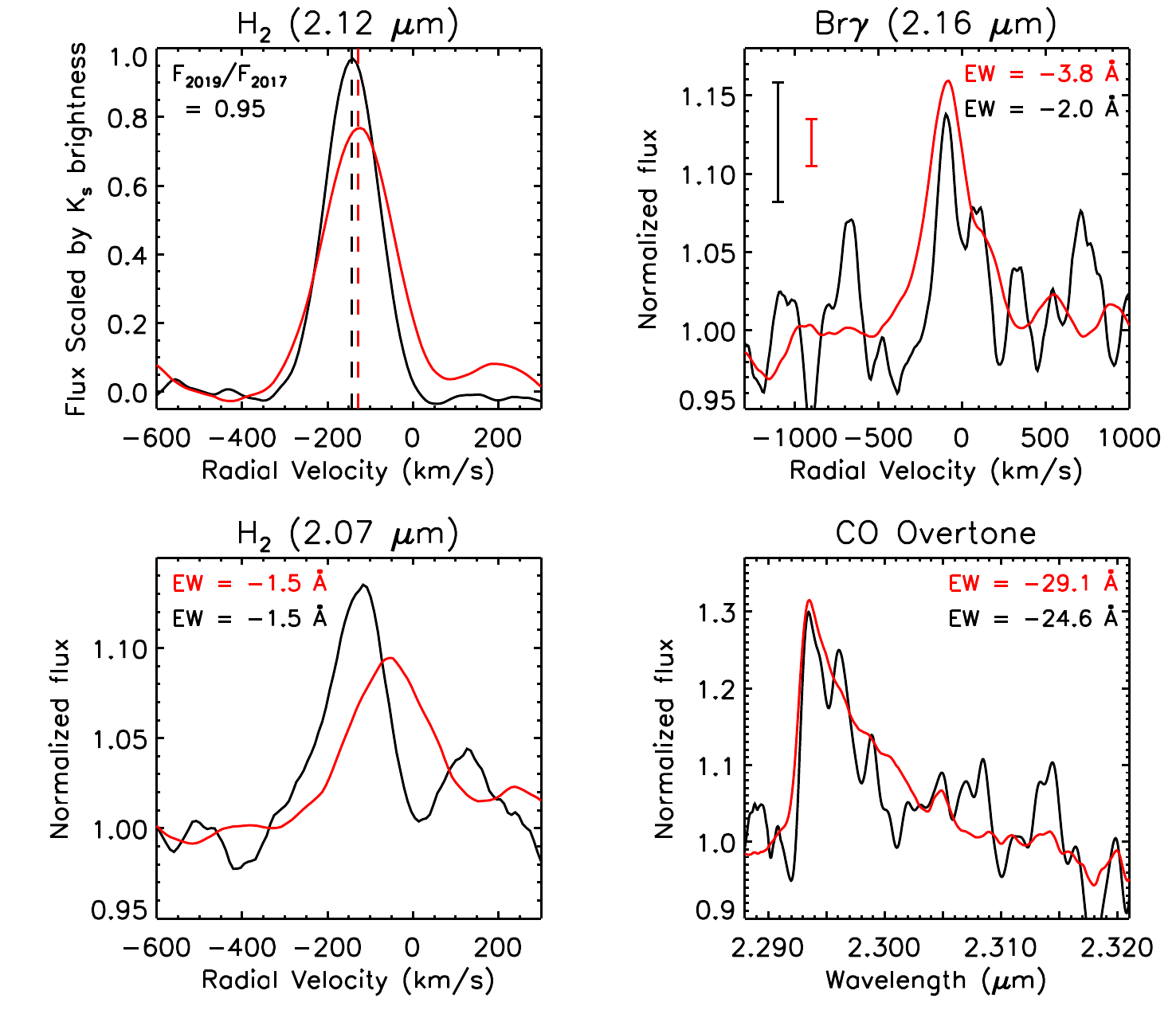}
\caption{Spectra of Stim5 in 2017 and 2019 epochs. {\it Left}: normalised $K_s$-band spectra of Stim5. The spectrum from 2019 epoch is de-reddened by the power-law near-infrared reddening curve from \citet{WangS2019} with $A_{K_s} = 1.2$~mag. The de-reddened spectrum is shown in black. {\it Right}: Spectral profiles of Stim5 in 2017 and 2019 epochs. The line profile of the H$_2$ 2.12 $\mu$m emission is scaled to match the $K_s$-band flux and the continuum is removed.  Line profiles of Br$\gamma$, H$_2$ 2.07 $\mu$m, and CO bandhead emission are normalised by continuum levels. Equivalent widths are shown on each panel. The error bars in the panel of Br$\gamma$ represent standard deviations of the continuum. This figure is available in the online material.}
\label{fig:Stim5lines}
\end{figure*}

Spectra of the emission line object Stim5 were obtained in 2017 and 2019. In this section, we study the spectroscopic variation of Stim5 over the two-year interval. The overall $K_s$ amplitude of Stim5 is 3.8~mag in the 2010 - 2019 interval of the VVV data. Starting from 2010, Stim5 experienced a rapid rise of $\Delta K_s > 2.2$~mag in three years. After a brief dimming ($\Delta K_s \sim 1.4$ mag) in 2014, Stim5 rose further to its photometric maximum, $m_{K_s} = 12.3$~mag, in 2015 (see the lower-middle panel of Figure~\ref{fig:2mass}). It then stayed at a bright plateau in the light curve from 2015--2018 and started fading again in 2019. The intra-year variations are modest, $\Delta K_s \sim 0.3$~mag, in comparison to the inter-year variations. Between the two spectroscopic epochs, it faded by 1.2--1.5~mag in $K_s$, based on VVV photometry taken within 100 days of each of the two epochs.  The spectra of Stim5 in two epochs are shown in Figure~\ref{fig:Stim5lines}. Different $K_s$ continuum slopes are seen between two epochs, and the change is well fitted by adding the power law de-reddening curve towards the galactic centre ($A_{K_s} = $~1.2~mag and $\alpha = -2.07$, \citealt{WangS2019}) to the spectrum in the fainter epoch. The Br$\gamma$ emission had constant radial velocity in the two epochs, with broad profiles consistent with the accretion disc origin.  The equivalent width (EW) of the Br$\gamma$ line was slightly larger in the brighter 2017 epoch but the difference was within the error bar.  The CO overtone emission and H$_2$ 2-1 S(3) line at $2.07$ $\mu$m also had the similar EWs between the two epochs. Considering the spectroscopic variation and the low $\Delta W2$/$\Delta K_s$ ratio between 2017 and 2019 epochs, we conclude that the fading of Stim5 after 2017 is due to  increasing line-of-sight extinction.

The $H_2$ 1-0 S(1) line ($2.12$ $\mu$m) shows different variation behaviour against the continuum and other emission features as the EW of the $2.12$ $\mu$m line is stronger in the fainter 2019 epoch. We measured the flux change of this line by doing an approximate flux calibration based on the VVV $K_s$-band photometry taken closest to the spectroscopic epochs. After subtracting the reddened continuum, we found that the $2.12$ $\mu$m line flux in the fainter 2019 epoch is at least 20\% larger than the 2017 epoch despite the 1.2--1.5 mag decline in $K_s$ brightness. This indicates that the emission from the outflow arises on a larger spatial scale than the obscuration source causing the change in extinction towards the inner disc. This contrasts with the fading of the H$_2$ 2-1 S(3) $2.07$ $\mu$m line indicated by the constant equivalent width, noted above. In the literature, H$_2$~$2.07$ $\mu$m emission has been detected from hot components of HH objects \citep[e.g.][]{Pike2016} and the flux ratio between H$_2$ $2.07$ $\mu$m and $2.12$ $\mu$m lines was applied to determine the LTE temperature \citep{Gredel1995}. However, in Stim5, the different variation behaviours between these two emission lines suggest that they might have different physical origins. The $2.07$ $\mu$m line, with similar variation mode as Br$\gamma$ emission, could arise from the inner disc associating with mass accretion process. 

\section{AGB stars and symbiotic systems: v51, v319 and v370}
\label{sec:AGB}

In Paper I, the authors found that long period pulsating AGB stars (Miras) are more abundant than YSOs among the brighter end of the VVV variable samples. Miras are bright ($-8 < M_{K_s} < - 6$~mag) and have periodic variation driven by the pulsation of the photosphere \citep[e.g.][]{Whitelock1991}. The near-infrared spectra of AGB stars are dominated by molecular absorption features, for example $^{12}$CO $\Delta v = 3$ absorption bands in the $H$ bandpass \citep{Origlia1993} and $^{12}$CO $\Delta v = 2$ absorption bands in the $K$ bandpass \citep{Wallace1997}.  As a consequence of nucleosynthesis, strong $^{13}$CO $\Delta v = 2$ absorption lines are seen beyond 2.34 $\mu$m \citep{Ramstedt2014}. Water absorption bands are also commonly seen in AGB stars, including steam bands in the $J$, $H$ and $K$ bandpasses and the 3.1~$\mu$m water ice band. The key spectral features to distinguish AGB stars from FUors are strong $^{13}$CO absorption and $^{12}$CO $\Delta v = 3$ absorption.

Br$\gamma$ emission is observed in a few AGB stars, such as $\chi$ Cyg \citep{Hinkle1984}, NSV 11749 \citep{Rodriguez-Flores2014} and XID 6592 \citep{DeWitt2013}. Although shocks can excite low excitation emission lines during the expansion and contraction of the stellar envelope \citep{Hinkle1984}, H{\sc i} recombination lines in AGB stars are associated with binary systems, known as D-type symbiotic stars. Here the HI arises in the accretion flow from the AGB star on to its companion \citep{Kenyon1984, Mikolajewska2007}. In D-type symbiotic systems the near-infrared emission arises partly from the photosphere and partly from the circumstellar dust shell \citep{Whitelock1992, Belczynski2000, Phillips2007}, such that the near-infrared colours overlap with those of YSOs \citep{Corradi2008}. Due to variable dust obscuration, D-type symbiotic stars have more scattered light curves than most normal Miras \citep{Gromadzki2009}. Similar variable dust obscuration is also seen in the Miras with extreme mass loss rates that dominate the Mira population in VVV due to the effect of the survey saturation limit \citep{Contreras2017}.

In this work, three objects, v51, v319 and v370, show spectral features of AGB stars including water vapour absorption, $^{12}$CO and $^{13}$CO absorption. Br$\gamma$ emission is detected on v319 and v370 indicating that they are O-rich D-type symbiotic stars.  According to \citet{Lucas2017}, Mira variable stars in VVV and the United Kingdom Infrared Sky Survey typically have redder 1 to 5~$\mu$m infrared colours than YSOs, due to being completely enclosed in a shell of warm dust. However, both v319 and v370 are bluer ($W1 - W2 < 1.5$~mag) than the AGB selection region in that work, suggesting thinner dust shells around these two objects.  The {\it WISE} colours of v370 fall into the YSO region on the WISE colour-colour diagram ($W1 - W2$ vs. $W3 - W4$) in agreement with D-type symbiotic stars \citep[see Figure 1 from][]{Rodriguez-Flores2014}.
Periodic variations are measured with the VIRAC2 light curves of v319 and v370, with 
$P$ = 304 days and 319 days, respectively. There is a large scatter in the phase-folded light curve of v370, which is consistent with symbiotic stars. Given that we have reported several new high amplitude periodic YSOs in this work, the overlap in colour space between YSOs and some D-type symbiotic stars indicates that colours and time series photometry cannot reliably distinguish the two types of variable star. High quality spectroscopy is required for a clear identification. We note that X-ray emission is often detected from symbiotic stars but these two stars do not appear to have been the subject of sensitive X-ray observations and so no detection has been reported.

Strong $^{13}$CO ($\Delta \nu = 2$) absorption bandheads are detected in v51. A shallow absorption feature around 2.345 $\mu$m is observed in FUors \citep[e.g. Figure 5][]{Connelley2018} and M-type dwarf stars \citep{Rayner2009}. However, the $^{13}$CO bandhead at 2.345 $\mu$m is much deeper than the feature seen in FUors, comparable to $^{12}$CO bandheads. M-type giant stars show similarly strong $^{13}$CO \citep{Rayner2009}, having undergone the first dredge up. Consequently, we classified v51 as a post-main-sequence object. The {\it WISE} colours of v51 ($W1 - W2 = 1.66$ and $W3 - W4$ = 2.83) fall outside the empirical colour space of dusty Mira stars defined by \citet{Lucas2017}, and the light curve of v51 is dominated by two V-shape dips separated by about 2500 days. Each dip has $\Delta K_s \sim 2.7$~mag and lasts 600 to 1000 days. The ratio of infrared magnitude variations was observed during the second dip as $\Delta K_s/\Delta W2 = 0.26$, consistent with variable line-of-sight extinction. The spectrum of v51 has absorption features including water vapour in the $H$ bandpass, Br$\gamma$, $^{12}$CO overtone and $^{13}$CO in $K$-band.

According to \citet{Contreras2017}, Miras should lie at distances $d> 25$~kpc to fall below the saturation limit of VVV, in the absence of foreground extinction. We attempted to measure kinematic distances for v51, v319, and v370 based on the radial velocity of the CO overtone bands. Located at galactic longitude ($l = 339.16^{\circ}$), the radial velocity of v319 ($+$120~km s$^{-1}$) cannot be fitted by a Galactic rotation model, indicating that it is a halo object. For v51 and v370, the radial velocity-based kinematic distances are 12.5 and 14.6 kpc, consistent with the range of kinematic distances allowed by the VIRAC2 proper motions. These two stars are therefore identified as Galactic disc members.

\section{Spectra of 38 sources observed in this paper}
In this section, we present the near-infrared spectra of the 38 sources observed in the 2017 and 2019 spectroscopic epochs. The information of spectroscopic observation is listed in Table~\ref{tab:obs}.

\begin{figure*} 
\centering
\includegraphics[width=3.3in,angle=0]{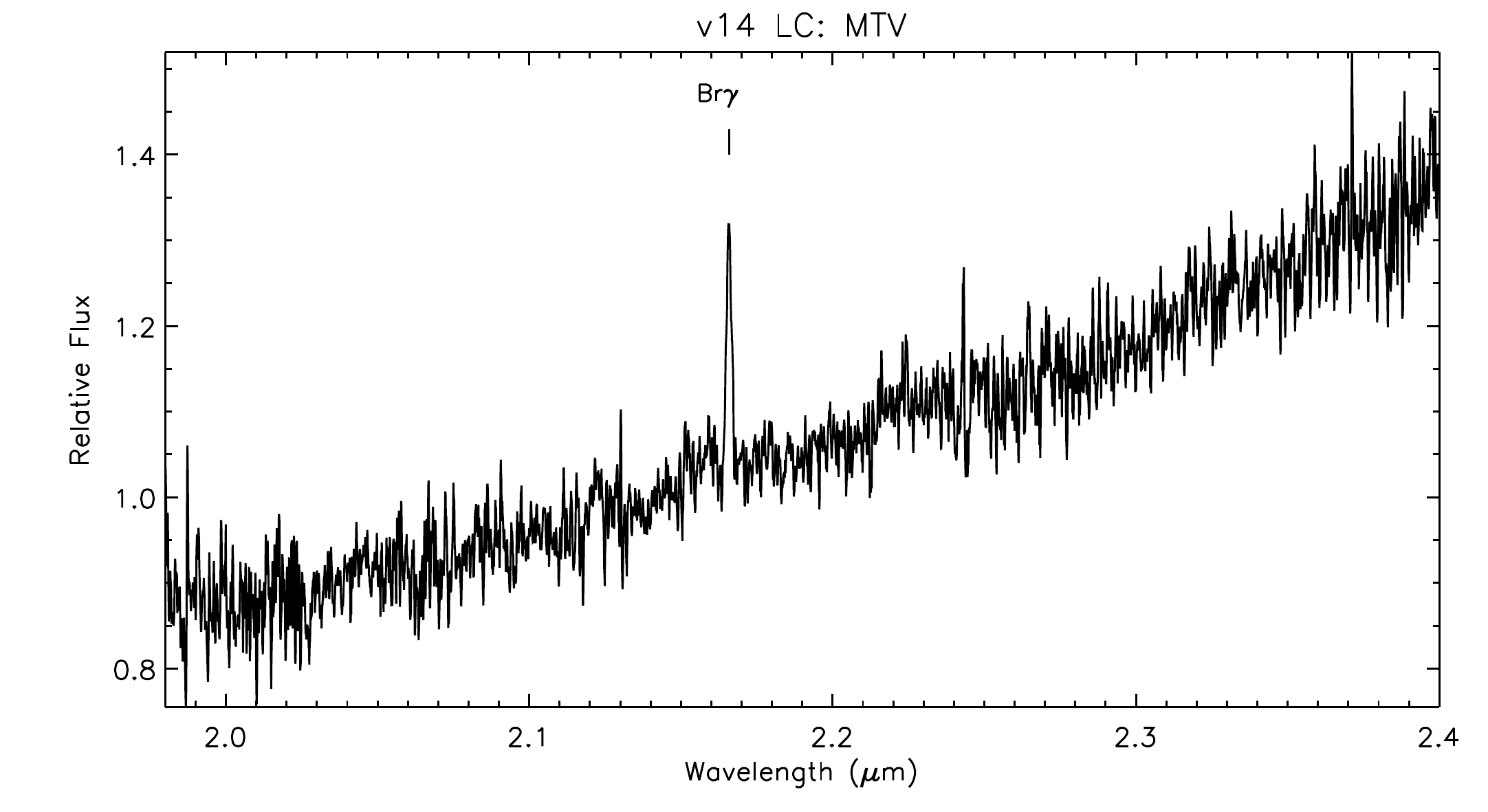}
\includegraphics[width=3.3in,angle=0]{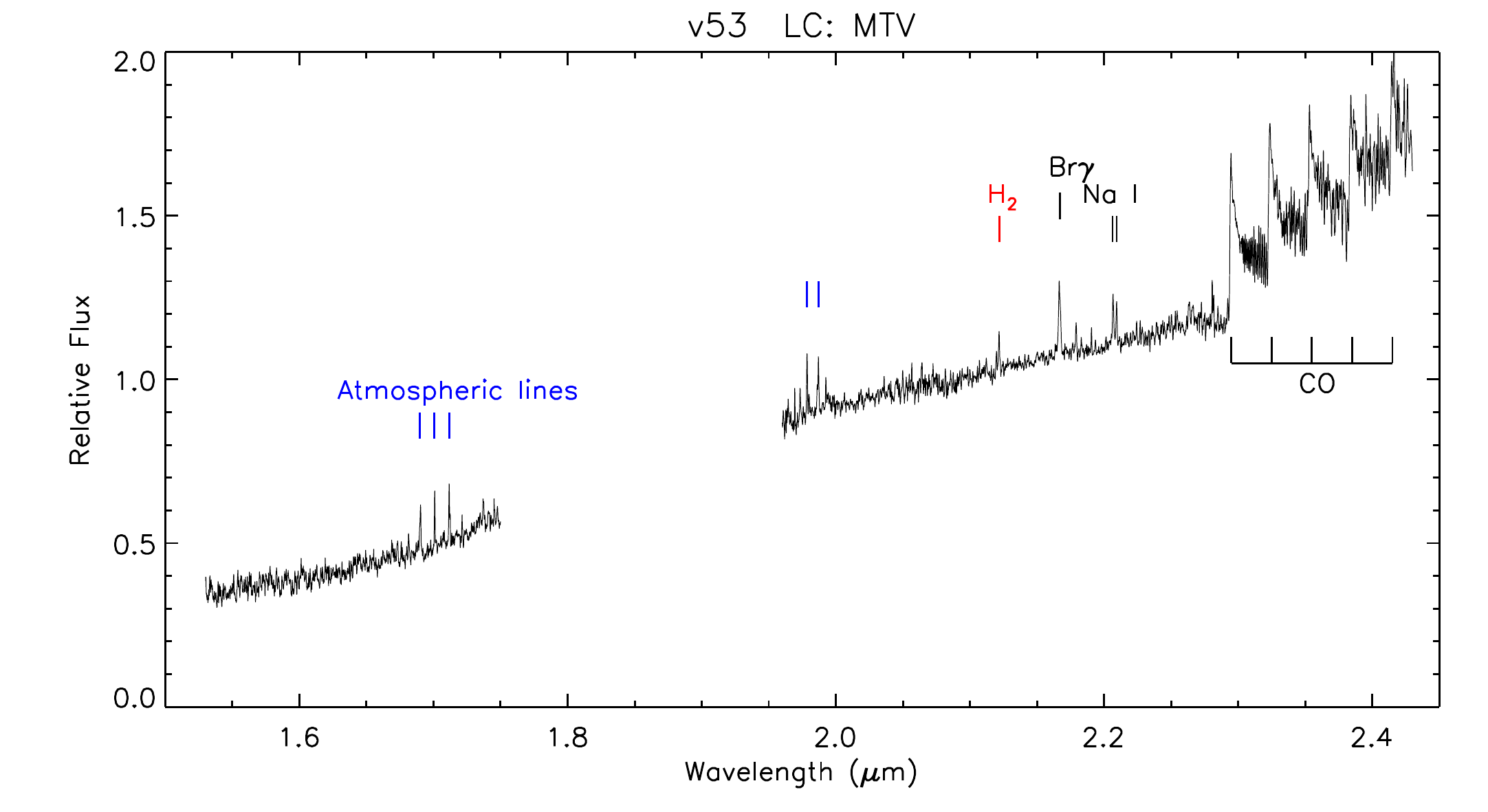}
\includegraphics[width=3.3in,angle=0]{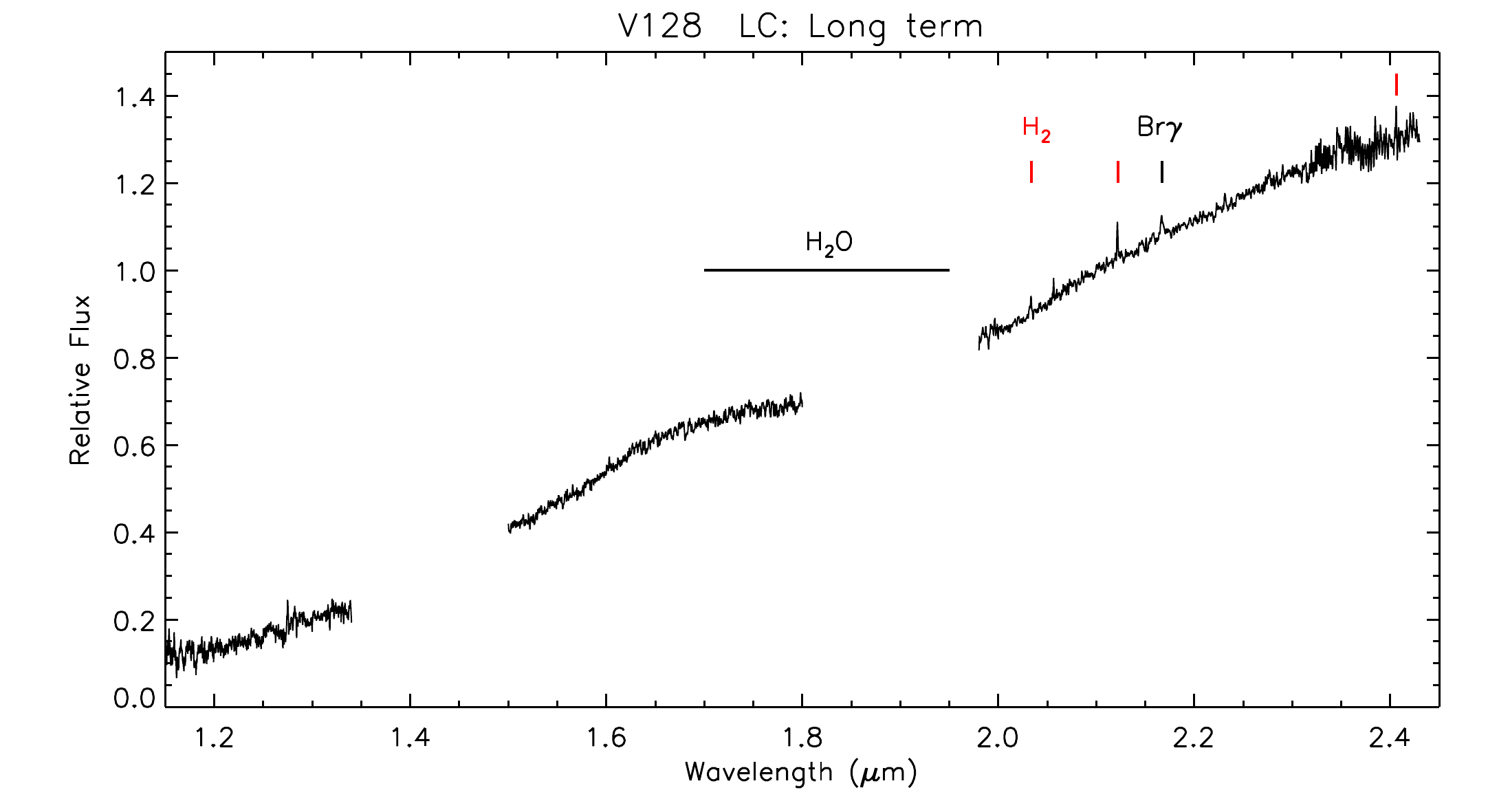}
\includegraphics[width=3.3in,angle=0]{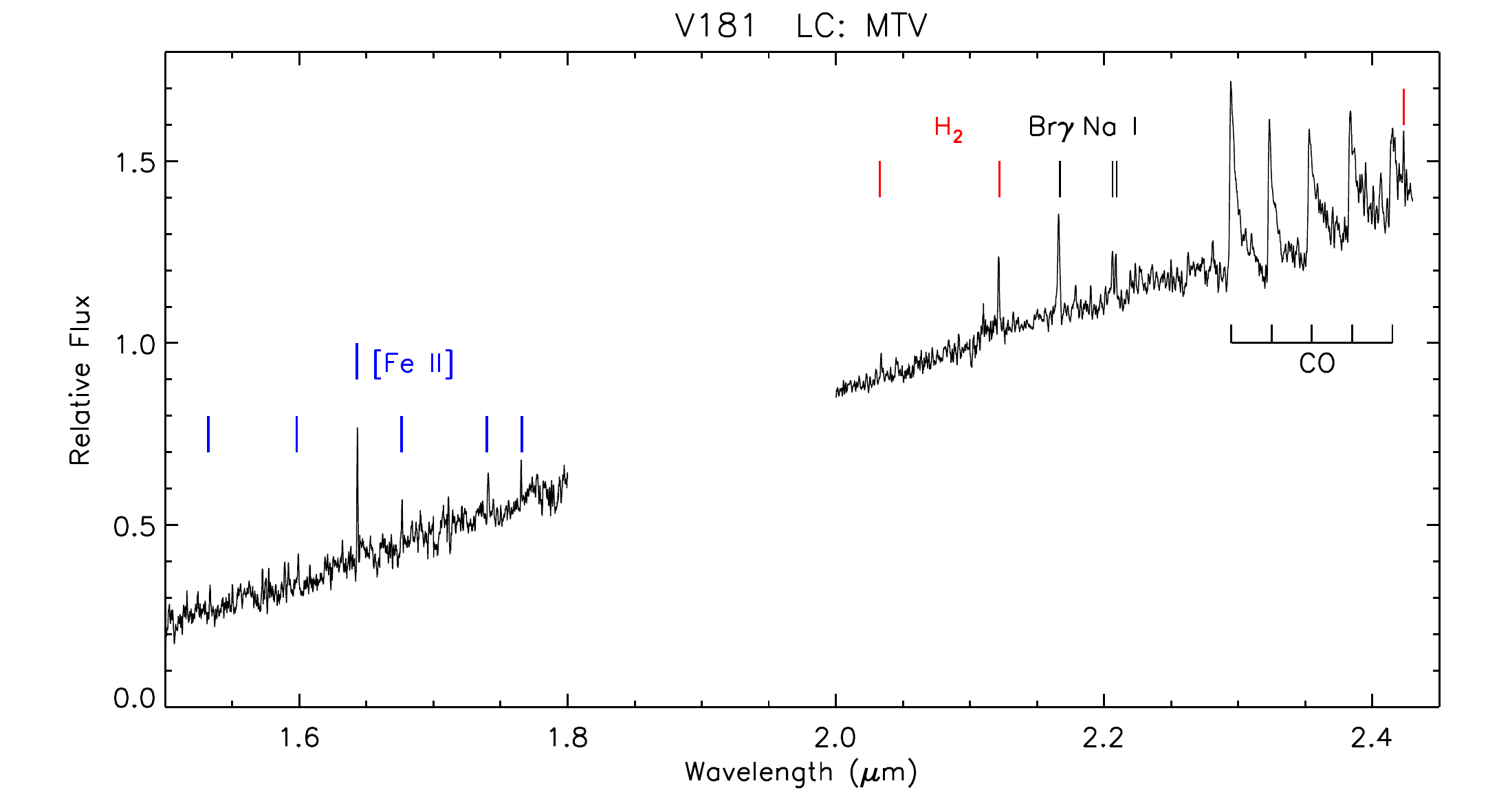}
\includegraphics[width=3.3in,angle=0]{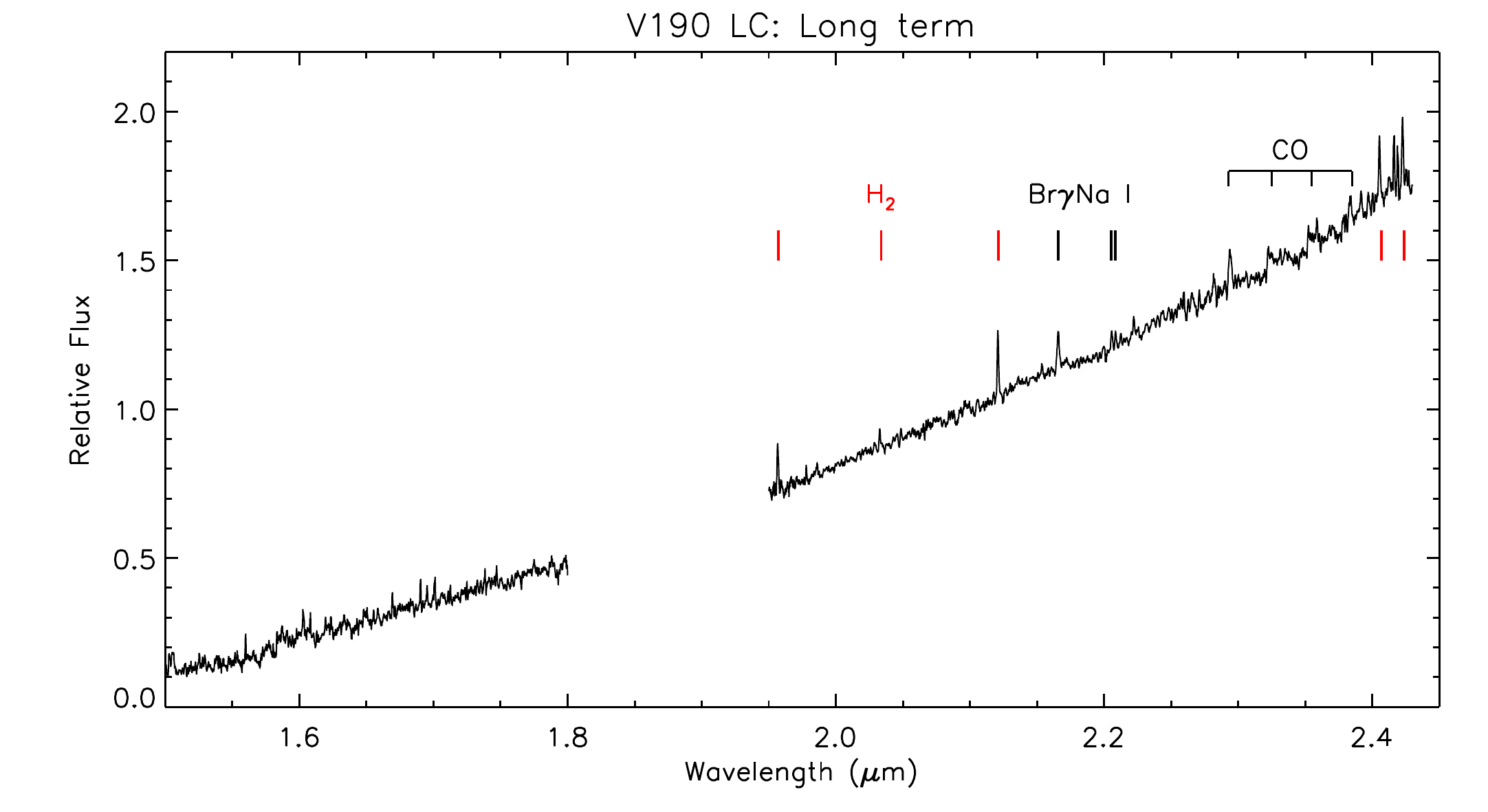}
\includegraphics[width=3.3in,angle=0]{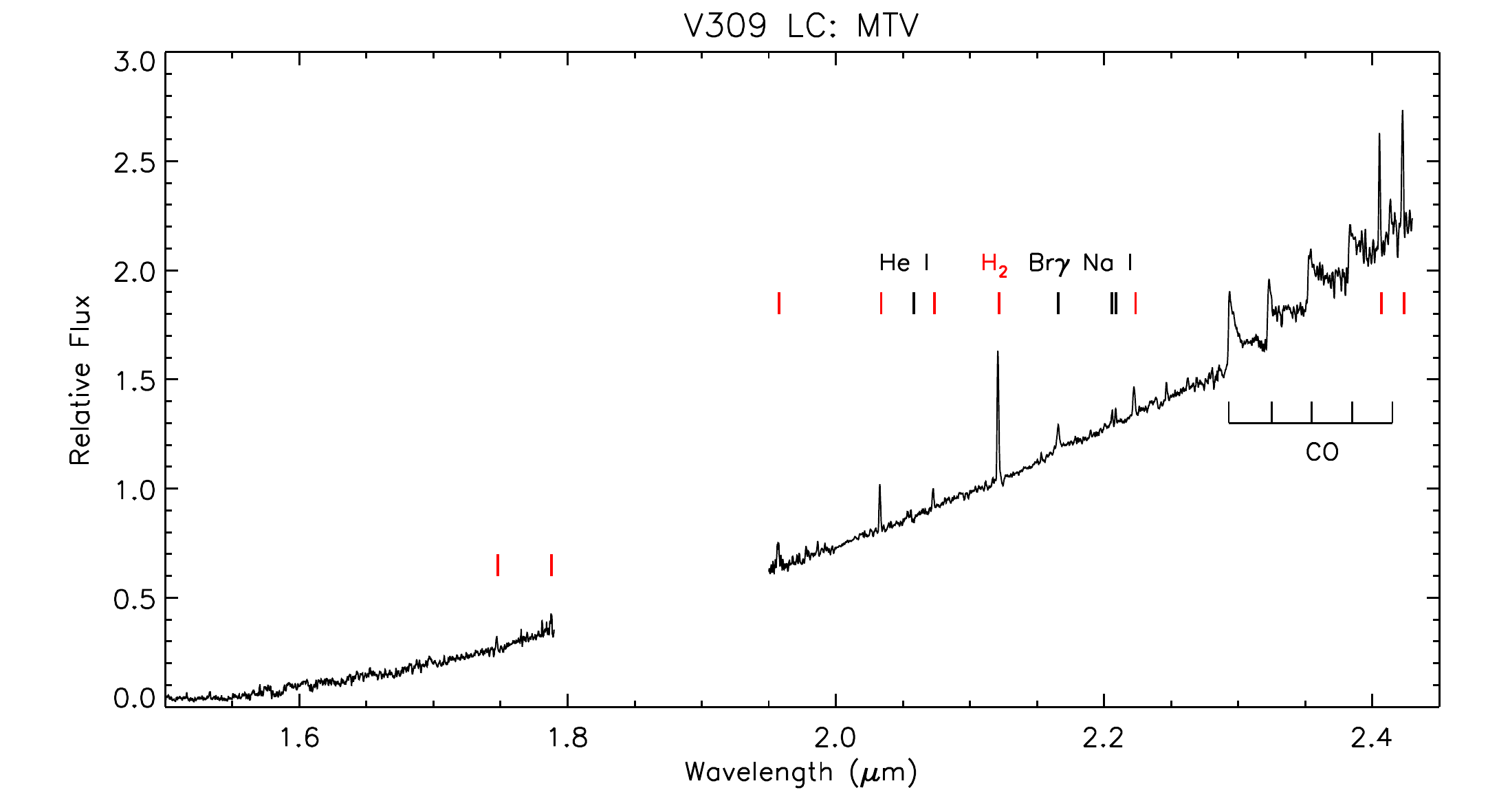}
\includegraphics[width=3.3in,angle=0]{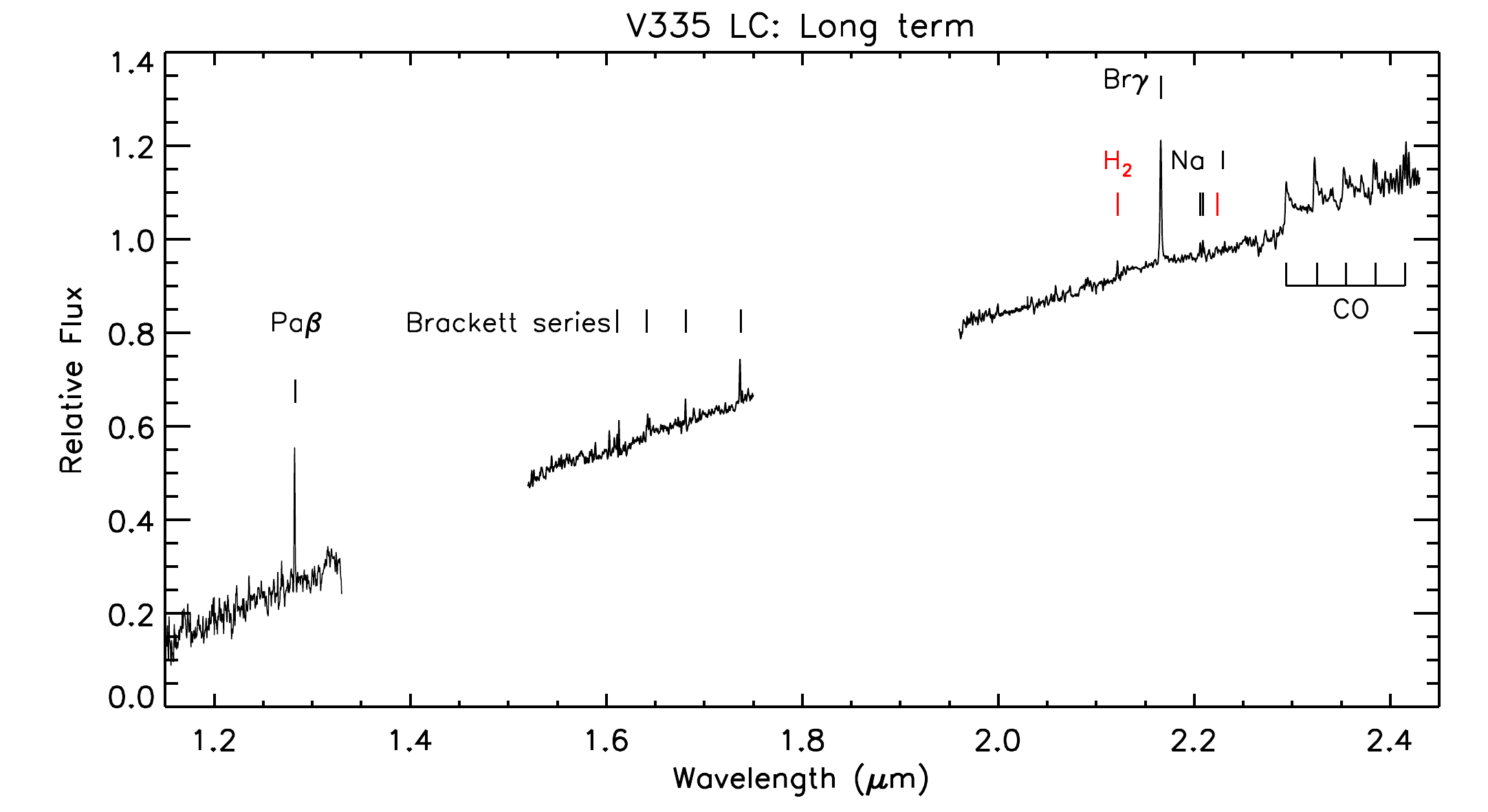}
\includegraphics[width=3.3in,angle=0]{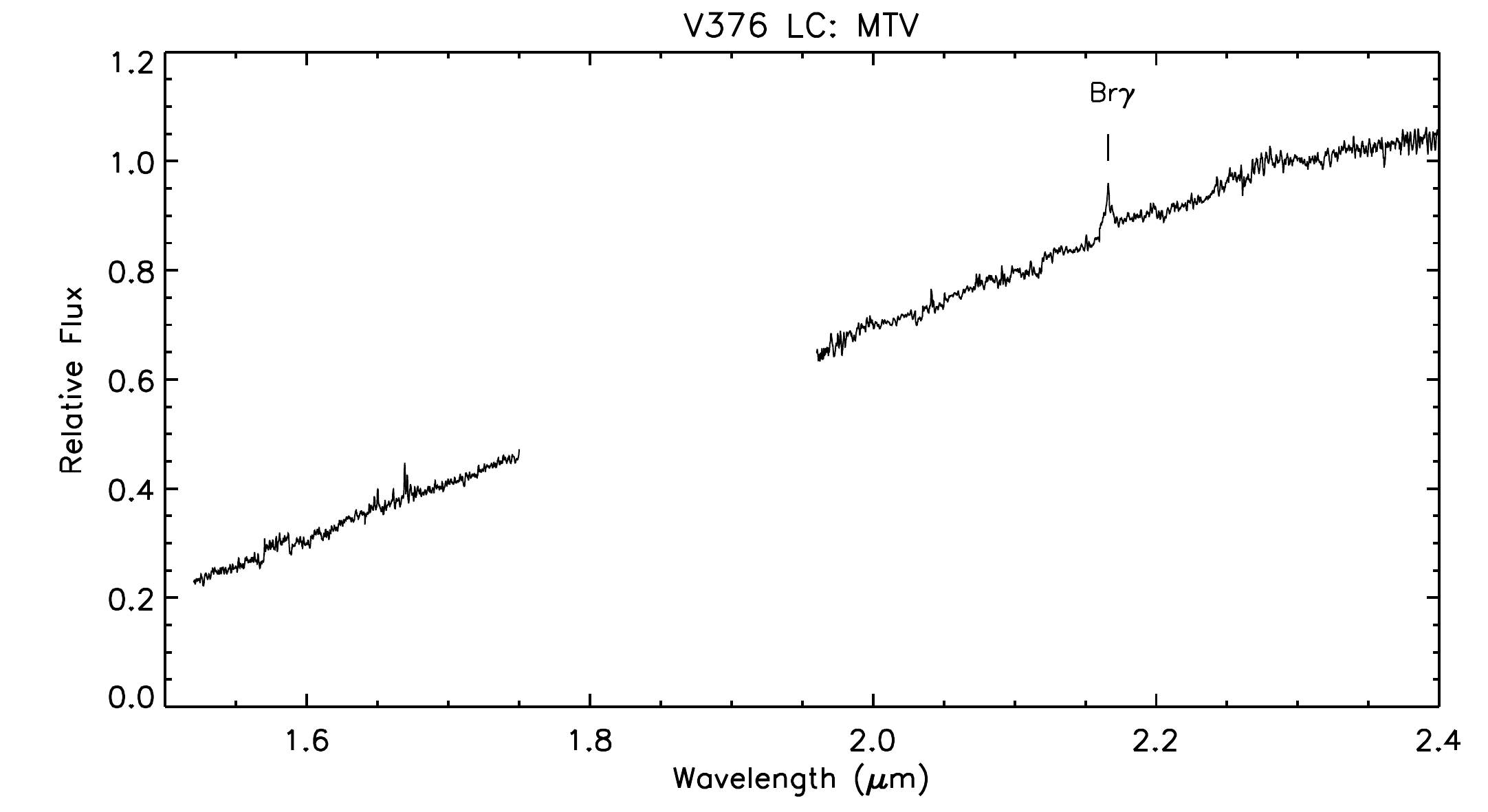}
\includegraphics[width=3.3in,angle=0]{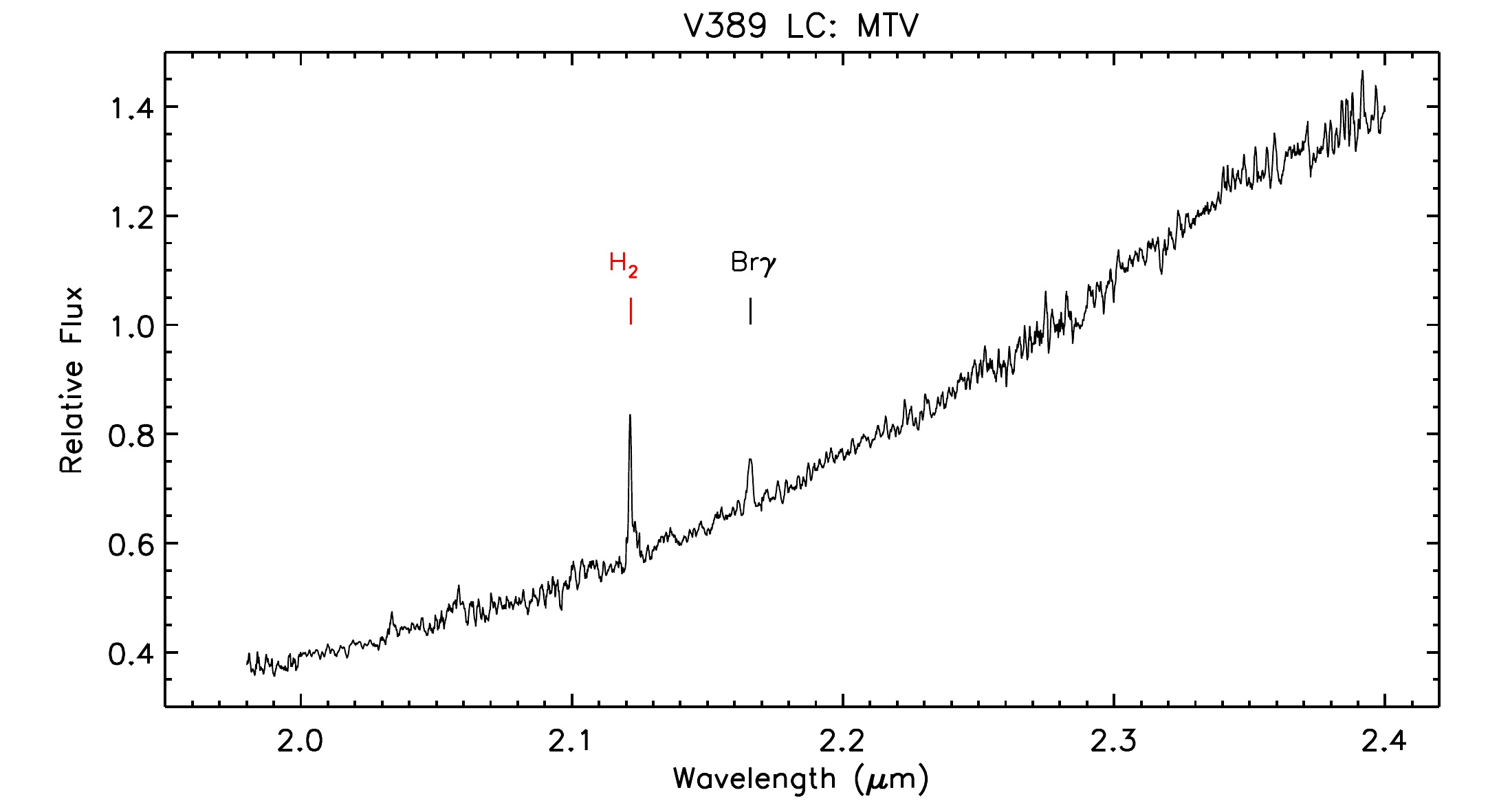}
\includegraphics[width=3.3in,angle=0]{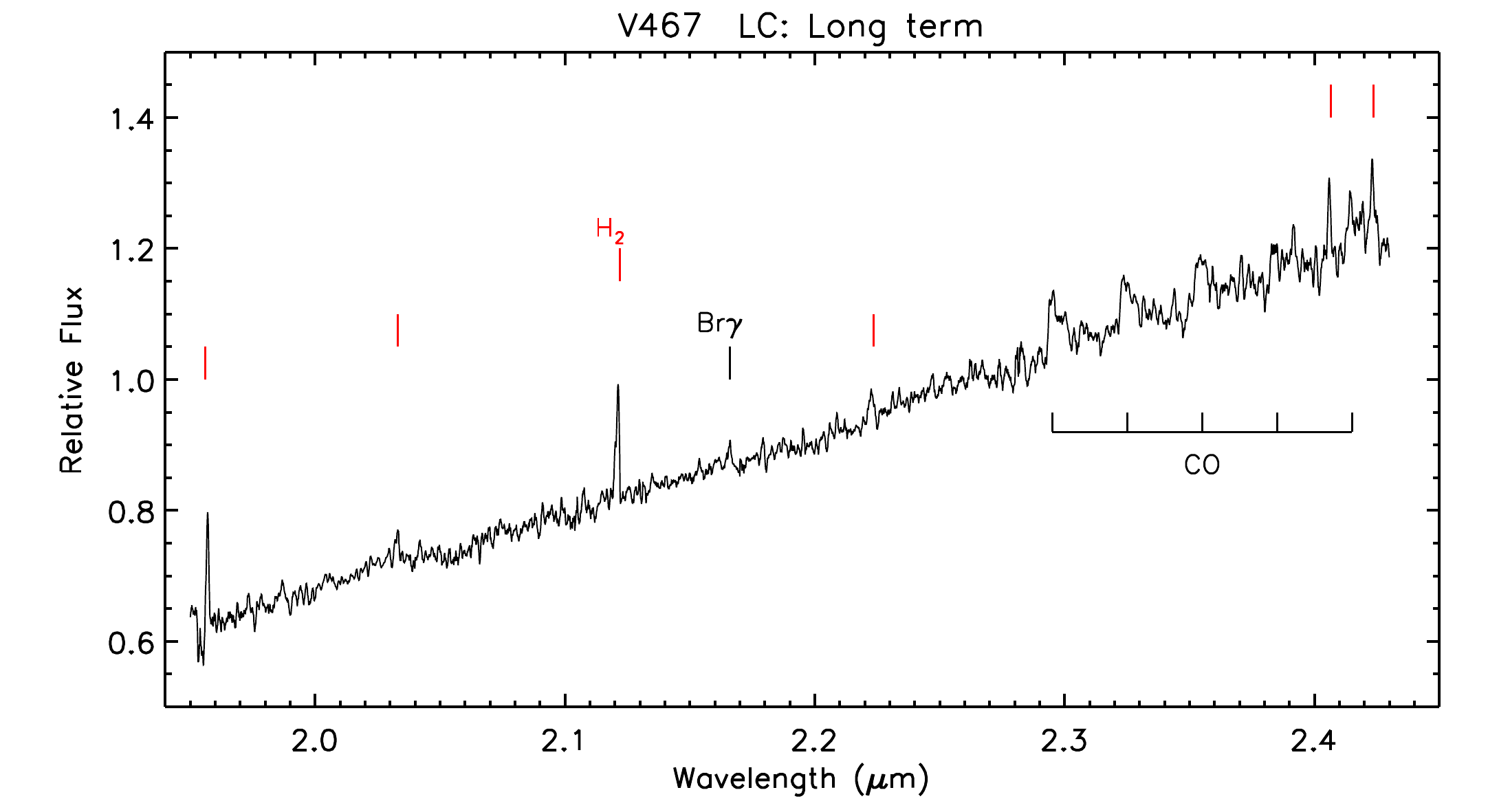}
\caption{Near-IR spectra of emission line YSOs observed in this work. These YSOs have signatures of magnetospheric accretion process. Spectroscopic features (e.g. molecular hydrogen emission, hydrogen recombination lines, sodium doublets, molecular bands) and atmospheric absorption regions are marked out on some plots as examples.}
\label{fig:spec_example}
\end{figure*}

\begin{figure*} 
\centering
\includegraphics[width=3.3in,angle=0]{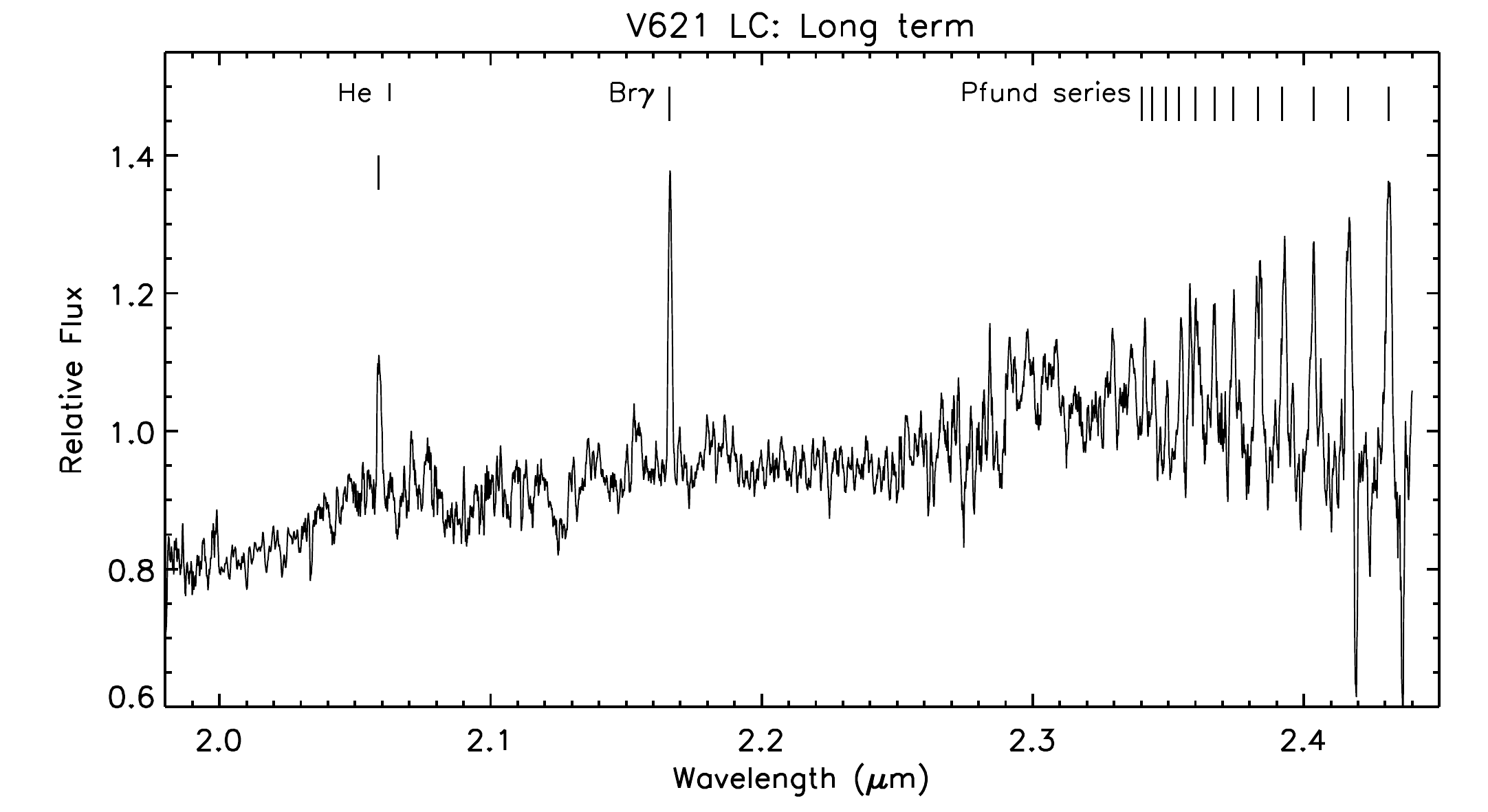}
\includegraphics[width=3.3in,angle=0]{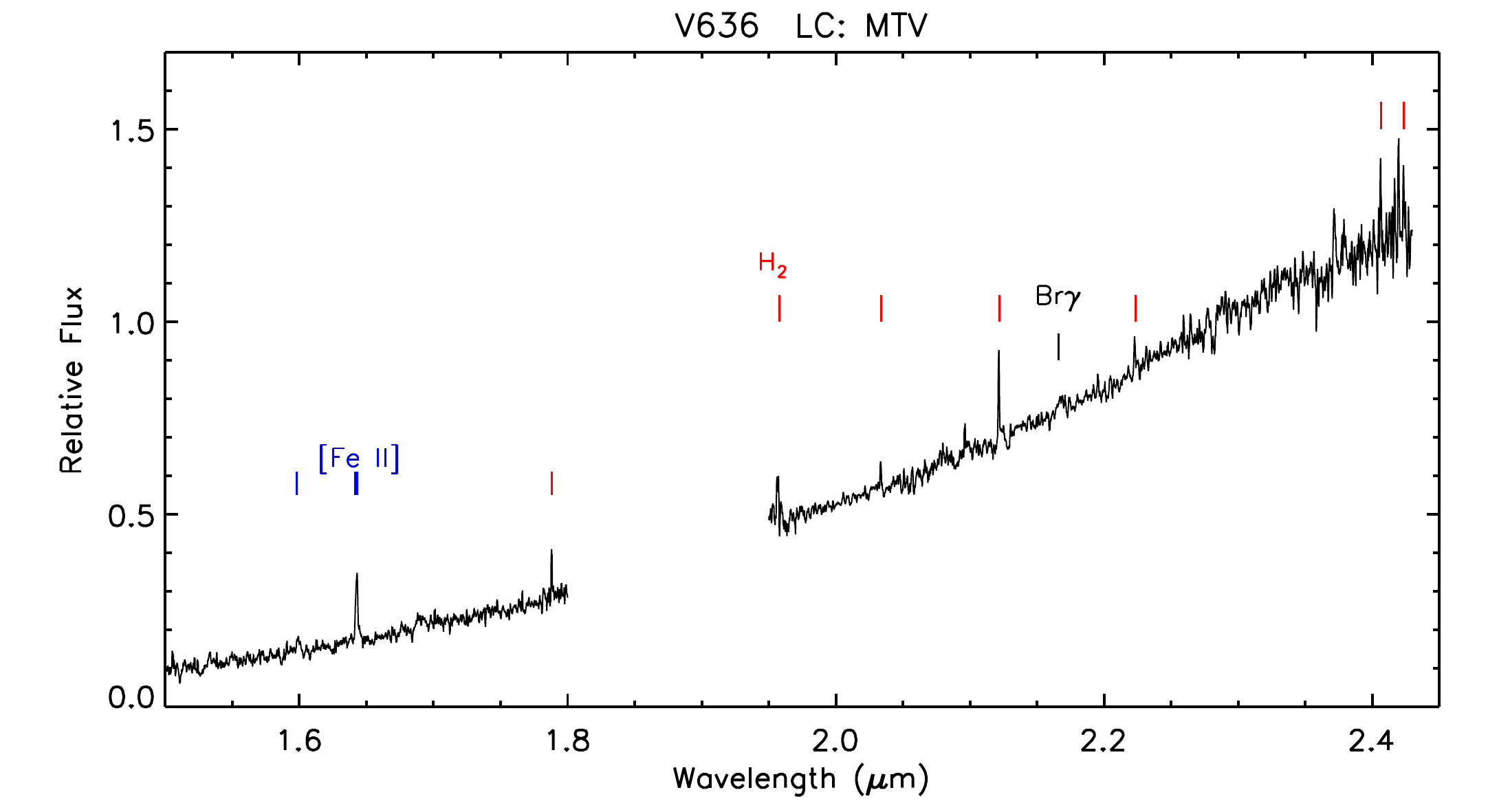}
\includegraphics[width=3.3in,angle=0]{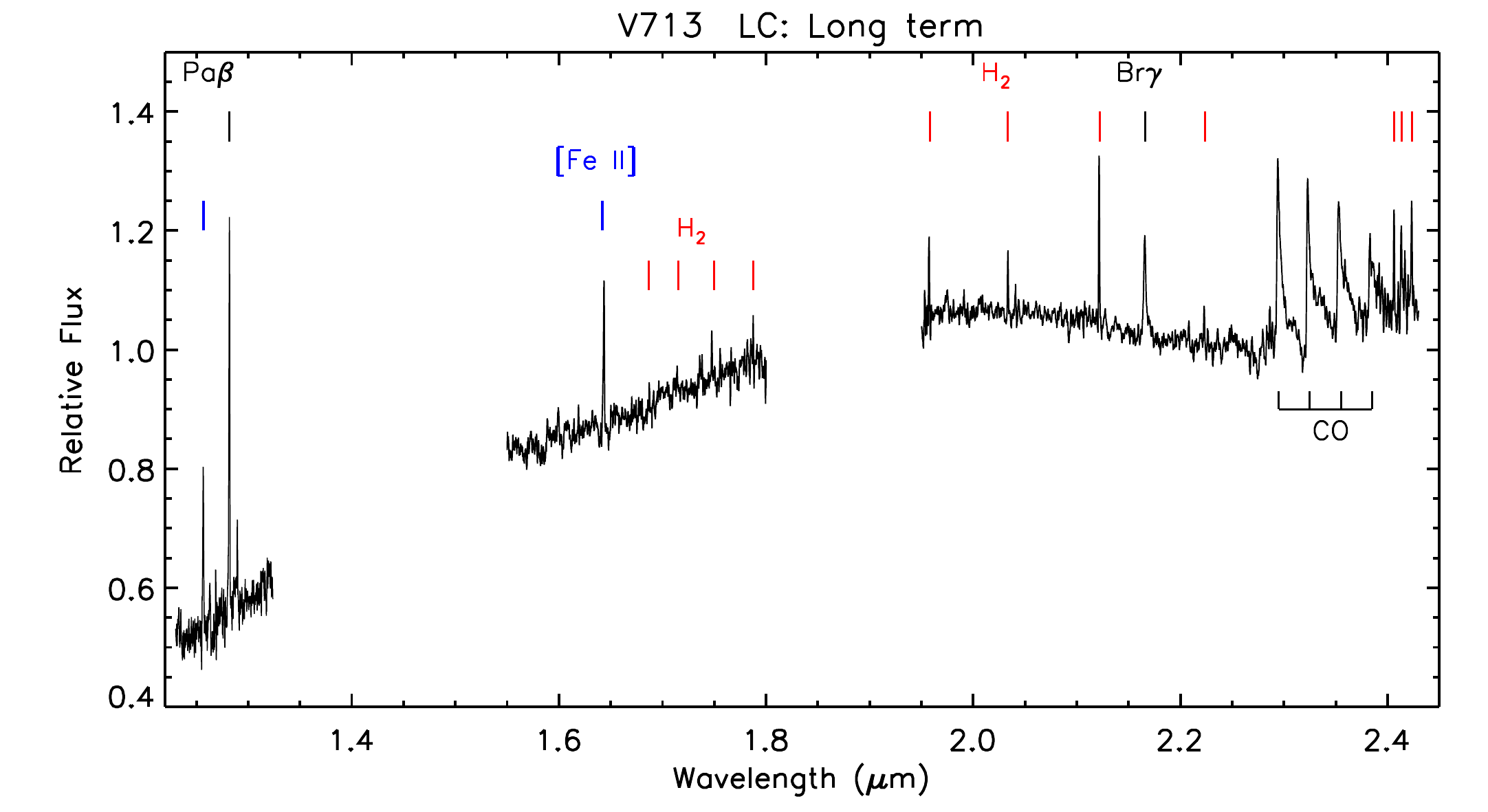}
\includegraphics[width=3.3in,angle=0]{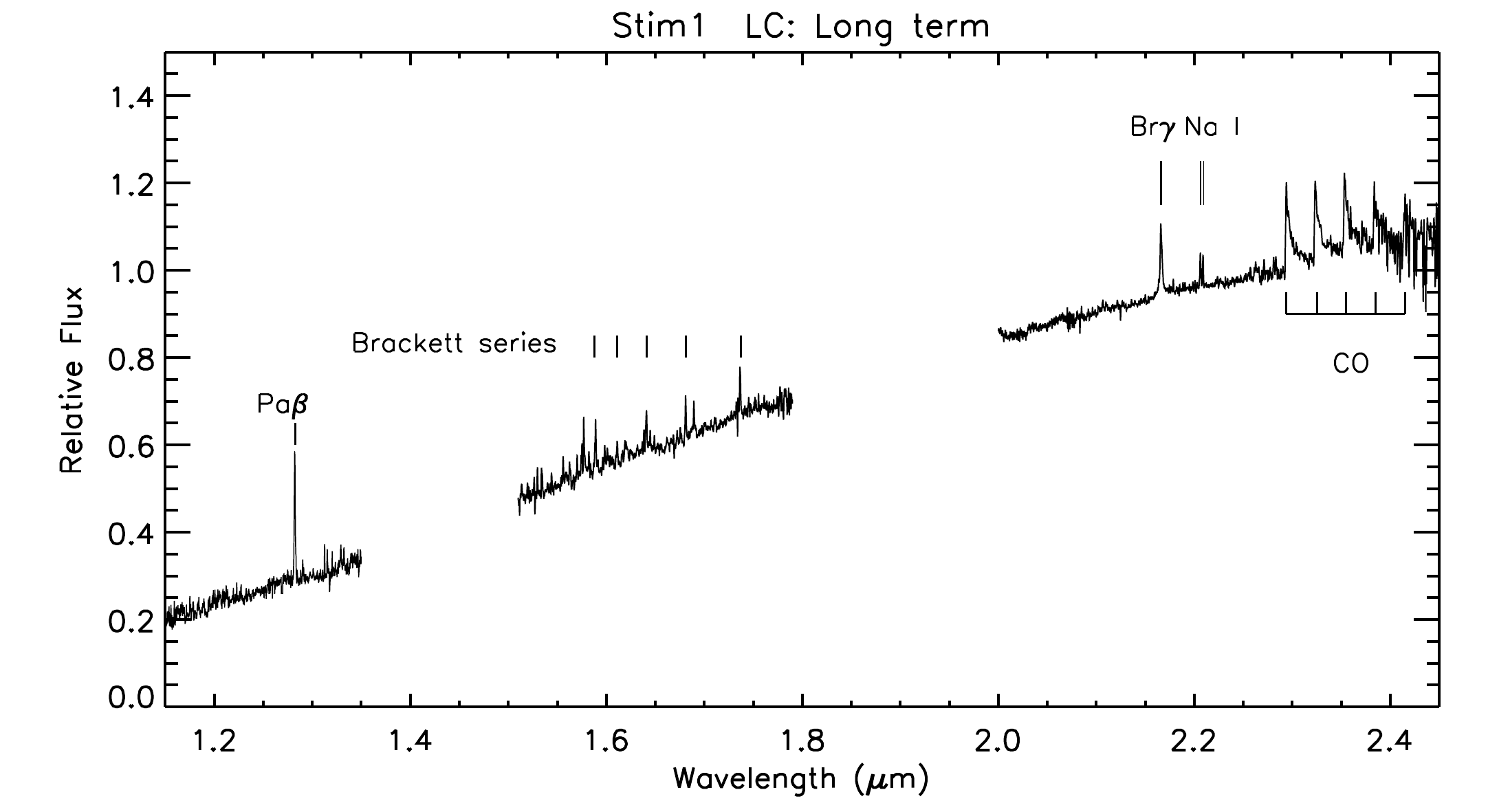}
\includegraphics[width=3.3in,angle=0]{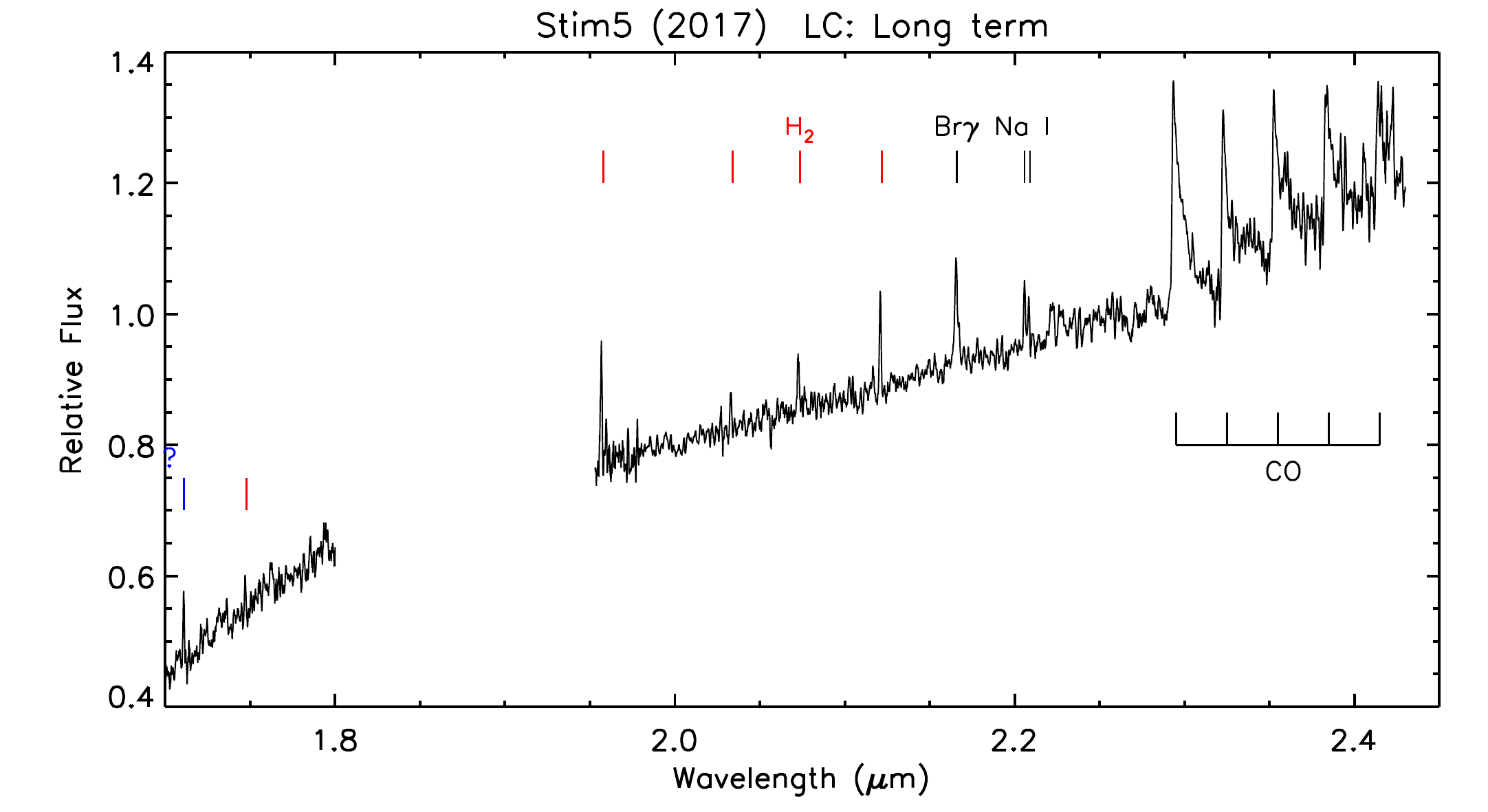}
\includegraphics[width=3.3in,angle=0]{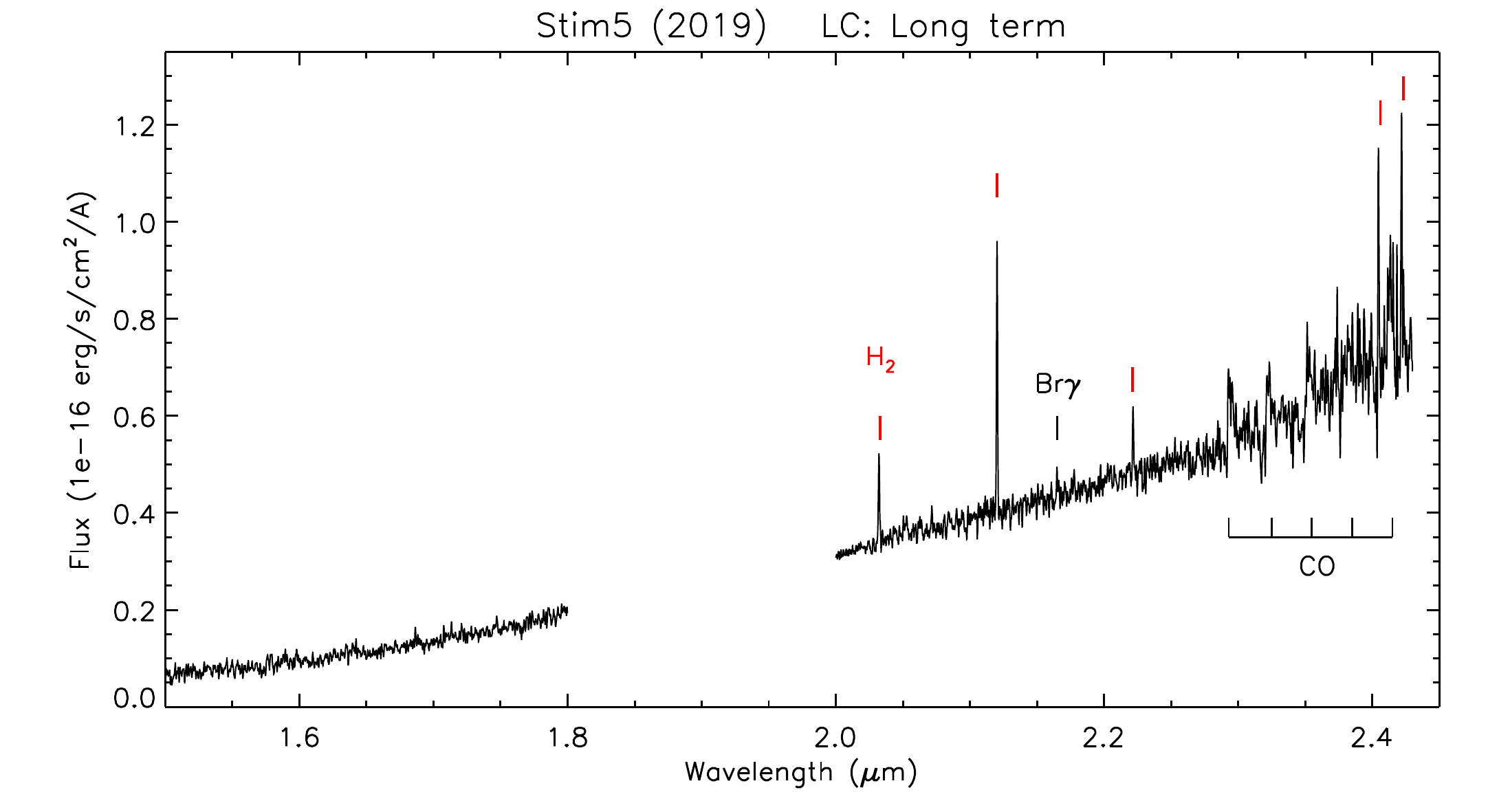}
\includegraphics[width=3.3in,angle=0]{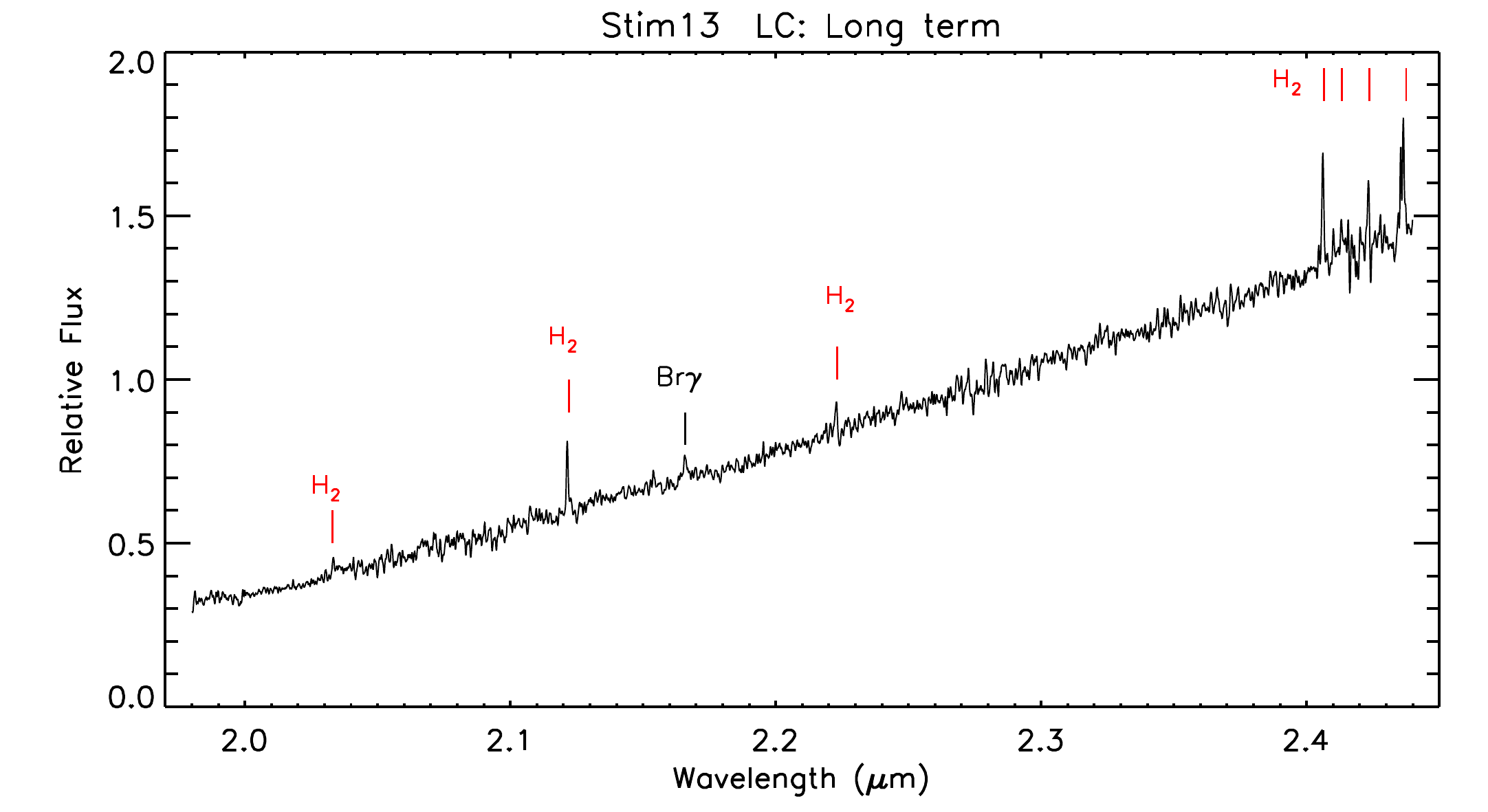}
\includegraphics[width=3.3in,angle=0]{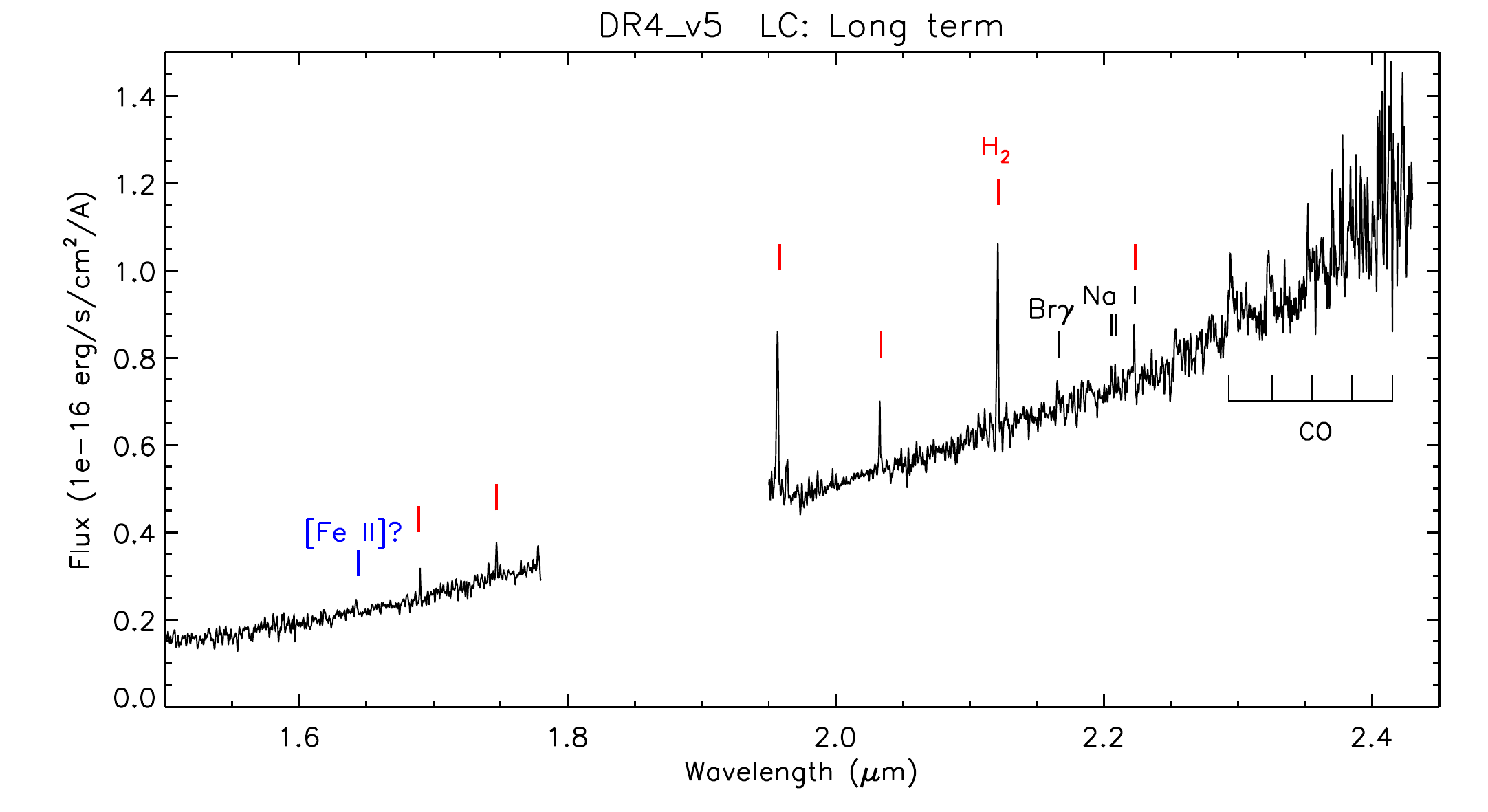}
\includegraphics[width=3.3in,angle=0]{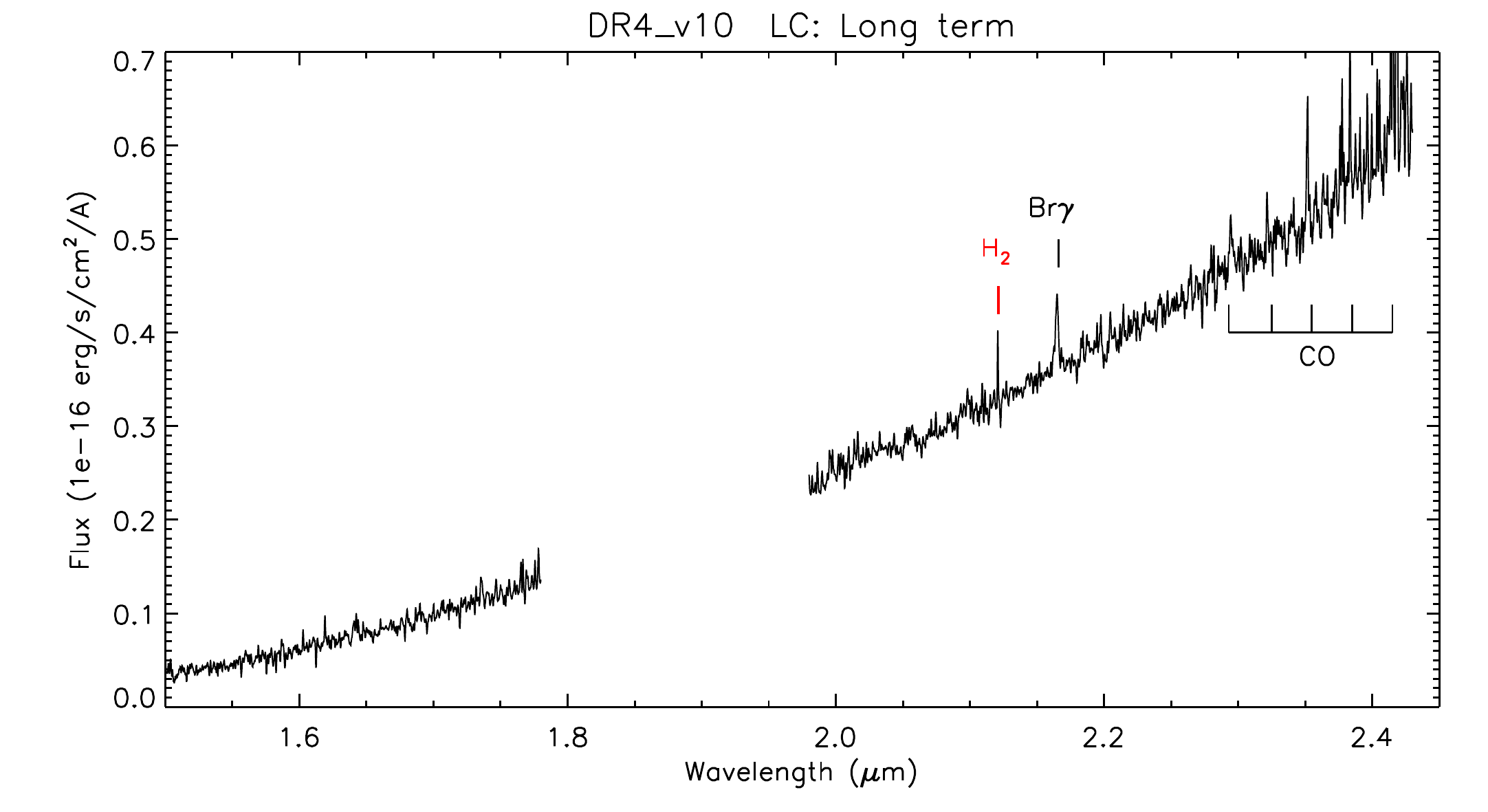}
\includegraphics[width=3.3in,angle=0]{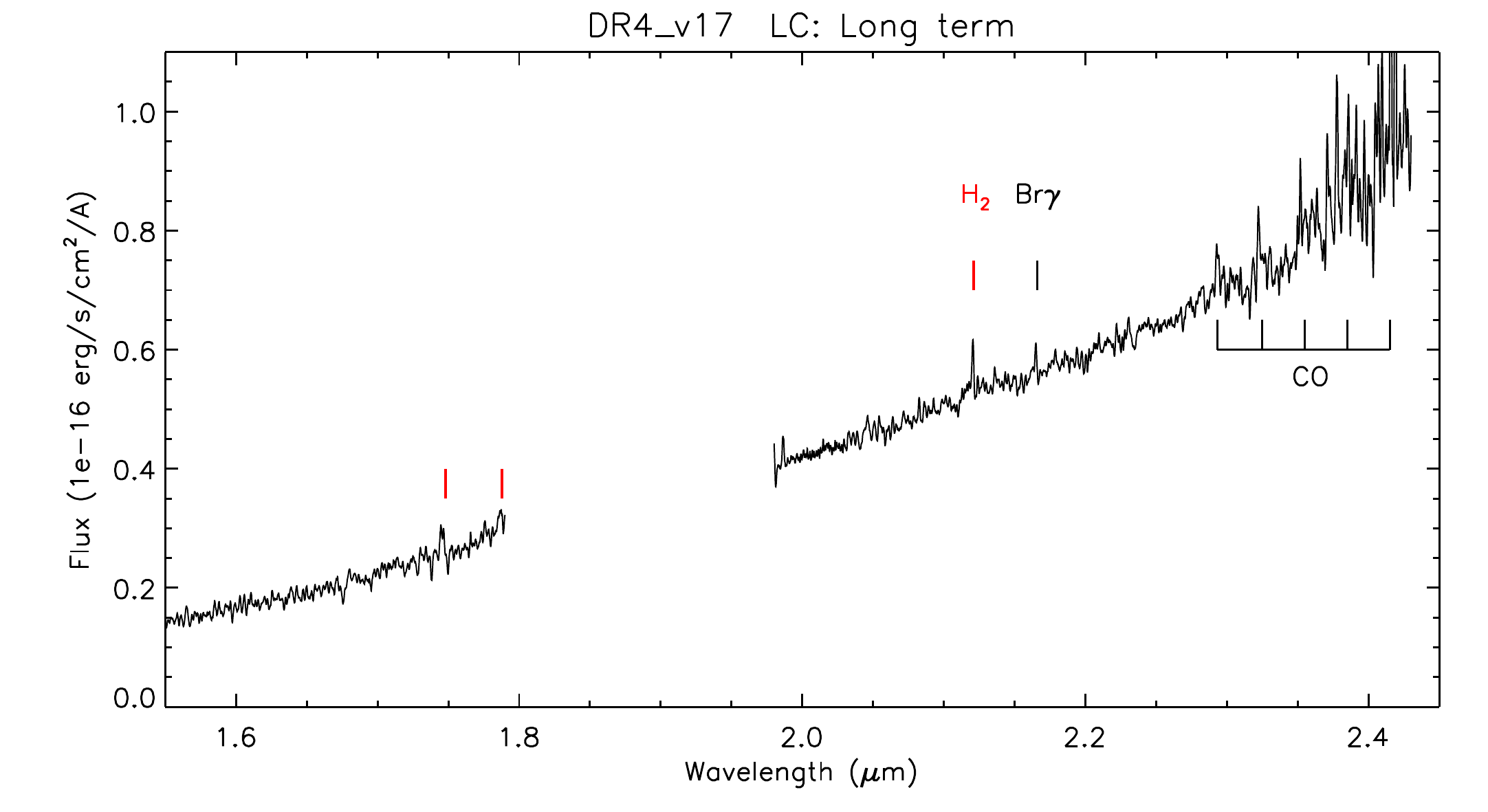}
\caption{Continued}
\end{figure*}

\begin{figure*} 
\centering

\includegraphics[width=3.3in,angle=0]{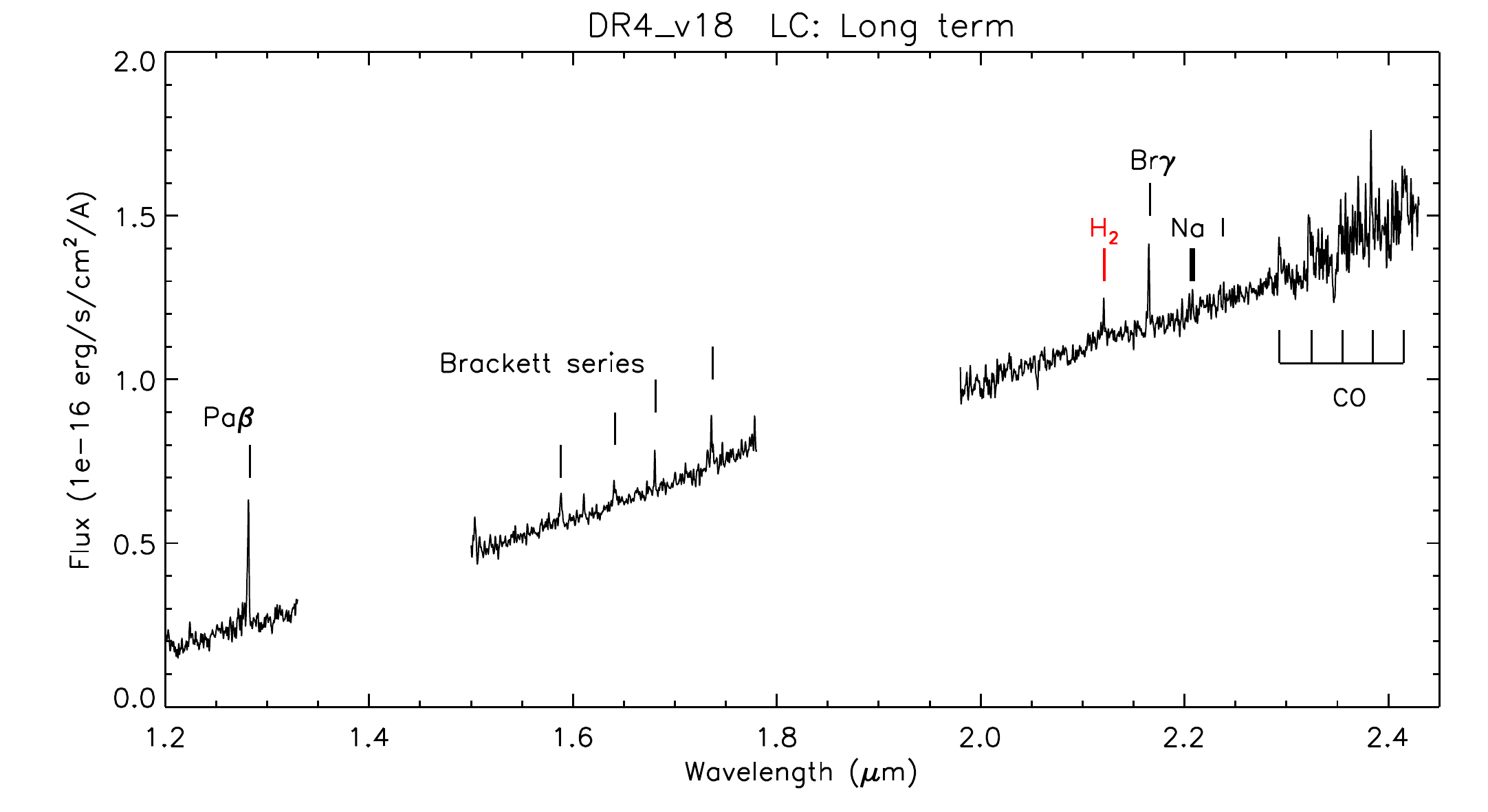}
\includegraphics[width=3.3in,angle=0]{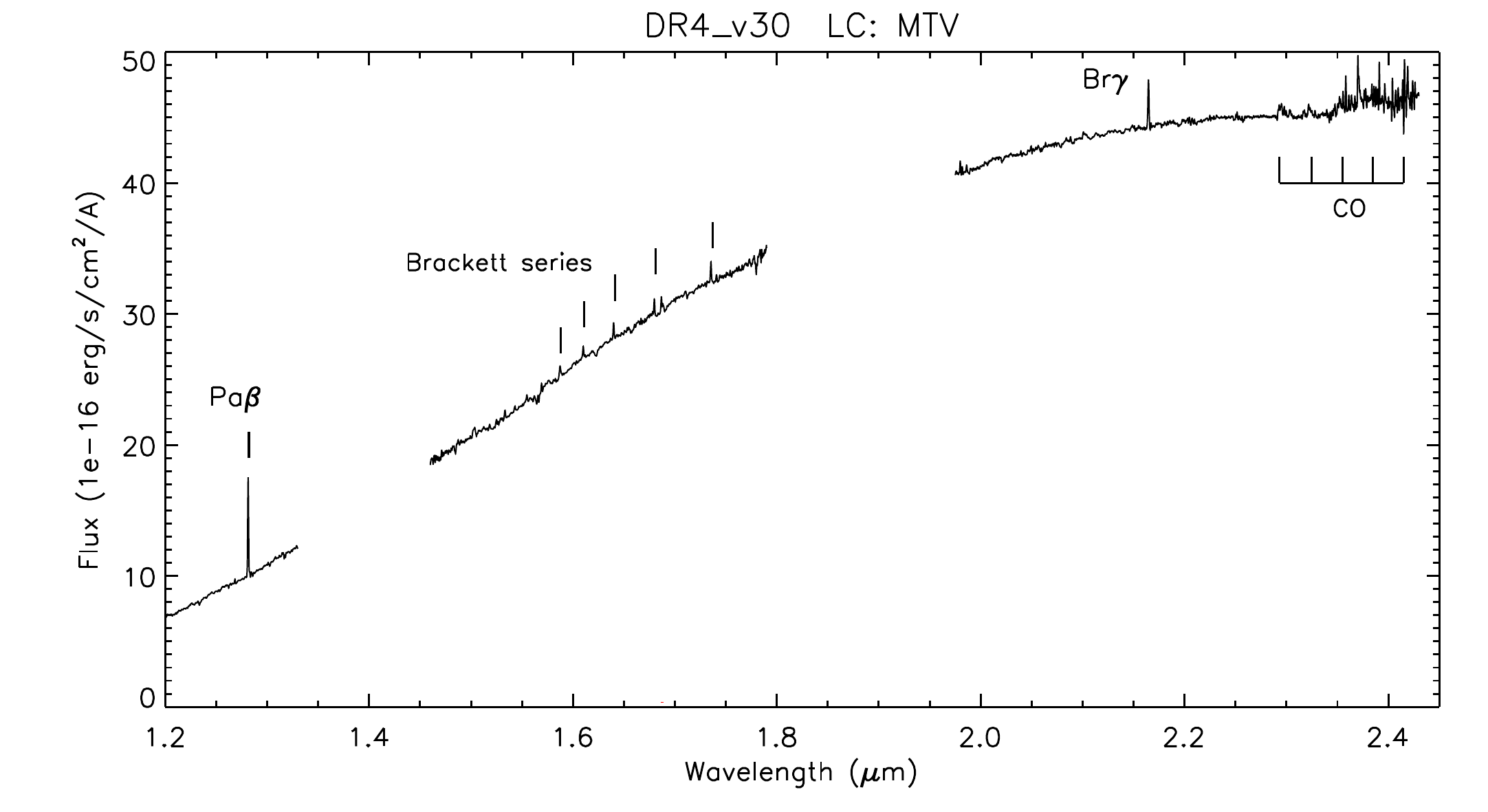}
\includegraphics[width=3.3in,angle=0]{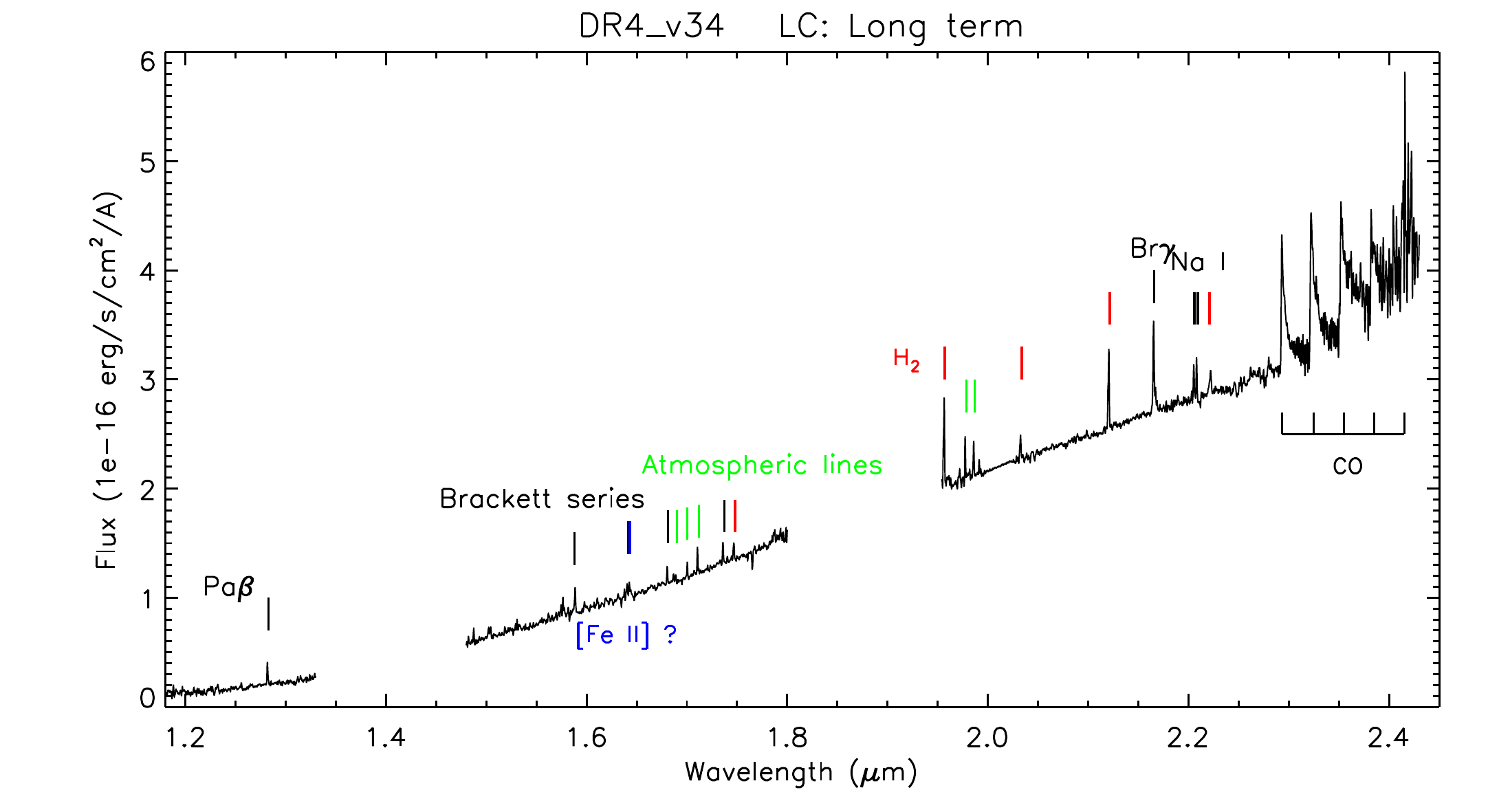}
\includegraphics[width=3.3in,angle=0]{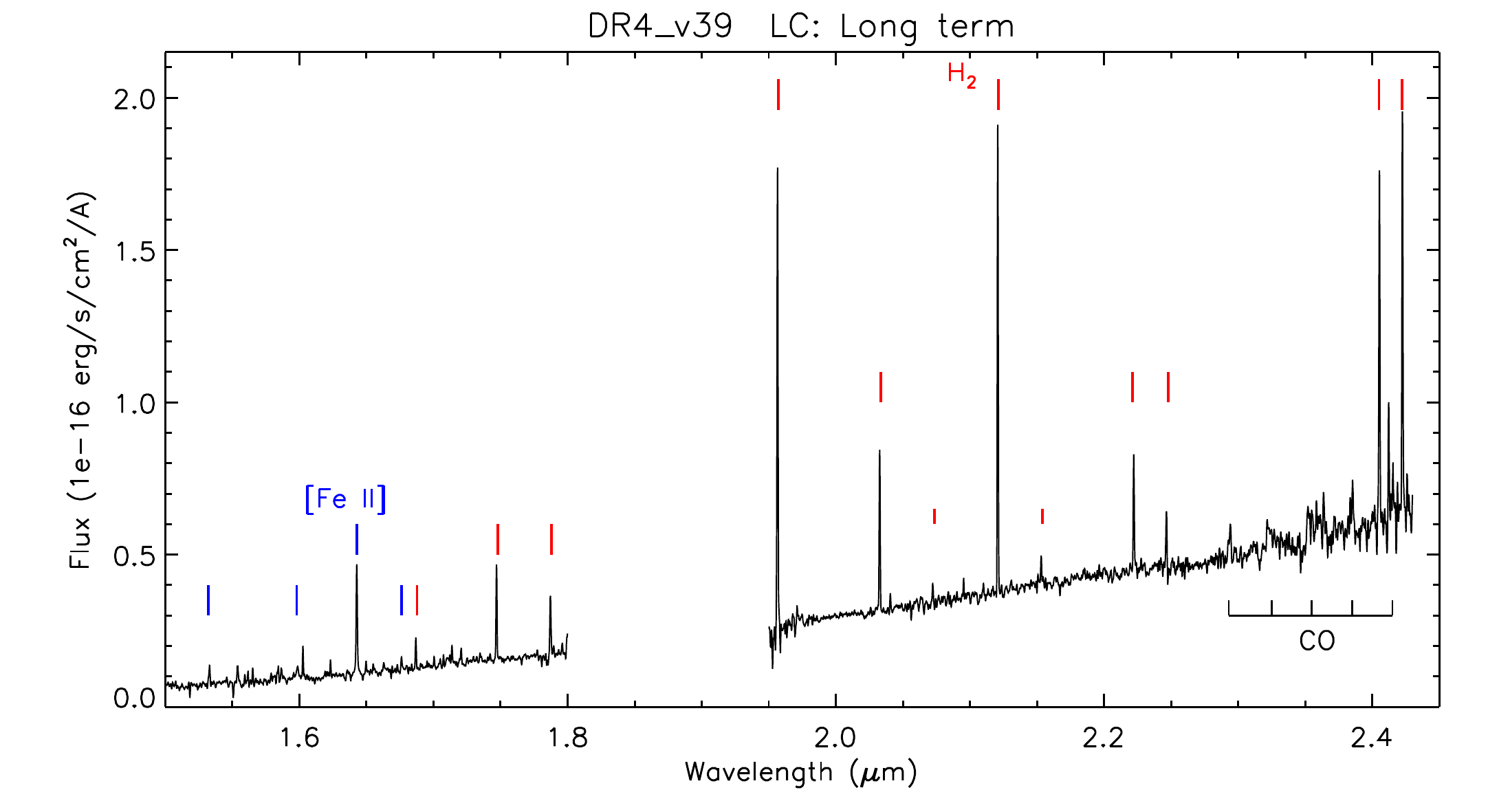}
\includegraphics[width=3.3in,angle=0]{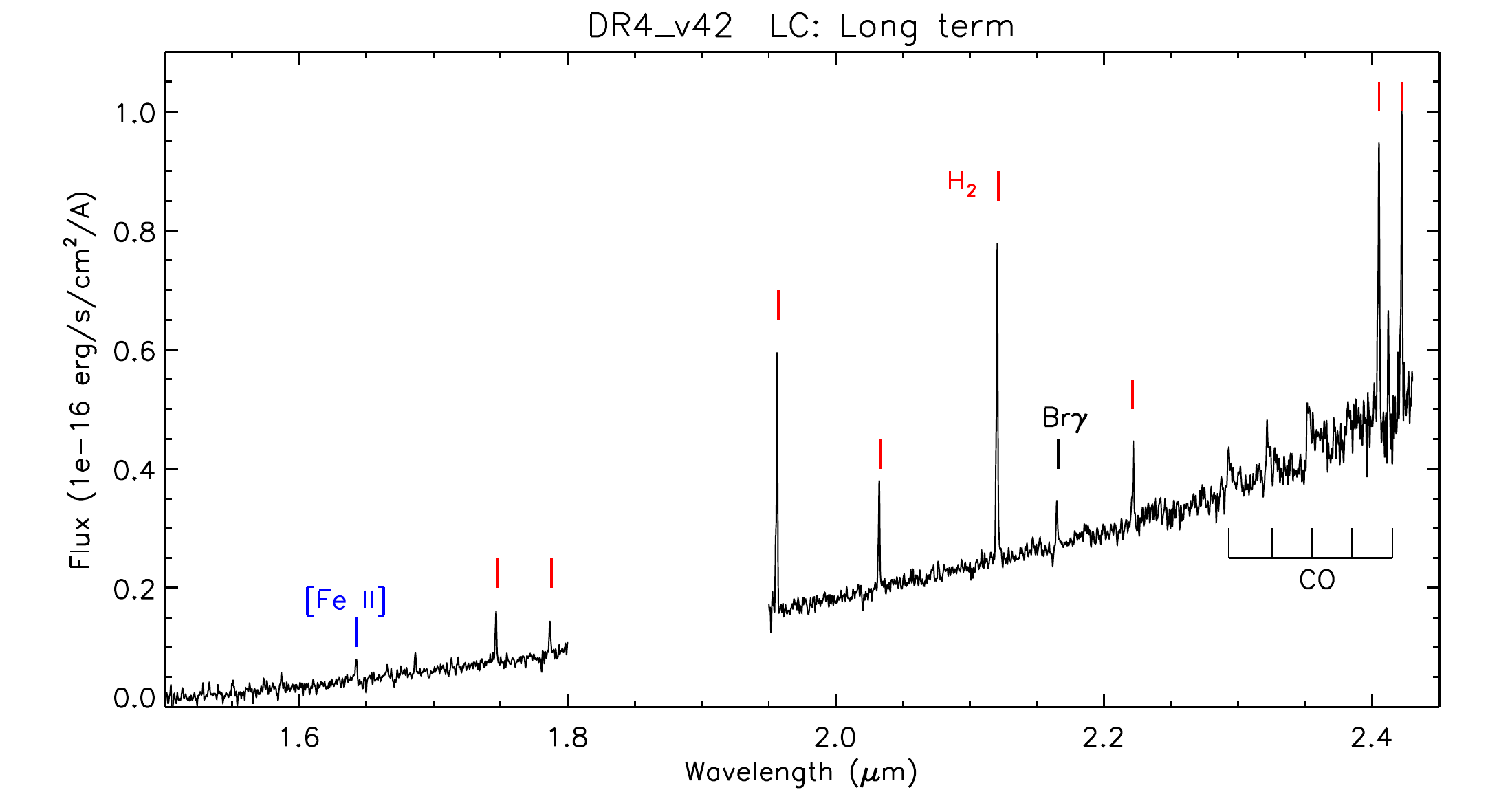}
\includegraphics[width=3.3in,angle=0]{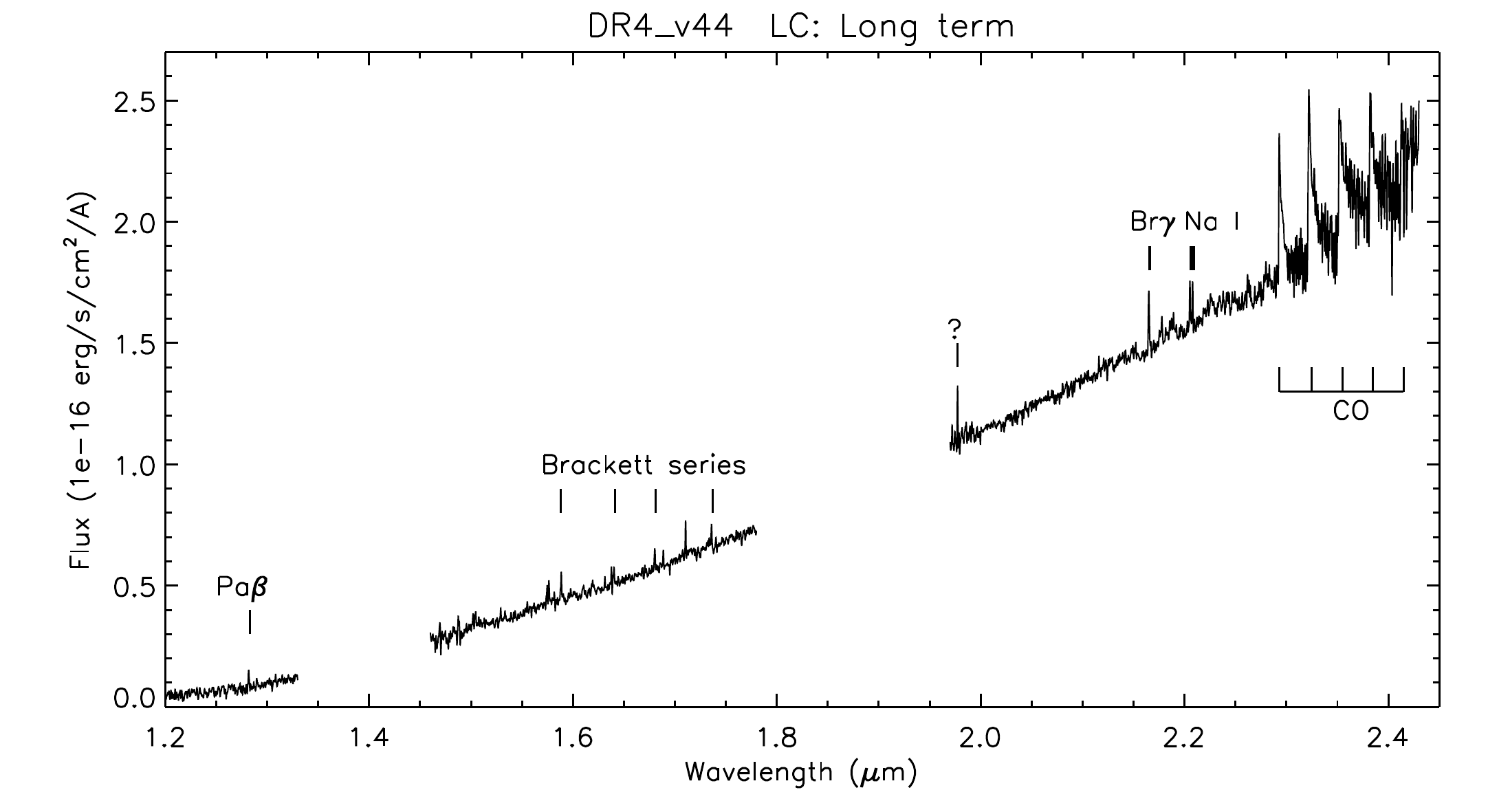}
\caption{Continued}
\end{figure*}

\begin{figure*} 
\centering
\includegraphics[width=3.3in,angle=0]{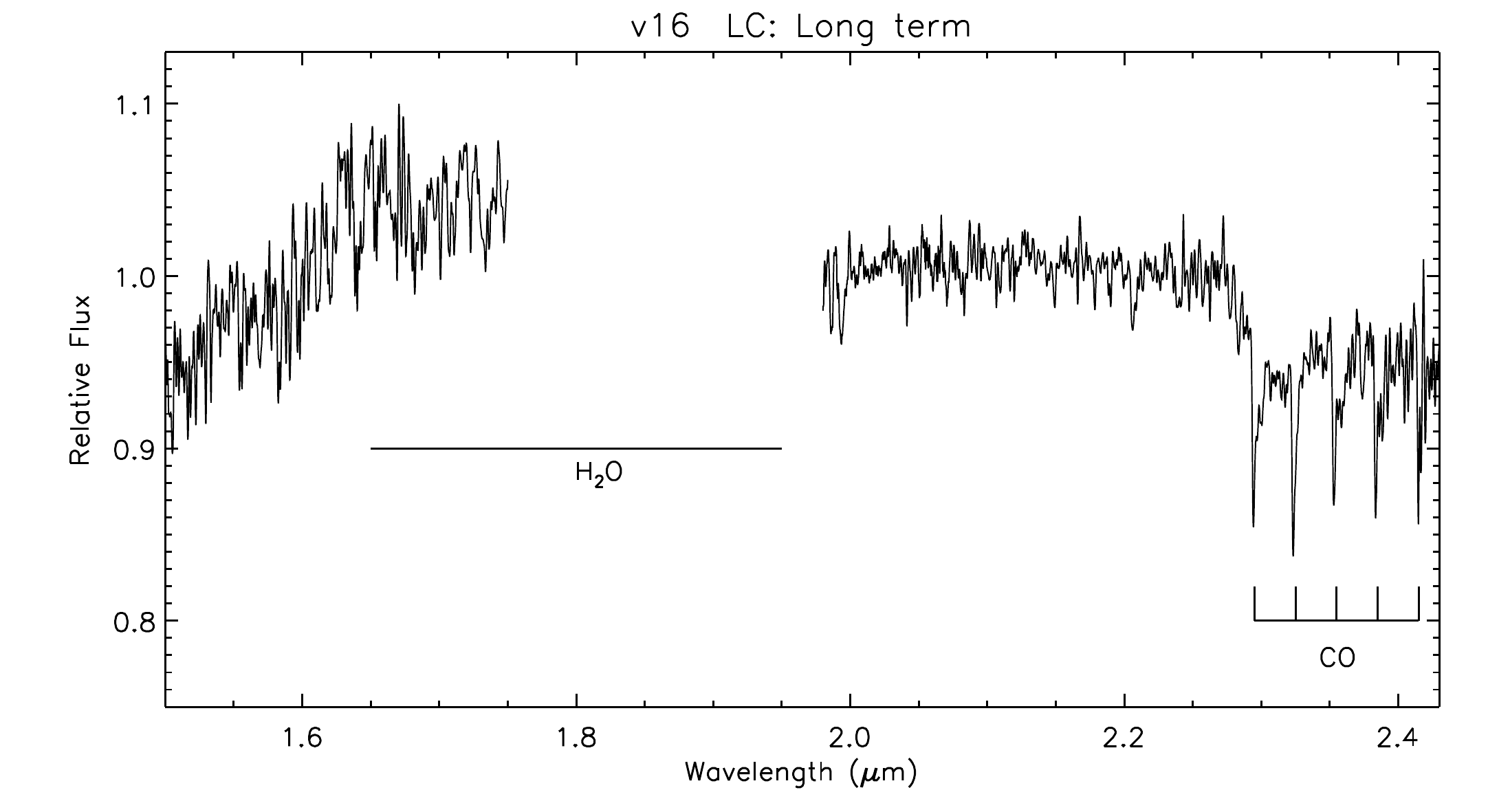}
\includegraphics[width=3.3in,angle=0]{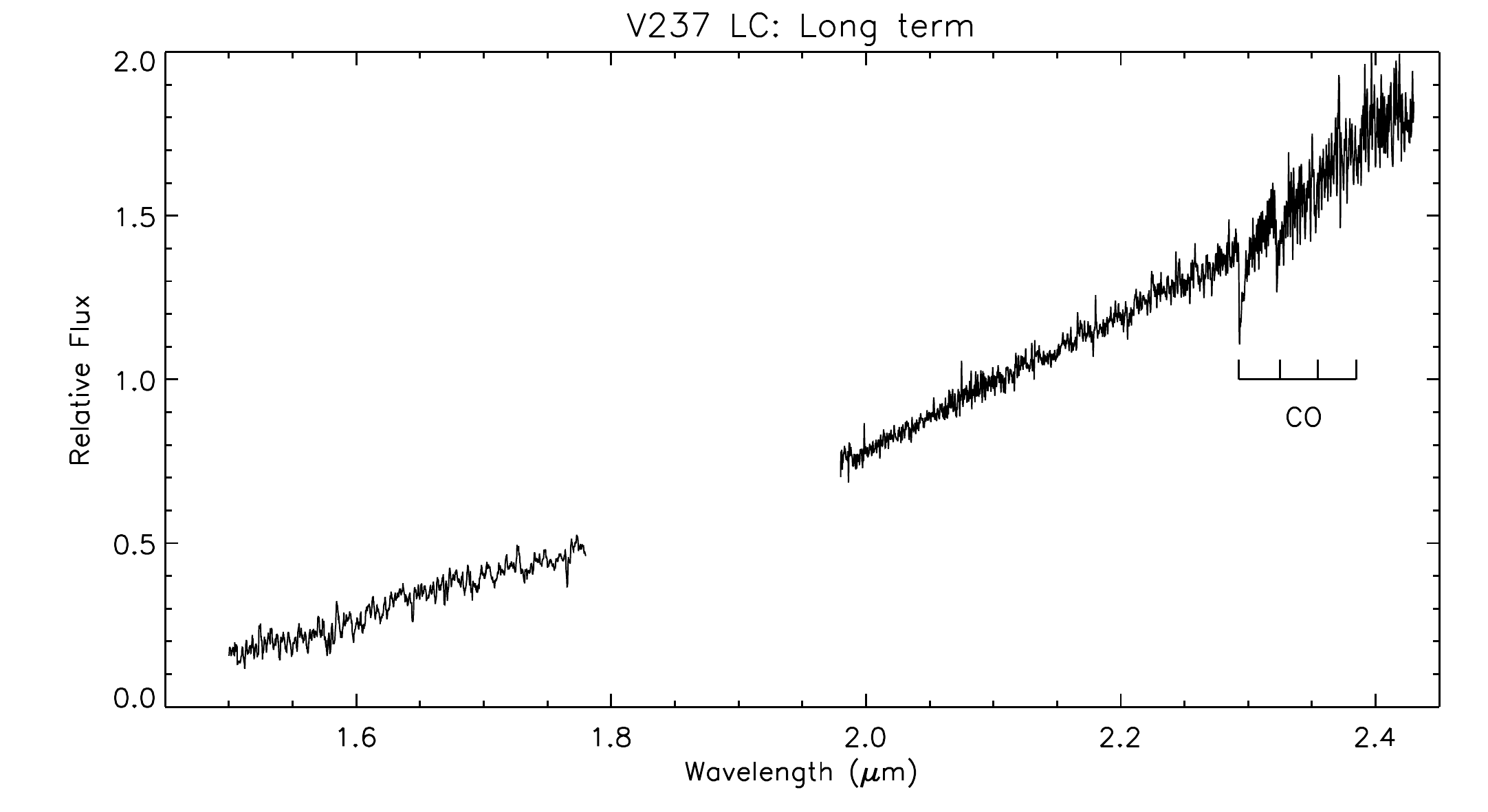}
\includegraphics[width=3.3in,angle=0]{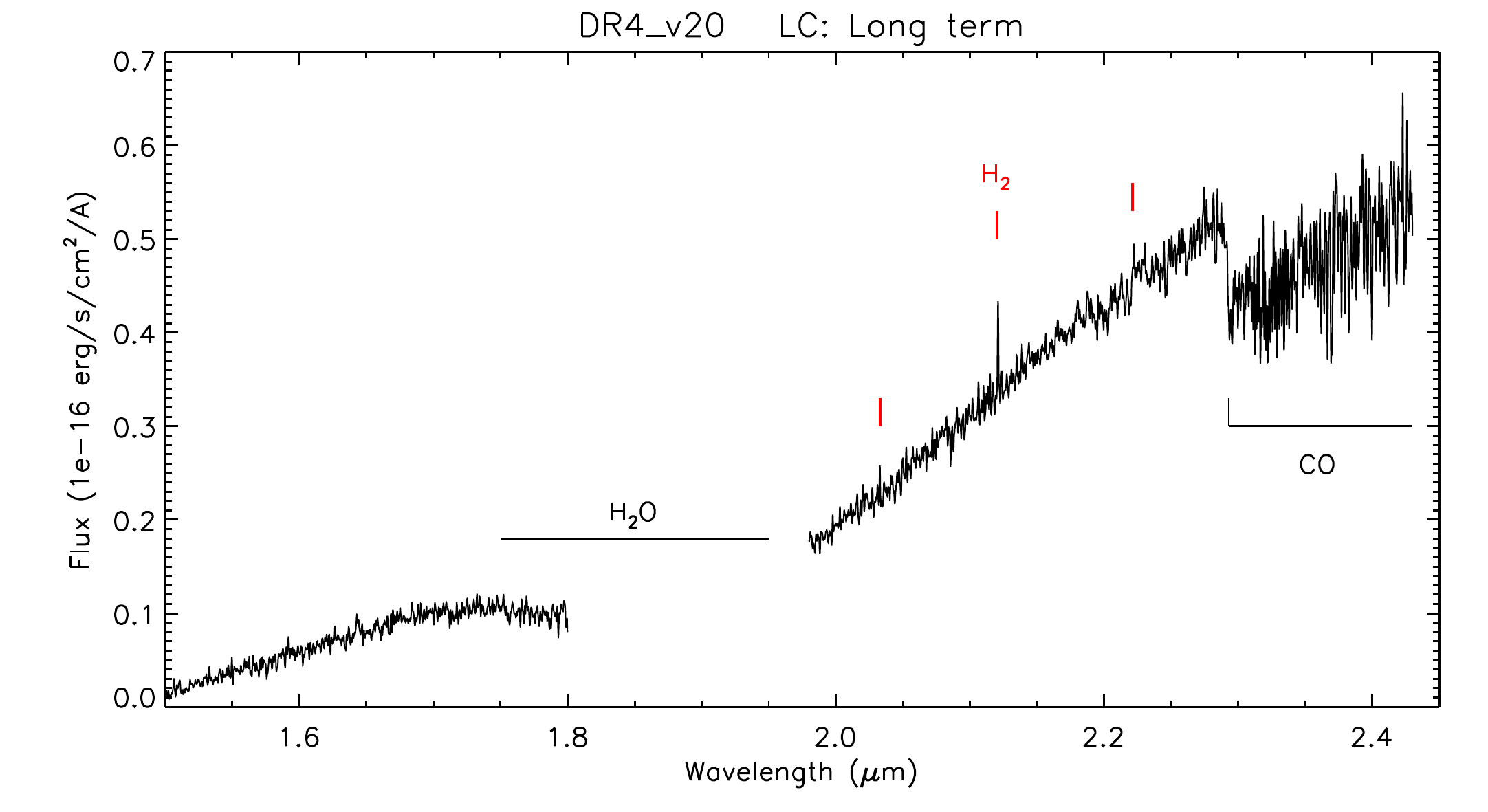}
\caption{Near-IR spectra of FUor-like YSOs with CO absorption features.}
\label{fig:co_absorption}
\end{figure*}

\begin{figure*} 
\centering
\includegraphics[width=3.3in,angle=0]{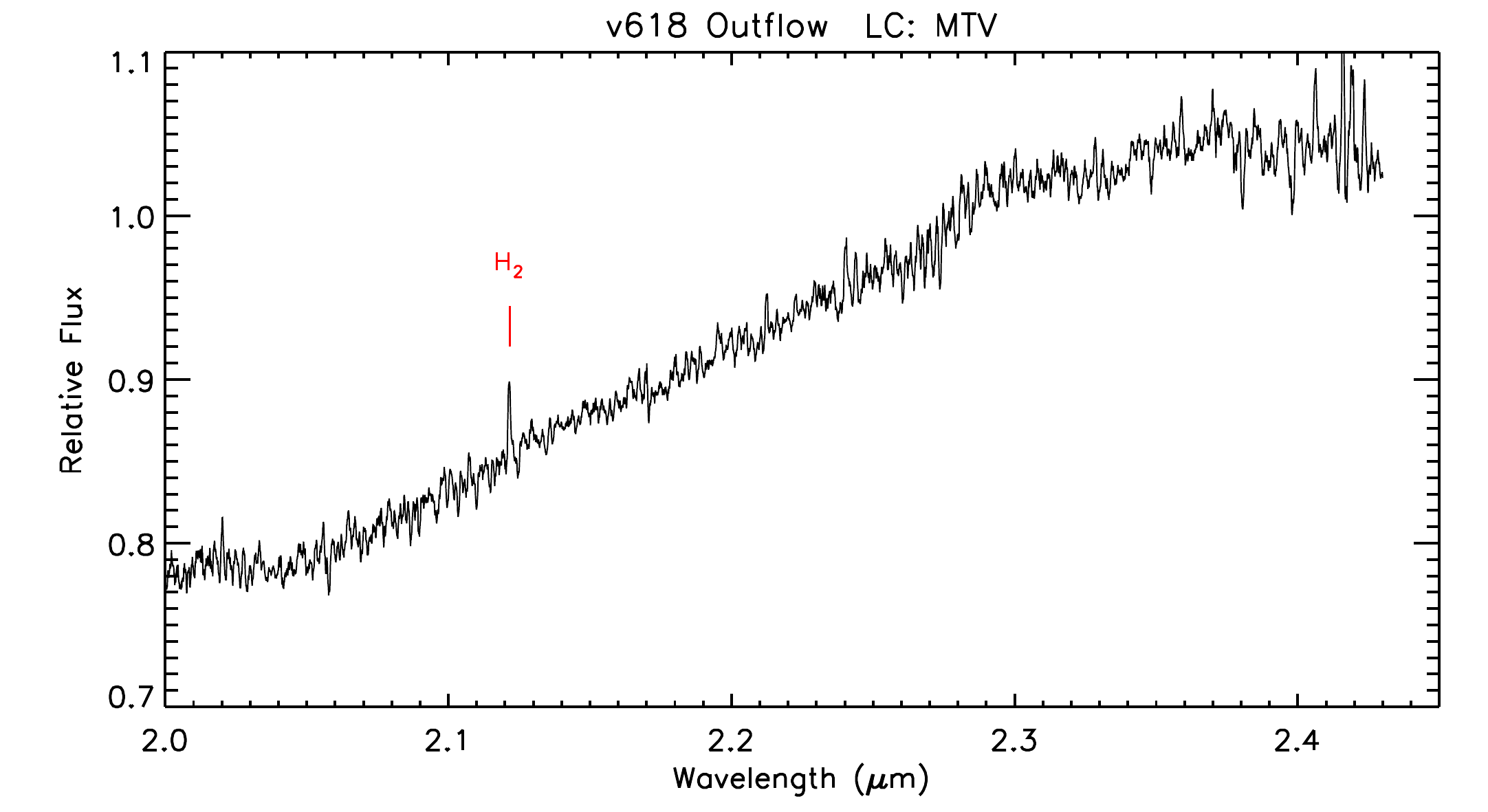}
\includegraphics[width=3.3in,angle=0]{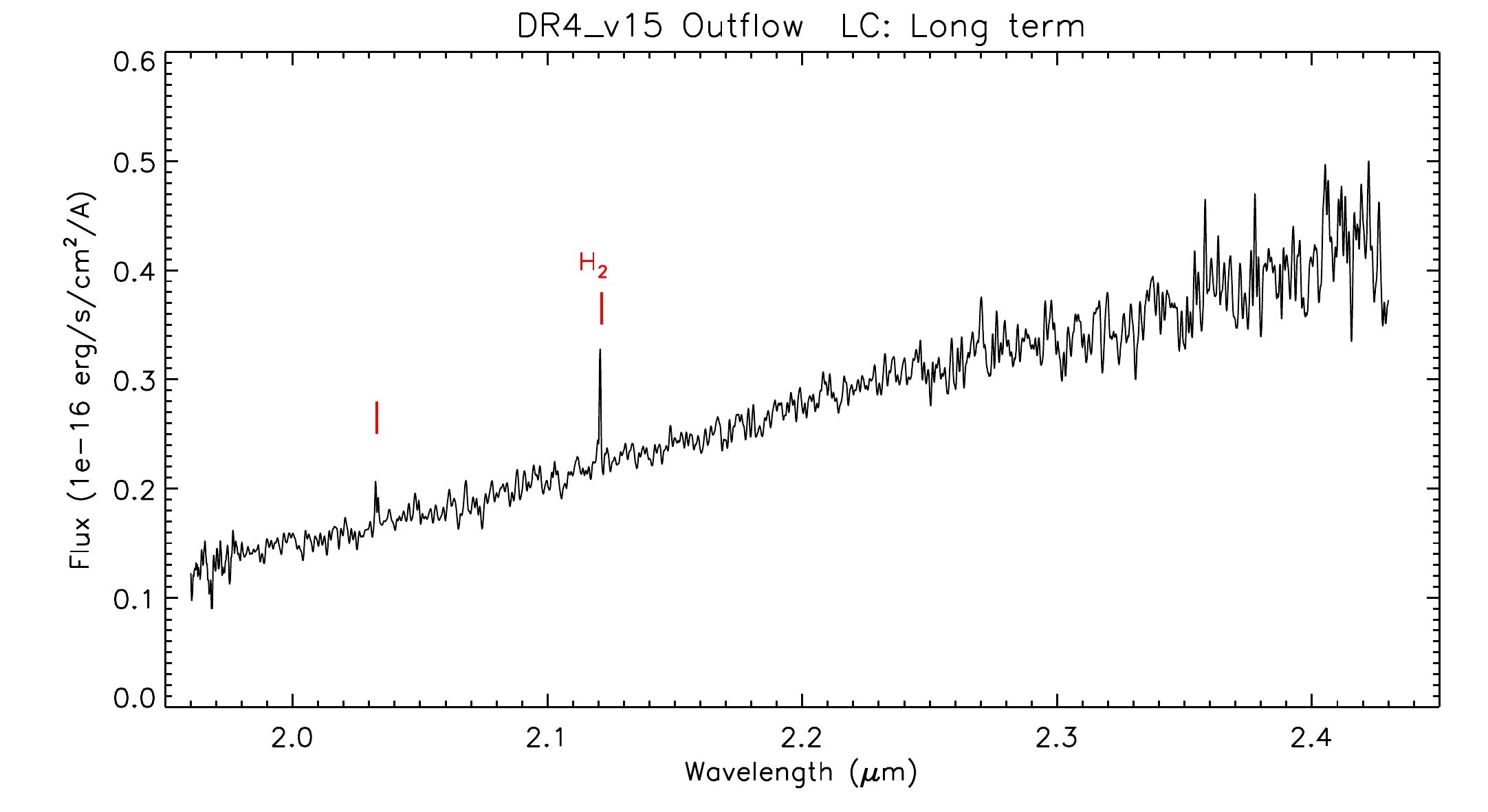}
\includegraphics[width=3.3in,angle=0]{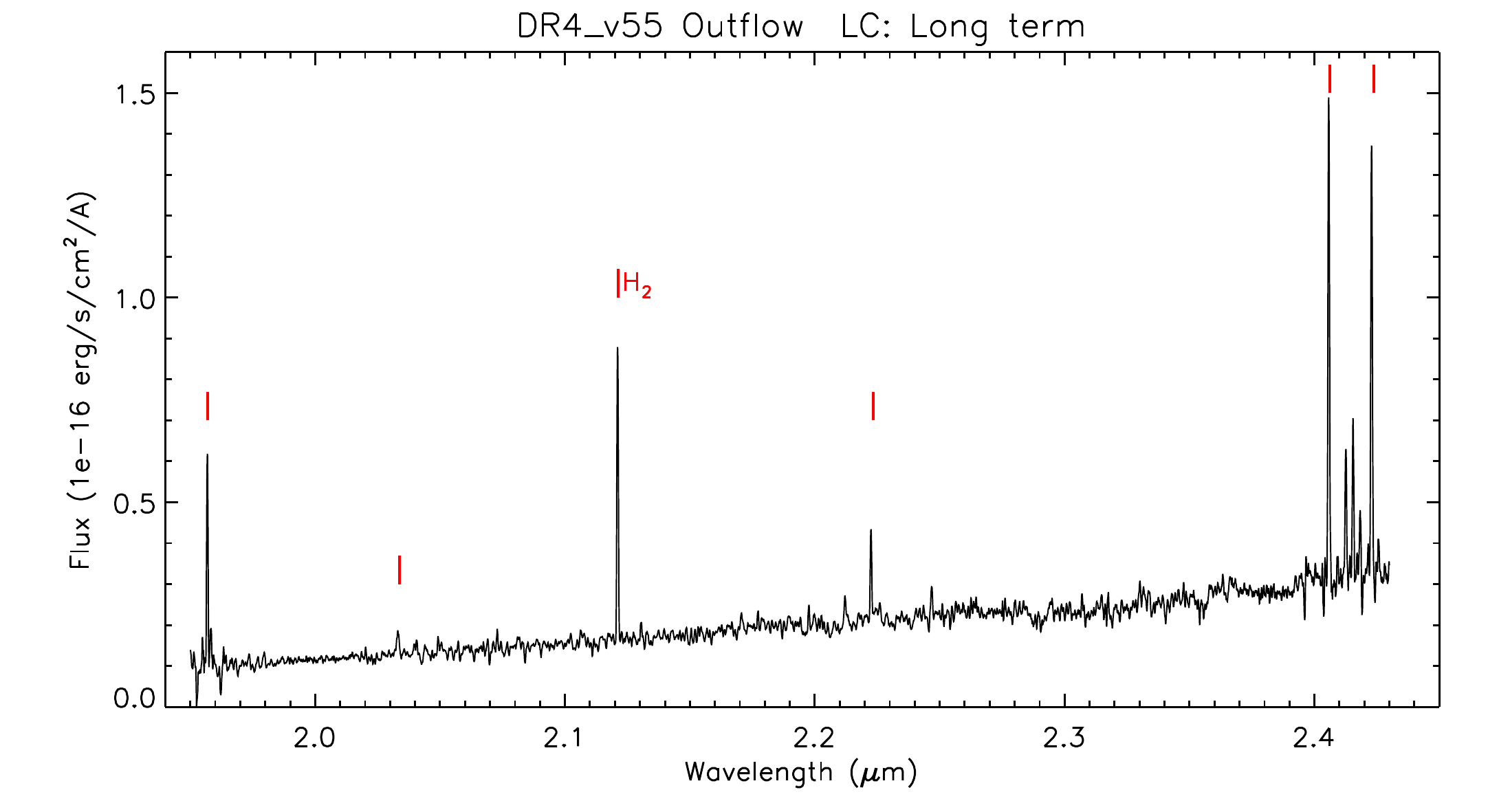}
\includegraphics[width=3.3in,angle=0]{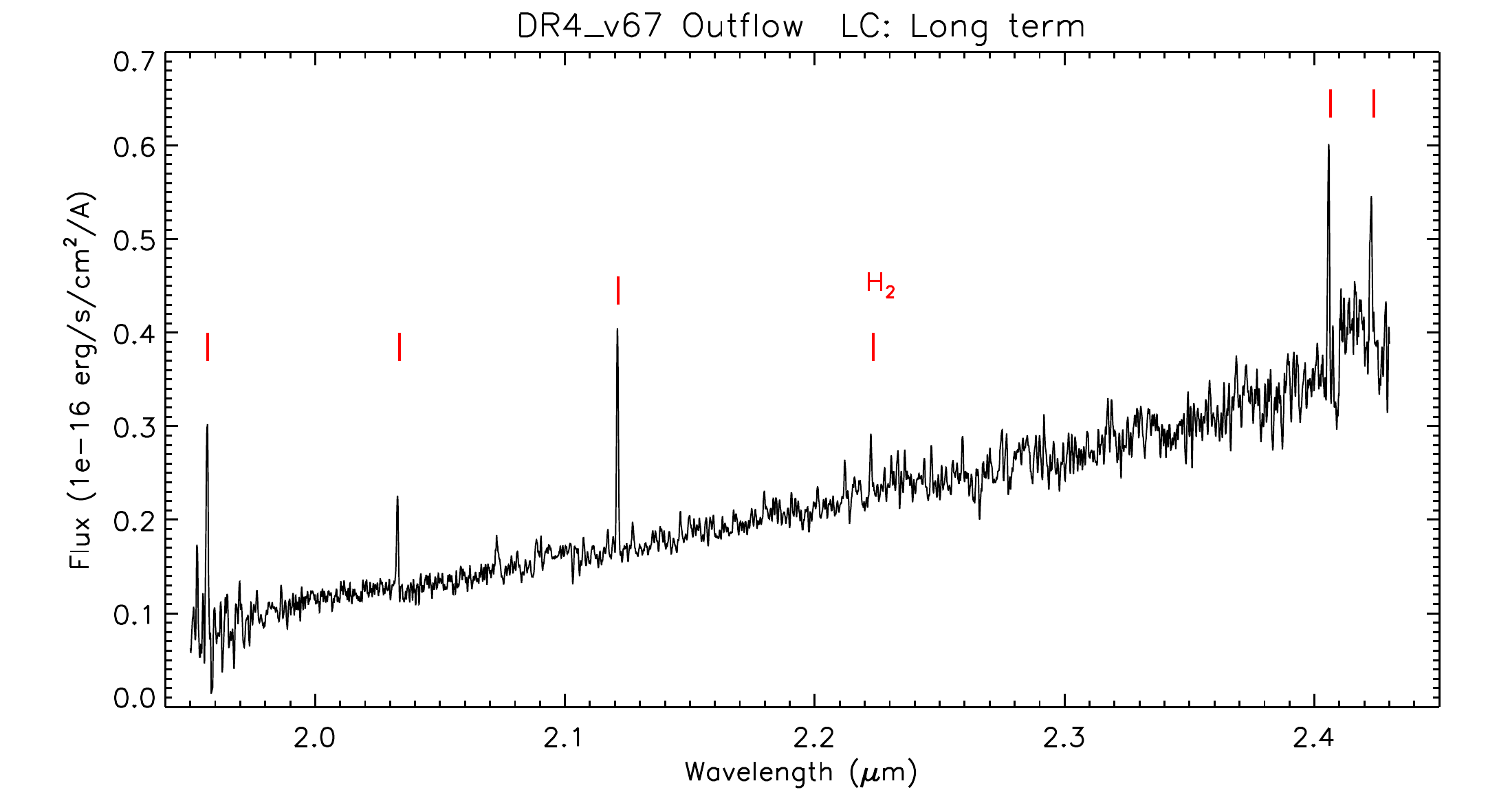}
\includegraphics[width=3.3in,angle=0]{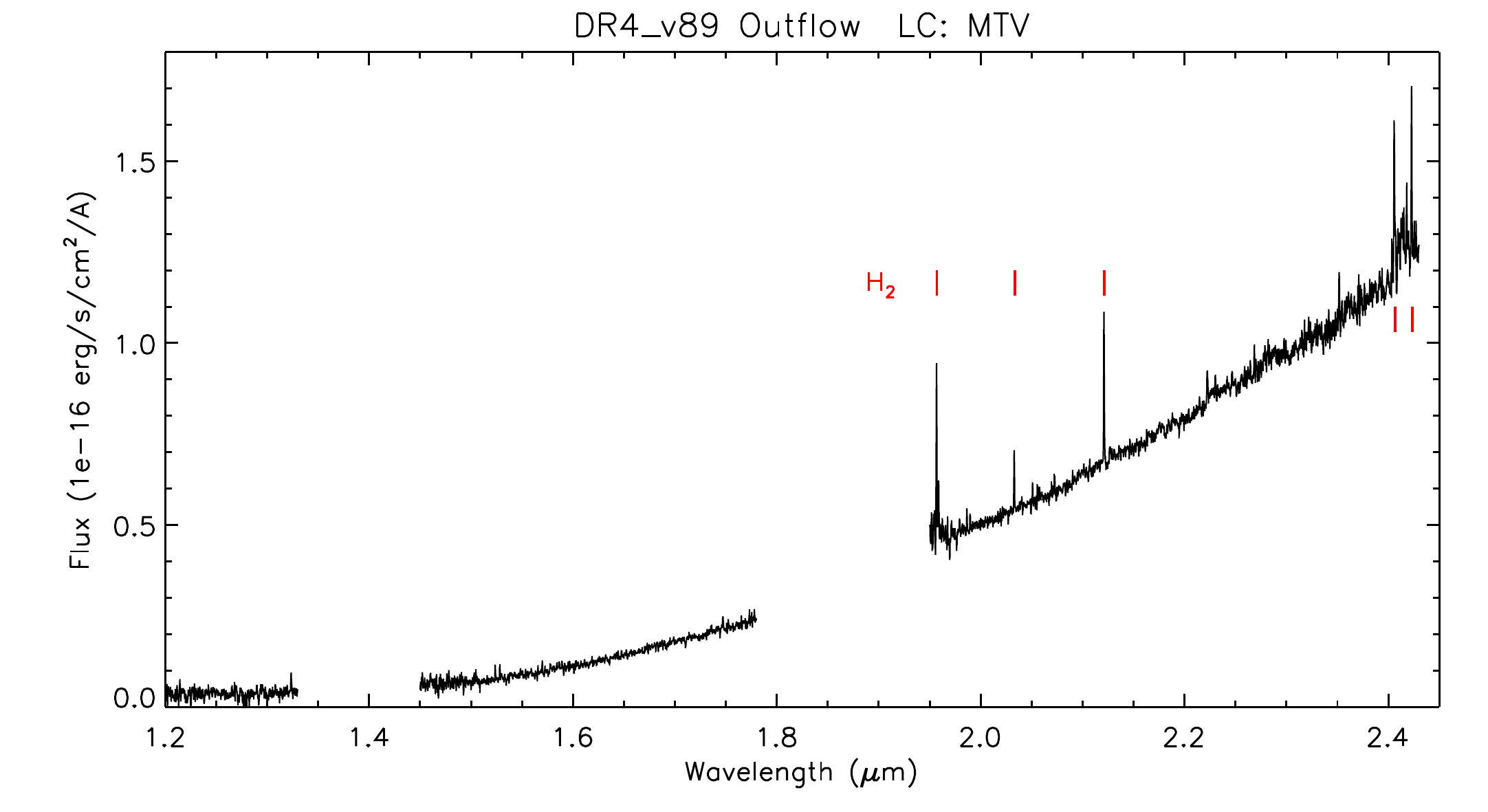}
\includegraphics[width=3.3in,angle=0]{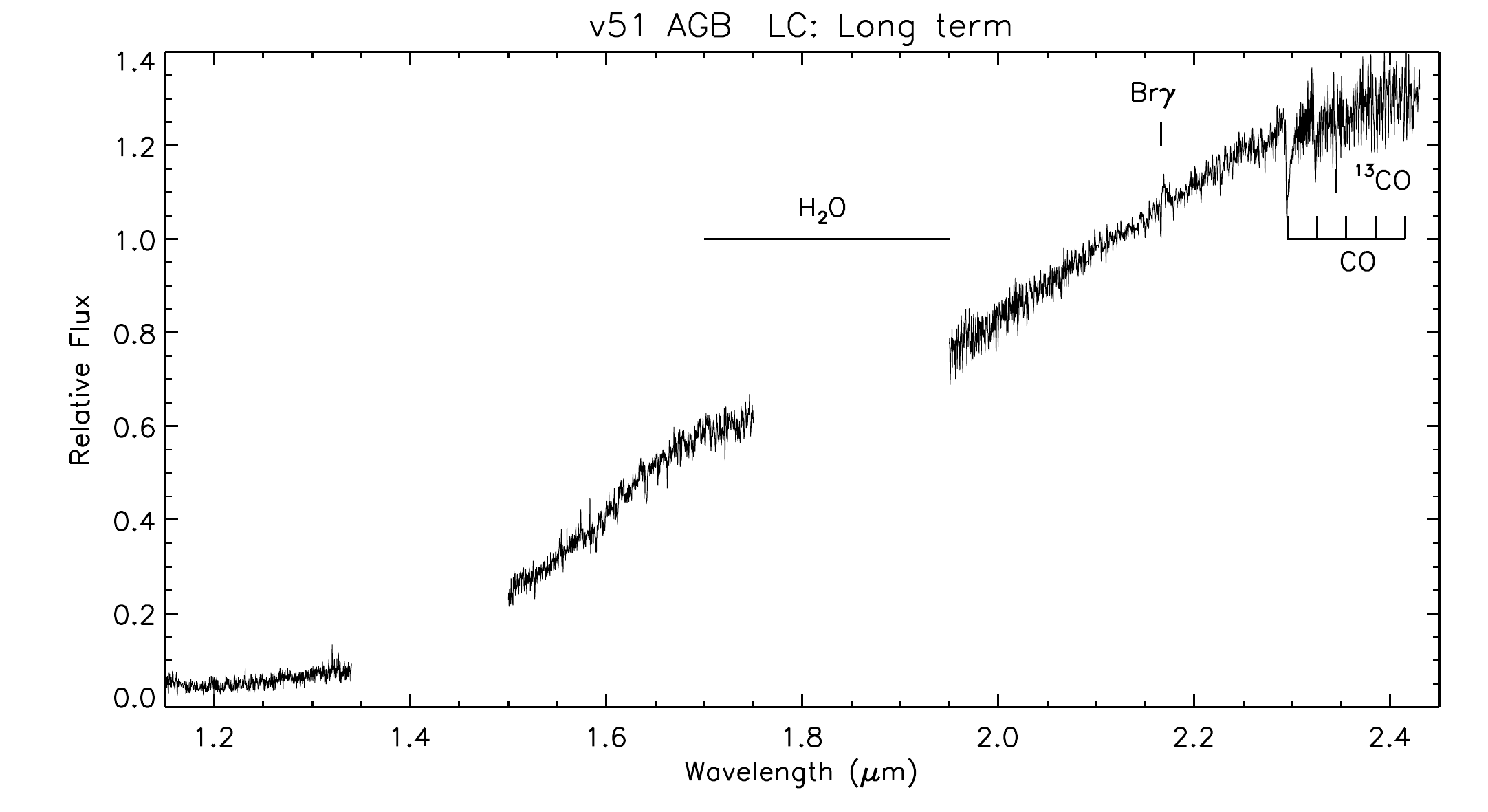}
\includegraphics[width=3.3in,angle=0]{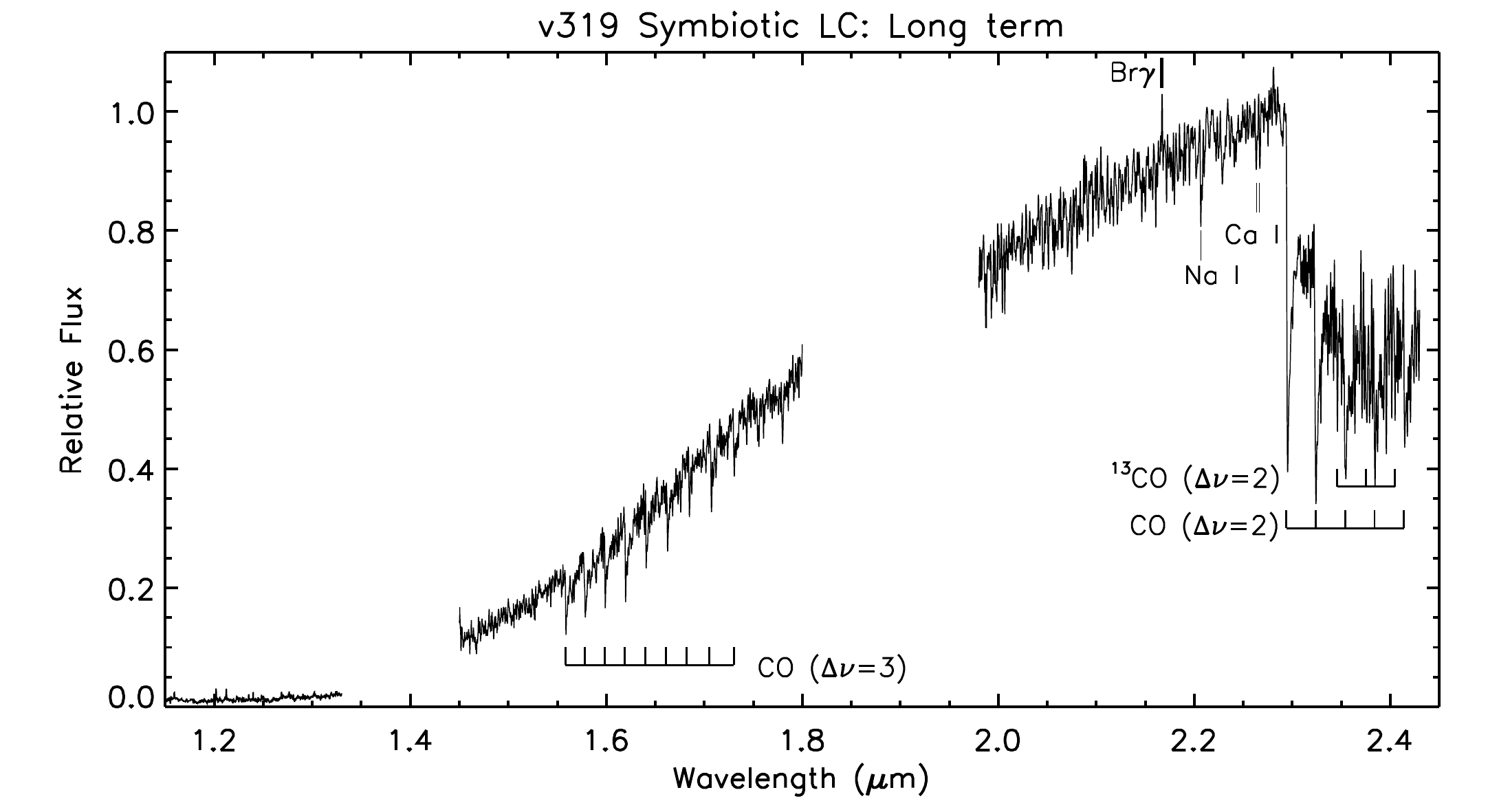}
\includegraphics[width=3.3in,angle=0]{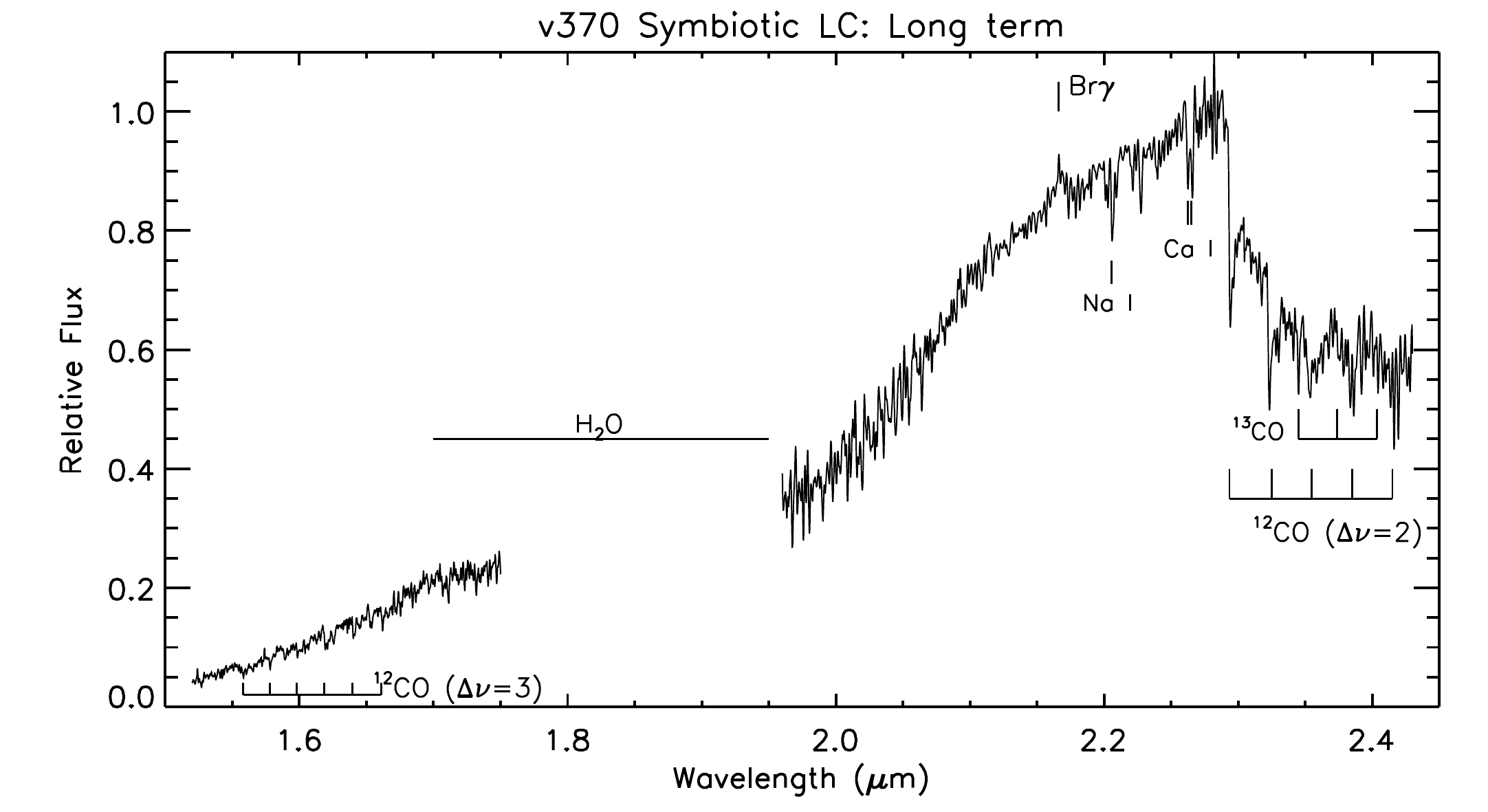}
\includegraphics[width=3.3in,angle=0]{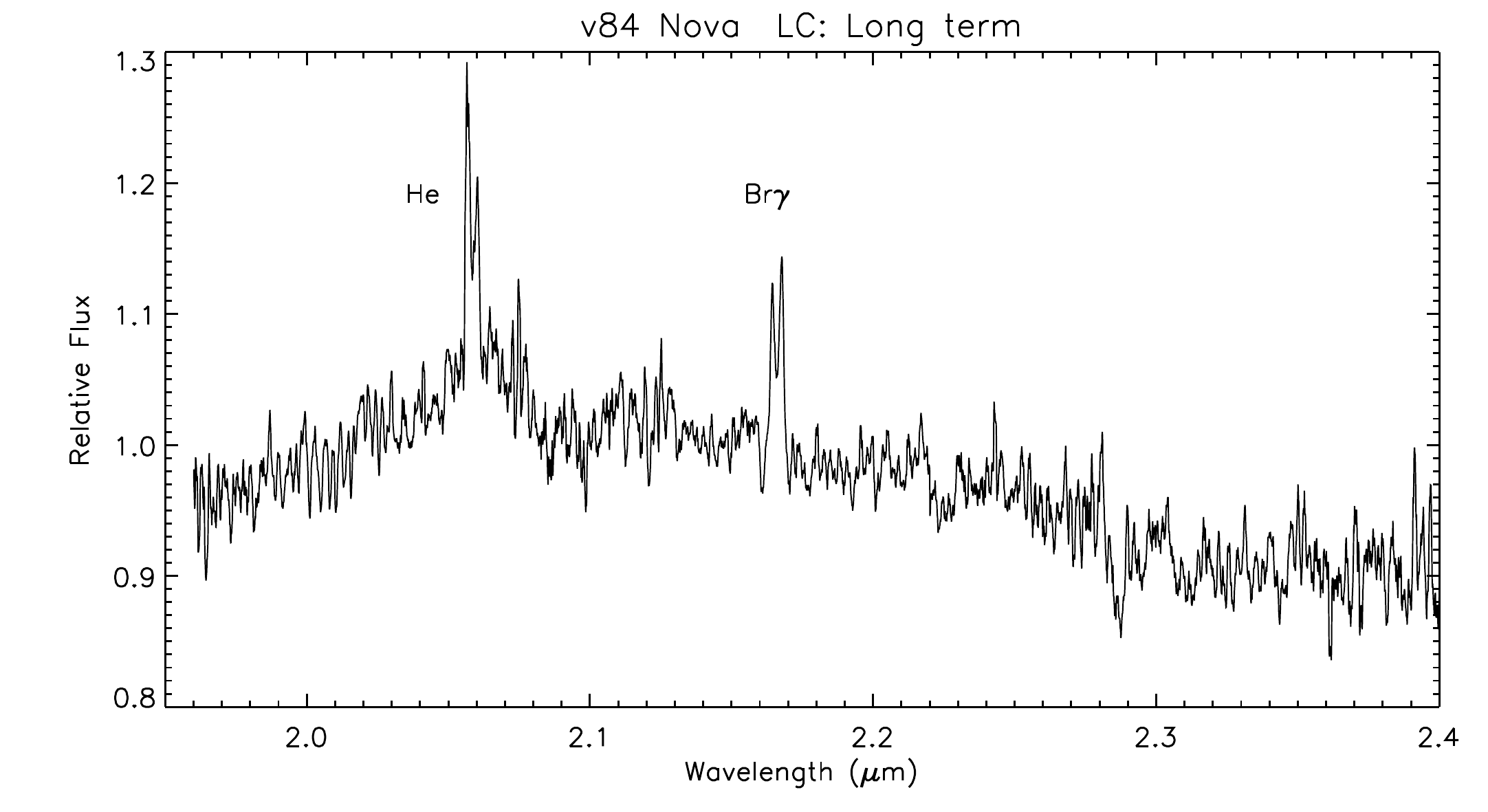}
\includegraphics[width=3.3in,angle=0]{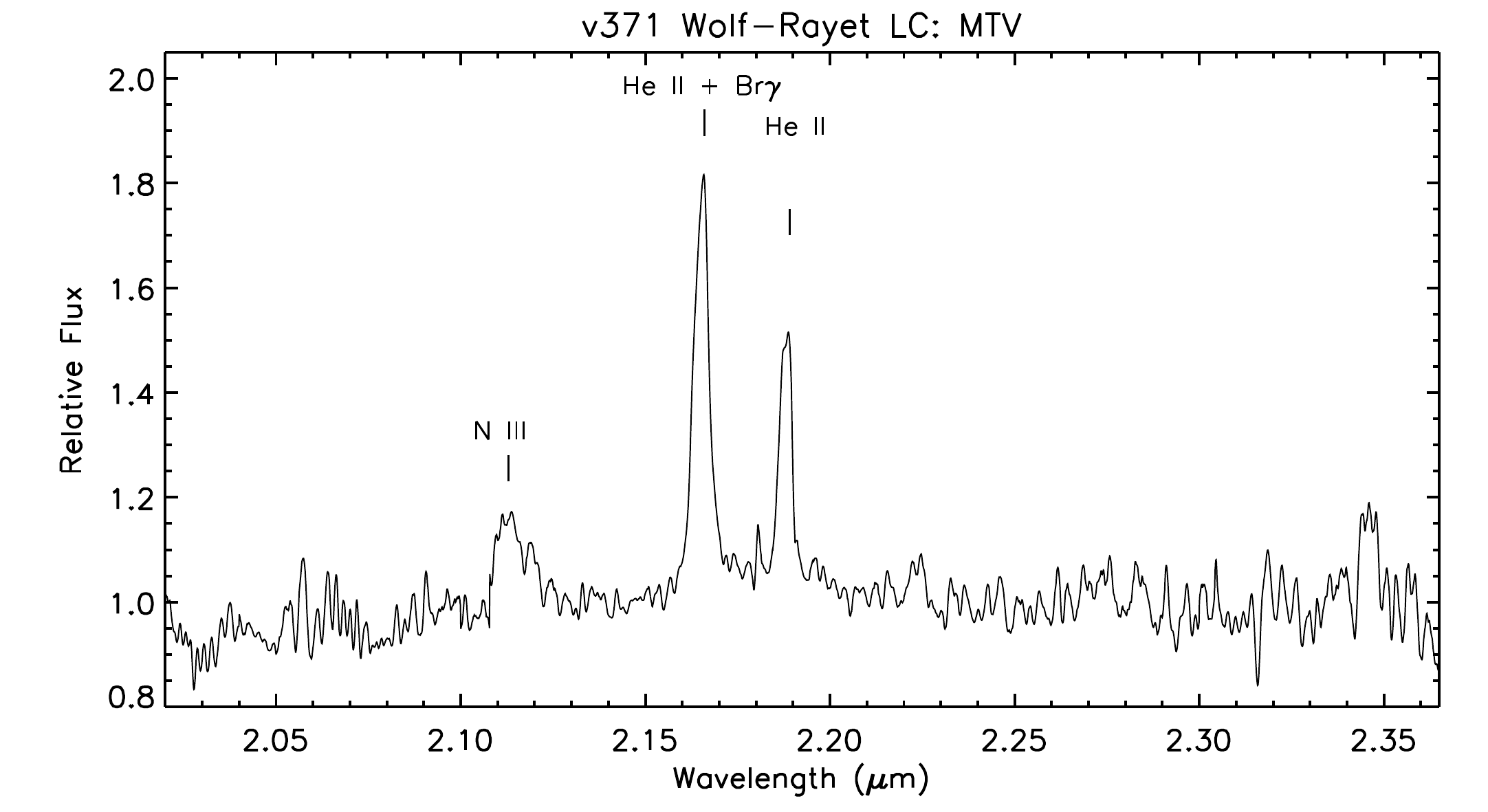}
\caption{YSOs with outflow dominated spectra and post-main-sequence objects.}
\end{figure*}

\end{document}